\documentclass[acmsmall, screen, nonacm]{acmart}

\title{Formulating Uncomputation Beyond Classical Ancilla Use}

\author{Bonan Su} 
\authornote{Work completed while visiting the University of Wisconsin-Madison.}
\orcid{0009-0009-7279-0658}
\email{sbn24@mails.tsinghua.edu.cn}
\affiliation{
    \institution{Tsinghua University}
    \city{Beijing}
    \country{China}
}

\author{Shengyuan Yang}
\email{syang686@wisc.edu}
\orcid{0009-0006-3298-7999}
\affiliation{
    \institution{University of Wisconsin-Madison}
    \city{Madison}
    \state{WI}
    \country{USA}
}

\usepackage{mathrsfs}
\usepackage{mathpartir}
\usepackage{extarrows}
\usepackage{cancel}
\usepackage{makecell}
\usepackage{stmaryrd}
\usepackage{lineno}
\usepackage{subcaption}
\usepackage{tcolorbox}
\usepackage{enumitem}
\usepackage{booktabs}
\usepackage{tabularx}
\usepackage{adjustbox}
\usepackage{longtable}
\usepackage[ruled,vlined,linesnumbered]{algorithm2e}
\usepackage{hyperref}
\usepackage{pgfplots}
\usepgfplotslibrary{groupplots}
\pgfplotsset{compat=1.18}
\newcolumntype{L}[1]{>{\raggedright\arraybackslash}p{#1}}

\usepackage{natbib}

\usepackage{amssymb}
\usepackage{listings}
\usepackage{tikz}
\usetikzlibrary{quantikz2}
\usetikzlibrary{arrows.meta,calc,decorations.pathreplacing,tikzmark,backgrounds,fit}

\newcommand{\bC}{\mathbb{C}}

\newcommand{\cS}{\mathcal{S}}

\newcommand{\cA}{\mathcal{A}}

\newcommand{\cL}{\mathcal{L}}

\newcommand{\dom}{\mathrm{dom}}

\newcommand{\Record}{\operatorname{Record}}
\newcommand{\Choice}{\operatorname{Choice}}
\newcommand{\State}{\operatorname{State}}

\newcommand{\cnorm}[1]{{\left\vert\kern-0.25ex\left\vert\kern-0.25ex\left\vert #1 \right\vert\kern-0.25ex\right\vert\kern-0.25ex\right\vert}}
\renewcommand{\>}{\rangle}
\newcommand{\<}{\langle}

\newcommand{\pare}[1]{\left ( #1 \right )}

\definecolor{darkgreen}{rgb}{0.0, 0.4, 0.0}

\newcommand{\bzero}{\mathbf{0}}

\newcommand{\ixRightarrowLabelGap}{0ex}
\newcommand{\ixRightarrowAboveSep}{0.2ex}
\makeatletter
\DeclareRobustCommand{\ixRightarrow}[1]{%
  \mathrel{\mathpalette{\ixarrow@overlay\Rightarrow\Rightarrowfill@}{#1}}%
}
\DeclareRobustCommand{\ixrightarrow}[1]{%
  \mathrel{\mathpalette{\ixarrow@overlay\rightarrow\rightarrowfill@}{#1}}%
}
\newcommand{\ixarrow@overlay}[4]{%
  \begingroup
    \setbox\z@=\hbox{$\m@th#3#1$}%
    \setbox\tw@=\hbox{$\m@th
      \ifx#3\displaystyle\scriptstyle
      \else\ifx#3\textstyle\scriptstyle
      \else\scriptscriptstyle\fi\fi #4$}%
    \setbox4=\hbox{$\m@th#3\mkern6mu$}%
    \dimen@=\wd\tw@
    \advance\dimen@ by\wd4
    \ifdim\dimen@>\wd\z@
      \setbox\z@=\hbox to\dimen@{#2#3}%
    \fi
    \dimen@=\ht\z@
    \advance\dimen@ by\dp\tw@
    \advance\dimen@ by\ixRightarrowLabelGap\relax
    \setbox4=\hbox{\ooalign{%
      \hfil\copy\z@\hfil\cr
      \hfil\raise\dimen@\copy\tw@\hfil\cr
    }}%
    \ifdim\dimexpr\ht\tw@+\dp\tw@\relax>\z@
      \advance\dimen@ by\ht\tw@
      \advance\dimen@ by\ixRightarrowAboveSep\relax
      \ifdim\dimen@>\ht4
        \ht4=\dimen@
      \fi
    \fi
    \dp4=\dp\z@
    \box4
  \endgroup
}
\makeatother

\usepackage{amsthm}
\usepackage{xcolor}
\usepackage{tcolorbox}
\tcbuselibrary{skins}
\tcbuselibrary{breakable}

\newtheorem{theorem}{Theorem}
\newtheorem{proposition}[theorem]{Proposition}
\newtheorem{lemma}[theorem]{Lemma}
\newtheorem{remark}[theorem]{Remark}

\theoremstyle{definition}
\newtheorem{definition}[theorem]{Definition}
\newtheorem{example}[theorem]{Example}

\definecolor{csRoleBlue}{HTML}{2B5C92}
\definecolor{csCleanOrange}{HTML}{A95116}
\definecolor{csPhaseGreen}{HTML}{267452}
\definecolor{csWrapGrey}{HTML}{717783}
\definecolor{ancillaRoleBeige}{HTML}{625440}

\tikzset{cscomp/.style={draw=csRoleBlue,dashed,fill=csRoleBlue!5},
 csuse/.style={draw=csPhaseGreen,dashed,fill=csPhaseGreen!7},
 csclean/.style={draw=csCleanOrange,line width=0.9pt,fill=csCleanOrange!8},
 cswrap/.style={draw=csWrapGrey,dashed,fill=csWrapGrey!5},
 cscomplabel/.style={text=csRoleBlue},csuselabel/.style={text=csPhaseGreen},
 cscleanlabel/.style={text=csCleanOrange},cswraplabel/.style={text=csWrapGrey}}

\newcommand{\csgateframe}[4]{%
 \gategroup[wires=#2,steps=#3,background,
 style={cs#1,rounded corners=2pt,inner xsep=-1.5pt,inner ysep=-1pt, yshift=0pt},
 label style={cs#1label,label position=above,anchor=south,yshift=-6pt}]{\footnotesize #4}}

\newcommand{\csgateframed}[4]{%
 \gategroup[wires=#2,steps=#3,background,
 style={cs#1,rounded corners=2pt,inner xsep=-1.5pt,inner ysep=-1pt, yshift=0pt},
 label style={cs#1label,label position=below,anchor=south,yshift=-19pt}]{\footnotesize #4}}

\NewDocumentEnvironment{csproof}{+b}{%
 \par\smallskip\begingroup\small
 \renewcommand{\arraystretch}{1.15}\setlength{\tabcolsep}{4pt}
 \noindent\begin{tabularx}{\linewidth}{@{}p{.46\linewidth}>{\raggedright\arraybackslash}X@{}}
 \toprule Judgment & Rules and extraction \\
 \midrule #1 \bottomrule
 \end{tabularx}\par\endgroup\smallskip}{}

\allowdisplaybreaks

\begin{document}

\begin{abstract}
    Ancilla qubits are routinely used to simplify quantum circuit construction and reduce implementation costs, but their use introduces the burden of designing proper uncomputation afterward. To alleviate this, prior work has developed automated uncomputation synthesis based on a computational-basis formulation of ancilla use.
    However, this formulation is tailored to classically structured ancilla computations. 
    For more general circuits, it may be undefined, yield unintended functionality, or fail to synthesize uncomputation. 
    It therefore falls short of providing a precise and general formulation of uncomputation beyond classical ancilla use.
    
    To address these limitations, we propose a new formulation of ancilla use that specifies the operation of each gate through two complementary aspects: (1) the functionality it implements and (2) the transformation it induces on ancilla roles.
    Based on this specification, we define uncomputation as restoring the initial ancilla role while implementing the identity functionality.
    This perspective enables a simple Use-first strategy for automatically synthesizing uncomputation in cases both within and beyond the scope of prior approaches.
    We further develop a complete inference system for reasoning about specifications, together with a derivation-directed procedure for automatically synthesizing efficient uncomputation, enabling systematic uncomputation synthesis even in challenging cases beyond the scope of our automatic strategy.
\end{abstract}

\maketitle

\section{Introduction}\label{sec:intro}
\emph{Ancilla qubits} are auxiliary qubits introduced alongside data qubits to simplify the implementation of quantum algorithms or reduce their cost.
Rather than carrying the primary input or output, they assist in realizing the desired functionality in various ways, for example by storing intermediate results~\cite{cuccaro-adder}, enabling phase kickback~\cite{Fourier-basis}, and coherently encoding spectral information~\cite{QSVT}.

However, unlike classical temporary variables, ancilla qubits cannot in general be directly discarded after computation: by the principle of implicit measurement~\cite[\S 4.4]{qcqi}, residual entanglement with the data qubits may compromise the correct realization of the intended functionality.
This necessitates the \emph{uncomputation} of ancilla qubits before they are discarded, which, intuitively, coherently restores them to their initial states without altering the functionality of the computation.

To relieve programmers of the additional burden of designing how ancilla qubits should be uncomputed after use, substantial prior work has investigated how to formulate and automatically synthesize uncomputation for quantum circuits and programs~\cite{silq,unqomp,reqomp,RwUn,qurts,modular}.
Across these works, the definition of uncomputation---which determines what uncomputation is considered correct and therefore what should be synthesized---largely follows a formulation rooted in Silq's notion of \texttt{qfree}-typed functions~\cite{silq}, tailored to \emph{classical ancilla use} as described below.

\subsection{Prior Formulation of Uncomputation for Classical Ancilla Use}
Specifically, given a unitary operator \(U\) representing the computation circuit, which acts on the joint system of the primary system \(\cS\) and the ancilla system \(\cA\), the corresponding uncomputation \(V\) is expected to be a unitary operator acting on the joint system and behaving as follows~\cite{silq,RwUn}:
\begin{equation}\label{eq:computation-uncomputation}
{\color{blue}|0\>_{\cA}}\tikzmarknode{eqone-input}{|\psi\>_{\cS}}\xmapsto[\text{unitary }U]{\text{computation}}
U\pare{|0\>_\cA|\psi\>_\cS}=
\sum_b\hspace{-2em}{\color{blue}\tikzmarknode{eqone-basis}{\underbrace{|b\>_{\cA}}_{\substack{\text{computational basis}\\
\text{of }\cA}}}}\hspace{-2em}|\psi_b\>_{\cS}
\xmapsto[\text{unitary }V]{\text{uncomputation}}{\color{blue}|0\>_{\cA}}\tikzmarknode{eqone-output}{\sum_b|\psi_b\>_{\cS}}.
\vphantom{\rule[-3.8em]{0pt}{3.8em}}
\begin{tikzpicture}[remember picture,overlay,draw=black,text=black,line width=0.6pt]
    \draw ([yshift=-2pt]eqone-input.south west) -- ([yshift=-2pt]eqone-input.south east);
    \draw ([yshift=-2pt]eqone-output.south west) -- ([yshift=-2pt]eqone-output.south east);
    \coordinate (eqone-function-start) at ([yshift=-2pt]eqone-input.south);
    \coordinate (eqone-function-end) at ([yshift=-5pt]eqone-output.south);
    \coordinate (eqone-function-level) at ([yshift=-5pt]eqone-basis.south);
    \draw[-{Latex[length=4pt,width=5pt]}]
        (eqone-function-start) -- (eqone-function-start |- eqone-function-level)
        -- node[pos=0,above right=2pt,font=\small\upshape,inner sep=0pt] {\scriptsize functionality \(W\)}
        (eqone-function-end |- eqone-function-level) -- (eqone-function-end);
\end{tikzpicture}
\end{equation}

\noindent
Operationally, for an arbitrary input state \(|\psi\>_\cS\) of the primary system \(\cS\), with the ancilla system \(\cA\) initialized to the ground state \(|0\>_\cA\), the uncomputation \(V\) is defined by first expressing the post-computation state \(U\pare{|0\>_\cA|\psi\>_\cS}\) in the computational basis of the ancilla, and then coherently erasing the ancilla.
The overall computation followed by uncomputation is then taken to implement the \emph{functionality} \(W:|\psi\>_\cS \mapsto \sum_b |\psi_b\>_\cS\) on the primary system~\cite{RwUn}.

For instance, consider the circuit in Fig.~\ref{fig:C3X}, which implements a three-controlled NOT gate (\(C^3X\)) using one ancilla qubit, with \(\cS=\{q_1,q_2,q_3,q_4\}\) and \(\cA=\{a\}\).
For a binary string \(b=(b_1,b_2,b_3,b_4)\in\{0,1\}^4\), define \(f(b)=(b_1,b_2,b_3,b_4\oplus (b_1\wedge b_2\wedge b_3))\in\{0,1\}^4\).
Then, for an arbitrary input state \(|\psi\>_\cS=\sum_b\lambda_b|b\>_\cS\), the computation and uncomputation are formulated as
{
\[
|0\>_a|\psi\>_\cS=
{\color{blue}|0\>_a}\sum_b\lambda_b|b\>_\cS
\xmapsto[\text{unitary }U]{\text{computation}}
\sum_b\lambda_b{\color{blue}|b_1\wedge b_2\>_a}|f(b)\>_\cS
\xmapsto[\text{unitary }V]{\text{uncomputation}}
{\color{blue}|0\>_a}\sum_b\lambda_b|f(b)\>_\cS,
\]}

\noindent
where, for every \(b\in\{0,1\}^4\), \(|b_1\wedge b_2\>_a\) is a computational-basis state of the ancilla qubit.
This procedure faithfully captures both the intended functionality of the \(C^3X\) gate on the primary system and the coherent restoration of the ancilla qubit.

\begin{figure}[t]
    \begin{minipage}{0.27\textwidth}
        \centering
        \begin{quantikz}[row sep={0.42cm,between origins},column sep=0.2cm]
            \lstick{\(q_1\)}    & \ctrl{2}\csgateframe{comp}{5}{2}{Computation \(U\)} & \qw      &\qw& \ctrl{2}\csgateframed{clean}{5}{1}{Uncomputation \(V\)} & \qw\\
            \lstick{\(q_2\)}    & \ctrl{1}                                            & \qw      &\qw& \ctrl{1} & \qw\\
            \lstick{\(a:|0\>\)} & \targ{}                                             & \ctrl{2} &\qw& \targ{} & \qw\rstick{\(|0\>\)}\\
            \lstick{\(q_3\)}    & \qw                                                 & \ctrl{1} &\qw& \qw & \qw\\
            \lstick{\(q_4\)}    & \qw                                                 & \targ{}  &\qw& \qw & \qw
        \end{quantikz}
        \vspace{-1em}
        \caption{Circuit for a \(C^3X\) gate with one ancilla qubit \(a\).}
        \label{fig:C3X}
    \end{minipage}
    \hspace{1em}
    \begin{minipage}{0.3\textwidth}
        \centering
        \hspace{-1.5em}
        \begin{quantikz}[row sep={0.52cm,between origins},column sep=0.4cm,wire types={q,b},classical gap=0.07cm]
            \lstick{\(a:|0\>\)}  & \gate{X}\gategroup[wires=2,steps=2,background,
 style={draw=csRoleBlue,dashed,fill=csRoleBlue!5,rounded corners=2pt,inner xsep=-1.5pt,inner ysep=1.5pt,yshift=-0.5pt},
 label style={cscomplabel,label position=above,anchor=south,yshift=-6pt}]{\footnotesize Computation \(U\)} & \gate{H} & \qw \\[0.3em]
            \lstick{\(\cS\)} & \qw & \qw & \qw
        \end{quantikz}
        \caption{A circuit with ancilla register \(\cA=\{a\}\) for which uncomputation for classical ancilla use does not exist, adapted from~\cite[Figure~6(a)]{unqomp}.}
        \label{fig:unqomp-toy}
    \end{minipage}
    \hspace{1em}
    \begin{minipage}{0.35\textwidth}
        \centering
        \begin{tikzpicture}[x=1.65cm, y=1.15cm, >=stealth, every node/.style={inner sep=2pt}]
            \node (top-left) at (-1.45,0.5) {$\cS$};
            \node (bottom-left) at (-1.45,-0.9) {$\cA\otimes \cS$};
            \node (top-right) at (0.6,0.5) {$\cS$};
            \node (bottom-right) at (0.6,-0.9) {$\cA\otimes \cS$};
            \draw[->] (top-left) -- node (R1label) [left] {\(R_1\)} (bottom-left);
            \draw[double=csPhaseGreen!4, double distance=0.8pt, ->] (top-left) -- node (Wlabel) [above] {$W$} (top-right);
            \draw[double=csRoleBlue!4, double distance=0.8pt, ->] (bottom-left) -- node (Ulabel) [below] {$U$} (bottom-right);
            \draw[->] (top-right) -- node (R2label) [right] {\(R_2\)} (bottom-right);

            \coordinate (intro-logical-separator) at (0,0.10);
            \coordinate (intro-physical-separator) at (0,-0.50);
            \coordinate (intro-view-center) at ($(top-left)!0.5!(top-right)$);
            \coordinate (intro-view-north-west) at ([xshift=-3pt]current bounding box.north west);
            \coordinate (intro-view-south-east) at ([xshift=3pt]current bounding box.south east);
            \begin{scope}[on background layer]
                \path[draw=none,fill=csPhaseGreen!5]
                    (intro-view-north-west)
                    rectangle (intro-view-south-east |- intro-logical-separator);
                \path[draw=none,fill=csCleanOrange!5]
                    (intro-view-north-west |- intro-logical-separator)
                    rectangle (intro-view-south-east |- intro-physical-separator);
                \path[draw=none,fill=csRoleBlue!5]
                    (intro-view-north-west |- intro-physical-separator)
                    rectangle (intro-view-south-east);
            \end{scope}
            \draw[dashed,csPhaseGreen,line width=0.8pt]
                (intro-view-north-west |- intro-logical-separator)
                -- (intro-view-south-east |- intro-logical-separator);
            \draw[dashed,csRoleBlue,line width=0.8pt]
                (intro-view-north-west |- intro-physical-separator)
                -- (intro-view-south-east |- intro-physical-separator);
            \node[anchor=south,yshift=2pt,text=csPhaseGreen,
                font=\footnotesize\itshape,inner sep=0pt]
                at (intro-view-center |- intro-logical-separator) {Logical functionality};
            \node[anchor=north,yshift=-2pt,text=csRoleBlue,
                font=\footnotesize\itshape,inner sep=0pt]
                at (intro-view-center |- intro-physical-separator) {Physical realization};
            \node[text=ancillaRoleBeige,font=\footnotesize\itshape,inner sep=0pt]
                at ($(intro-view-center |- intro-logical-separator)!0.5!
                      (intro-view-center |- intro-physical-separator)$) {Ancilla roles};

        \end{tikzpicture}
        \caption{Commutative diagram defining the specification \(U:R_1\ixRightarrow{W}R_2\), with double arrows denoting unitaries.}
        \label{fig:commutative-diagram}
    \end{minipage}
\end{figure}

\subsection{Limitation of Formulation for Classical Ancilla Use}\label{sec:intro-limitation}
The above formulation is primarily motivated by circuits implementing classical Boolean functions, which constitute a broad and practically important class of ancilla-assisted quantum algorithms.
Its limitation, however, is also clear: it assumes that ancillas store temporary intermediate classical information in the computational basis, and is therefore insufficient to capture circuits involving \emph{non-classical ancilla use}, as exemplified in the next paragraph.


A direct manifestation of this limitation is that uncomputation is not well-defined for certain circuits, despite the existence of natural and reasonable solutions, as already noted in \cite[Fig.~6]{unqomp} and \cite[\S~3]{qurts}.
Before examining more practical examples in Section~\ref{sec:limitation}, we illustrate this issue with the simple circuit in Fig.~\ref{fig:unqomp-toy}.
According to the formulation for classical ancilla use in Eq.~\eqref{eq:computation-uncomputation}, the uncomputation \(V\) is required to act as
\begin{equation}\label{eq:limitation}
{\color{blue}|0\>_\cA}|\psi\>_\cS\xmapsto[\text{unitary }U]{\text{computation}}
{\color{blue}|-\>_\cA}|\psi\>_\cS=
{\color{blue}|0\>_\cA}\frac{|\psi\>_\cS}{\sqrt{2}}+{\color{blue}|1\>_\cA}\frac{-|\psi\>_\cS}{\sqrt{2}}\xmapsto[\text{unitary }V]{\text{uncomputation}}{\color{blue}|0\>_\cA}\pare{\frac{|\psi\>_\cS}{\sqrt{2}}+\frac{-|\psi\>_\cS}{\sqrt{2}}}=\bzero.
\end{equation}

However, no such unitary \(V\) exists.
This does not imply that a physically meaningful uncomputation is impossible: simply reversing the operation on the ancilla restores it while implementing the identity on the primary system.
Rather, the failure arises from the formulation itself, which assumes classical ancilla use and defines the induced functionality on the primary system by coherently summing the corresponding primary-system states over the computational basis of the ancilla.


\subsection{Research Questions and Overview of Our Approach}\label{sec:intro-approach}
To address the limitations of prior formulation, two natural research questions arise:
\begin{enumerate}
    \item \textbf{Formulation:} What constitutes a proper formulation of uncomputation for quantum circuits involving non-classical ancilla use?
    \item \textbf{Synthesis:} How can uncomputation be synthesized for such circuits under this formulation?
\end{enumerate}

\subsubsection{Our Answer to the Formulation Question}
We revisit the fundamental meaning of uncomputation: intuitively, it should restore the ancilla qubits to their initial ground state without altering the intended functionality. 
Prior work has largely focused on the former requirement, while addressing the latter by imposing a uniform interpretation of functionality based on classical ancilla use, which may consequently leave uncomputation or functionality undefined.


Based on this insight, the central idea of our formulation is to \emph{specify} the operation of a unitary \(U\) on \(\cA\otimes \cS\) from two complementary perspectives:
(1) what functionality it implements given the current state of ancillas, and
(2) how it transforms the state of the ancilla system. That is,
\begin{center}
    \boxed{
        \begin{aligned}
            \!\!
            \begin{array}{c}
                \text{A specification of }U,\\[.3em]
                \text{denoted by }U:R_1\ixRightarrow{W}R_2
            \end{array}
            \!\!\mathrel{\scalebox{1.5}[1.5]{$=$}}
            \begin{array}{c}
                \text{Implementing functionality }W\text{ under the current ancilla role }R_1\\[-.2em]
                +\\[-.2em]
                \text{Transforming the ancilla role from }R_1\text{ to }R_2
            \end{array}
        \end{aligned}
    }
\end{center}
Here, the \emph{functionality} \(W\) is a unitary operator acting on the primary system \(\cS\), representing the logical operation we intend to implement, while the \emph{ancilla roles} \(R_1,R_2\) are isometries from \(\cS\) to the joint system \(\cA\otimes\cS\).
The specification is valid if and only if the diagram in Fig.~\ref{fig:commutative-diagram} commutes.
We next explain the intuition underlying this notion and the specification.

\paragraph{Ancilla Role}
The ancilla role is a central notion in our formulation, abstracting different forms of ancilla use and capturing how ancillas participate in computation.
It characterizes how states of the primary system are extended with the ancilla system and embedded into the joint system.

For example, the ancilla system typically begins its participation in computation in the ground state, characterized by the ancilla role \(R=|0\>_\cA\otimes I_\cS\), which maps a primary-system state \(|\psi\>_\cS\) to \(R|\psi\>_\cS=|0\>_\cA\otimes|\psi\>_\cS\).
A more interesting example is \(R=\frac{1}{\sqrt{2}}(|0\>_\cA\otimes I_\cS+|1\>_\cA\otimes G_\cS)\) for some unitary \(G\) on \(\cS\).
Intuitively, this role characterizes an ancilla qubit coherently controlling the application of \(G\) to the primary system, creating entanglement between the ancilla and the primary system.

\paragraph{Specification} 
The commutativity in Fig.~\ref{fig:commutative-diagram} defines the validity of a specification. 
It separates the underlying \textcolor{csPhaseGreen}{logical functionality} \(W\) from its \textcolor{csRoleBlue}{physical realization} \(U\), with ancilla roles bridging the gap between logical intentions and physical implementations.
Depending on the form of ancilla use, as captured by \textcolor{ancillaRoleBeige}{ancilla roles} \(R_1\) and \(R_2\), the same unitary \(U\) may admit different valid specifications.

The intuition behind this commutativity is as follows.
A programmer intends to implement a functionality \(W\) on the primary system \(\cS\), but realizing it directly as a quantum circuit acting solely on \(\cS\) may be difficult or even intractable.
By introducing an ancilla system prepared in an appropriate role \(R_1\), the same functionality can instead be realized through a more readily implementable ancilla-assisted circuit \(U\) acting on the joint system.
Executing \(U\) then simultaneously implements the intended functionality \(W\) on the primary system and transforms the ancilla role from \(R_1\) to \(R_2\).

\paragraph{Uncomputation}
Separating functionality from ancilla-role transformation brings two key benefits.
First, it yields a more general formulation of \emph{uncomputation} that accommodates non-classical ancilla use: an uncomputation \(V\) is defined with respect to an ancilla role \(R\), rather than a computation circuit \(U\) as in prior work.

This is motivated by two considerations:
(1) Intuitively, uncomputation concerns only the current role of the ancilla system, regardless of how this role was established or what functionality \(W\) was previously implemented by \(U\);
and (2) Formally, defining uncomputation with respect to \(U\) would implicitly require \(U\) to determine the functionality \(W\) as well, yet no single strategy for inferring \(W\) from \(U\) can accommodate all forms of ancilla use.
For example, prior work adopts classical ancilla use as a uniform paradigm, effectively fixing \(W=\pare{\sum_b\<b|_\cA\otimes I_\cS}U\pare{|0\>_\cA\otimes I_\cS}\) as in Eq.~\eqref{eq:computation-uncomputation}, thereby giving rise to the ill-definedness limitation discussed above.

Motivated by these considerations, we define a valid uncomputation \(V\) as a unitary that restores the ancilla system while implementing the identity functionality, thereby leaving the primary system unaffected.
Formally, this is expressed by the following specification:
\begin{center}
    \boxed{
        \begin{aligned}
            \!\!
            \begin{array}{c}
                \text{An uncomputation \(V\) for ancilla role }R\\[.3em]
                \text{admits a specification }V:R\ixRightarrow{I}|0\>_\cA\otimes I_\cS
            \end{array}
            \mathrel{\scalebox{1.5}[1.5]{$=$}}
            \begin{array}{c}
                \text{Implementing the trivial functionality }W=I\\[-.2em]
                +\\[-.2em]
                \text{Restoring the ancilla role from }R\text{ to }|0\>_\cA\otimes I_\cS
            \end{array}
        \end{aligned}
    }
\end{center}
For the example in Eq.~\eqref{eq:limitation}, the computation admits a specification \(U:|0\>_\cA\otimes I_\cS\ixRightarrow{I}|-\>_\cA\otimes I_\cS\).
Any unitary \(V\) admitting \(V:|-\>_\cA\otimes I_\cS\ixRightarrow{I}|0\>_\cA\otimes I_\cS\) is then a valid uncomputation.

\begin{figure}[t]
\centering
\begingroup
\thickmuskip=6mu\relax
\[
\tikzmarknode{intro-chain-start}{R_0}
\mathrel{\tikzmarknode{intro-chain-1-2}{\xRightarrow{\ I\ }}} \tikzmarknode{intro-chain-1-3}{R_1}
\xRightarrow{W_1} \tikzmarknode{intro-chain-1-5}{R_1}
\xRightarrow{\ I\ }\tikzmarknode{intro-prepare-r2}{R_2}
\,\hspace{2pt}\tikzmarknode{intro-chain-1-7}{\ \cdots\ }\hspace{2pt}
\tikzmarknode{intro-prepare-rnm1}{R_{n-1}}
\xRightarrow{\ I\ } \tikzmarknode{intro-chain-1-9}{R_n}
\xRightarrow{W_n}\tikzmarknode{intro-chain-1-11}{R_n}
\mathrel{\tikzmarknode{intro-clean-arrow}{\xRightarrow{\ I\ }}}\tikzmarknode{intro-clean-rnm1}{R_{n-1}}
\,\tikzmarknode{intro-chain-1-15}{\cdots}
\,\tikzmarknode{intro-clean-r2}{R_2}
\xRightarrow{\ I\ }\tikzmarknode{intro-chain-1-17}{R_1}
\mathrel{\tikzmarknode{intro-chain-1-18}{\xRightarrow{\ I\ }}}\tikzmarknode{intro-clean-end}{R_0}.
\]
\noindent\begin{tikzpicture}[
    remember picture,overlay,
    intro stage frame/.style={rounded corners=2pt,blend mode=multiply,draw opacity=0.6,inner sep=0pt,outer sep=0pt},
    intro stage label/.style={font=\footnotesize,anchor=south,inner sep=0pt,text height=1.5ex,text depth=0.25ex},
    intro stage value/.style={font=\footnotesize,anchor=north,inner sep=0pt,text height=2ex,text depth=0.6ex}
]
    \begin{scope}[on background layer]
        \node[fit=(intro-chain-start)(intro-chain-1-2)(intro-chain-1-3)
                (intro-chain-1-5)(intro-prepare-r2)(intro-prepare-rnm1)(intro-chain-1-9)(intro-chain-1-11)
                (intro-clean-arrow)(intro-clean-rnm1)(intro-clean-r2)(intro-chain-1-17)(intro-chain-1-18)(intro-clean-end),
            inner xsep=0pt,inner ysep=3pt,outer sep=0pt] (intro-stage-bounds) {};

        \node[cscomp,intro stage frame,
            fit={([xshift=-3pt]intro-chain-start.west |- intro-stage-bounds.north)
                 ([xshift=1pt]intro-chain-1-3.east |- intro-stage-bounds.south)}]
            (intro-prepare-one) {};
        \node[csuse,intro stage frame,
            fit={([xshift=1pt]intro-prepare-one.north east)
                 ([xshift=1pt]intro-chain-1-5.east |- intro-stage-bounds.south)}]
            (intro-use-two) {};
        \node[cscomp,intro stage frame,
            fit={([xshift=1pt]intro-use-two.north east)
                 ([xshift=1pt]intro-prepare-r2.east |- intro-stage-bounds.south)}]
            (intro-prepare-three) {};
        \node[cscomp,intro stage frame,
            fit={([xshift=-1pt]intro-prepare-rnm1.west |- intro-stage-bounds.north)
                 ([xshift=1pt]intro-chain-1-9.east |- intro-stage-bounds.south)}]
            (intro-prepare-n-minus-one) {};
        \node[csuse,intro stage frame,
            fit={([xshift=1pt]intro-prepare-n-minus-one.north east)
                 ([xshift=1pt]intro-chain-1-11.east |- intro-stage-bounds.south)}]
            (intro-use-n) {};
        \node[csclean,intro stage frame,
            fit={([xshift=1pt]intro-use-n.north east)
                 ([xshift=2pt]intro-clean-end.east |- intro-stage-bounds.south)}]
            (intro-clean-frame) {};
    \end{scope}
    \node[cscomplabel,intro stage label]
        at ([yshift=2pt]intro-prepare-one.north) {{Prepare}};
    \node[csuselabel,intro stage label]
        at ([yshift=2pt]intro-use-two.north) {{Use}};
    \node[cscomplabel,intro stage label]
        at ([yshift=2pt]intro-prepare-three.north) {{Prepare}};
    \node[cscomplabel,intro stage label]
        at ([yshift=2pt]intro-prepare-n-minus-one.north) {{Prepare}};
    \node[csuselabel,intro stage label]
        at ([yshift=2pt]intro-use-n.north) {{Use}};
    \node[cscleanlabel,intro stage label]
        at ([yshift=2pt]intro-clean-frame.north) {Uncomputation = \(\text{{\textcolor{csRoleBlue}{Prepare}}}^{\textcolor{black}{\dagger}}\)};
    \node[cscomplabel,intro stage value]
        at ([yshift=-1pt]intro-prepare-one.south) {$U_1$};
    \node[csuselabel,intro stage value]
        at ([yshift=-1pt]intro-use-two.south) {$U_2$};
    \node[cscomplabel,intro stage value]
        at ([yshift=-1pt]intro-prepare-three.south) {$U_3$};
    \node[text=black,intro stage value]
        at ([yshift=-1pt]intro-chain-1-7.center |- intro-stage-bounds.south) {$\cdots$};
    \node[cscomplabel,intro stage value]
        at ([yshift=-1pt]intro-prepare-n-minus-one.south) {$U_{n-1}$};
    \node[csuselabel,intro stage value]
        at ([yshift=-1pt]intro-use-n.south) {$U_n$};
    \node[cscleanlabel,intro stage value]
        at ([yshift=-1pt]intro-clean-frame.south) {$V=U_{n-1}^\dagger;\cdots ;U_3^\dagger;U_1^\dagger$};
\end{tikzpicture}\par
\endgroup
\caption{Specify gates as {\textcolor{csRoleBlue}{Prepare}} or {\textcolor{csPhaseGreen}{Use}}, and reverse the {\textcolor{csRoleBlue}{Prepare}} portion to generate uncomputation.}
\label{fig:uncomputation}
\end{figure}
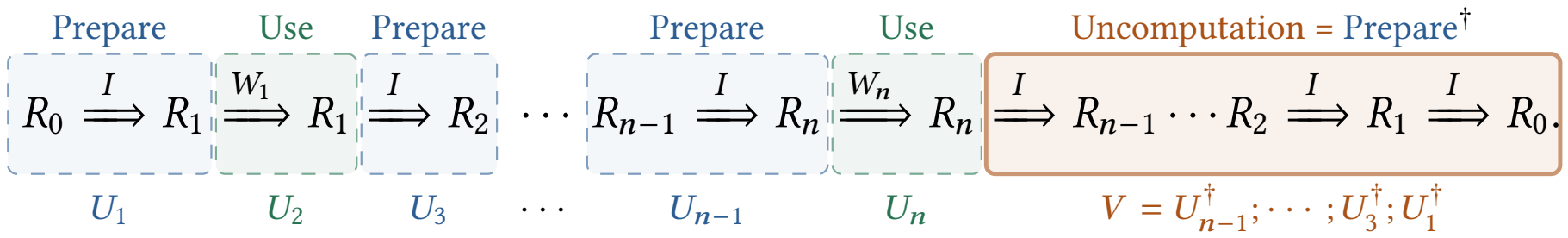

\subsubsection{Our Answer to the Synthesis Question.}
The second benefit of separating functionality from ancilla-role transformation is that it naturally motivates a principled strategy for uncomputation synthesis: reversing only the ancilla-role transformation while preserving the implemented functionality, thereby restoring the ancilla system without affecting the primary system.

This strategy operates by automatically deriving specifications for computation circuits.
More specifically, we identify two forms of specifications that characterize gate behaviors frequently encountered in practical ancilla-assisted circuits:
\begin{enumerate}
\item {\color{csRoleBlue}Prepare} the ancilla. \(U:R_1\ixRightarrow{I}R_2\): changes the ancilla role, implementing trivial functionality;
\item {\color{csPhaseGreen}Use} the ancilla. \(U:R\ixRightarrow{W}R\): preserves the ancilla role, implementing nontrivial functionality.
\end{enumerate}
Intuitively, each gate in an ancilla-assisted circuit is preferably specified to {\color{csPhaseGreen}Use} the ancilla to implement the intended functionality, unless the current ancilla role is inappropriate, necessitating a {\color{csRoleBlue}Prepare} operation.
This naturally leads to a simple \emph{Use-first} strategy for automatic uncomputation synthesis: we examine each gate in the computation circuit sequentially, classifying it as {\color{csPhaseGreen}Use} if the corresponding specification is valid, and as {\color{csRoleBlue}Prepare} otherwise.
Uncomputation is then constructed by reversing the {\color{csRoleBlue}Prepare} portion, as illustrated in Fig.~\ref{fig:uncomputation}.
For instance, in the \(C^3X\) circuit shown in Fig.~\ref{fig:C3X}, the first and second gates are classified as {\color{csRoleBlue}Prepare} and {\color{csPhaseGreen}Use}, respectively.
Thus, uncomputation is obtained by simply reversing the first gate.

Despite its simplicity, this strategy is remarkably effective.
We construct a benchmark suite combining benchmarks from prior work~\cite{reqomp,unqomp,RwUn} with additional general and practical ancilla-assisted circuits.
Prior automatic tools successfully solve 12 of the 19 cases, with all successful cases drawn from existing benchmarks.
In comparison, our strategy demonstrates broad applicability to existing benchmarks while extending to more general cases, successfully solving 17 of the 19 cases.

Furthermore, to enable complete reasoning about specifications and systematic uncomputation synthesis beyond the capabilities of the automatic strategy, we develop a sound and complete inference system for specifications, establishing the equivalence between validity and derivability:
\[
U:R_1\ixRightarrow{W}R_2 \mbox{ is valid} \iff U:R_1\ixrightarrow{W}R_2 \mbox{ is derivable in the inference system}.
\]
Derivations can serve as a formal language for programmers to annotate their ancilla-assisted circuits, making the intent of each gate explicit and thereby guiding uncomputation synthesis.

We further develop an automatic procedure that recursively constructs an efficient uncomputation from a given derivation, with circuit size linear in the length of the derivation.
This derivation-directed approach handles challenging cases beyond the capabilities of both prior work and our preceding strategy, demonstrating its generality and systematic nature.

\subsection{Contributions and Organization}
The remainder of this paper is organized around our main contributions as follows.
\begin{itemize}
   \item[\S~\ref{sec:limitation}] \textbf{Contribution 1}: We identify, through practical examples, three levels of limitations in prior formulations and synthesis approaches for uncomputation tailored to classical ancilla use: uncomputation may be undefined, may alter the intended functionality, or may remain beyond the capabilities of existing synthesis tools.
    \item[\S~\ref{sec:formulation}] \textbf{Contribution 2}: We introduce the notion of ancilla roles and develop a novel formulation that specifies gates in terms of functionality implementation and ancilla-role transformation, characterizing uncomputation as restoring the original ancilla role while implementing the identity functionality.
    \item[\S~\ref{sec:strategy}] \textbf{Contribution 3}: We develop a Use-first strategy to automatically synthesize uncomputation for circuits involving non-classical ancilla use, and evaluate its effectiveness on a benchmark suite comprising examples from prior work and practical ancilla-assisted quantum circuits.
    The results demonstrate that our strategy handles a broader and more general class of uncomputation synthesis problems than prior approaches.
    \item[\S~\ref{sec:reasoning}] \textbf{Contribution 4}: We develop a sound and complete inference system for specifications, together with an automatic, derivation-directed procedure for systematic uncomputation synthesis, and apply this approach to challenging cases beyond the capabilities of the preceding strategy.
\end{itemize}
Finally, we discuss related work in \S~\ref{sec:rw}.

\section{Three Levels of Limitations in Prior Uncomputation Formulations}\label{sec:limitation}

In this section, we continue the discussion from Section~\ref{sec:intro-limitation} by using practical examples to identify three levels of limitations in prior formulations of uncomputation for classical ancilla use, and analyze the fundamental causes underlying these limitations.

\begin{figure}[t]
    \begin{minipage}{0.54\textwidth}
        \centering
        \tikzset{
            phase gate label/.style={font=\small},
            phase gate box/.style={inner ysep=-2pt},
            phase gate middle/.style={phase gate box,yshift=-0.6pt}
        }
        \newcommand{\phasegatetext}[1]{\kern-2pt #1\kern-2pt}
        \begin{adjustbox}{max width=\linewidth}
        \begin{quantikz}[row sep={0.60cm,between origins},column sep=0.24cm,wire types={b,q,q,q},classical gap=0.07cm]
        \lstick{\(\cS\)} & \qw\csgateframe{comp}{4}{3}{Computation \(U\)} & \qw & \gate[wires=4]{\mathrm{ADD}} & \qw\csgateframe{clean}{4}{2}{Uncomputation \(V\)} & \qw  &\qw \\
        \lstick{\(|0\>\)}\lstick[wires=3,braces={transform canvas={xshift=-1.3em}},label style={xshift=-1.1em}]{\(\cA\)} & \gate{H} & \gate[style={phase gate box},label style={phase gate label}]{\phasegatetext{P\pare{-\frac{\pi}{4}}}} & & \gate[style={phase gate box},label style={phase gate label}]{\phasegatetext{P\pare{\frac{\pi}{4}}}} & \gate{H} & \qw\rstick{\(|0\>\)} \\
        \lstick{\(|0\>\)} & \gate{H} & \gate[style={phase gate middle},label style={phase gate label}]{\phasegatetext{P\pare{-\frac{\pi}{2}}}} & & \gate[style={phase gate middle},label style={phase gate label}]{\phasegatetext{P\pare{\frac{\pi}{2}}}} & \gate{H} & \qw \rstick{\(|0\>\)} \\
        \lstick{\(|0\>\)} & \gate{H} & \gate[style={phase gate box},label style={phase gate label}]{\phasegatetext{P\pare{-\pi}}}           & & \gate[style={phase gate box},label style={phase gate label}]{\phasegatetext{P(\pi)}}   & \gate{H} & \qw \rstick{\(|0\>\)}
        \end{quantikz}
        \end{adjustbox}
        \caption{Phase kickback circuit using a Fourier-basis ancilla, adapted from~\cite[Figure 2.]{Fourier-basis}.}
        \label{fig:Fourier-basis}
    \end{minipage}
    \hspace{2em}
    \begin{minipage}{0.35\textwidth}
        \begin{quantikz}[row sep={0.60cm,between origins},column sep=0.24cm]
            \lstick[wires=2,braces={decoration={pre=moveto,pre length=2.5pt,post=moveto,post length=2.5pt}}]{\(\cS\)} & \ctrl{2}\csgateframe{comp}{3}{2}{Computation \(U\)} & \qw & \ctrl{2}\csgateframed{clean}{3}{1}{Uncomputation \(V\)} &  \\
            & \ctrl{1} & \qw & \ctrl{1} &  \\
            \lstick{\(\cA:|0\>\)} & \gate{-iX} & \gate{T}& \gate{iX} & \rstick{\(|0\>\)}
        \end{quantikz}
        \caption{A controlled-\(T\) circuit using one ancilla qubit, adapted from~\cite[Eq.~(14)]{controlled-T}.}
        \label{fig:controlled-T}
    \end{minipage}
\end{figure}
\paragraph{Limitation 1: The uncomputation and functionality are not defined, even when a reasonable solution exists.}
In addition to the toy circuit in Fig.~\ref{fig:unqomp-toy}, a practical example exhibiting this limitation is phase kickback using a Fourier-basis ancilla~\cite{Fourier-basis}.
Fig.~\ref{fig:Fourier-basis} shows an example instance with both \(\cS\) and \(\cA\) comprising three qubits, where
\[
P(\theta)=|0\>\<0|+e^{i\theta}|1\>\<1|\mbox{ is the phase gate, and }
\operatorname{ADD}:|x\>_\cS|y\>_\cA\mapsto|x\>_\cS|y+x\mbox{ mod }8\>_\cA.
\]
It can be verified that the first two layers of gates prepare the ancilla system in a Fourier-basis state \(|\gamma\>_\cA\), while the computation portion of the circuit implements the following transformation:
\[
{\color{blue}|0\>_\cA}|\psi\>_\cS\xmapsto[\text{unitary }U]{\text{Computation}} {\color{blue}|\gamma\>_\cA} W|\psi\>_\cS,\quad\mbox{where }
|\gamma\>_\cA=\frac{1}{\sqrt{8}}\sum_{y=0}^{7}e^{-\frac{i\pi y}{4}}|y\>_\cA,\mbox{ and }W=\sum_{x=0}^7e^{\frac{i\pi x}{4}}|x\>\<x|.
\]
Since the primary and ancilla systems remain unentangled after computation, a natural and valid uncomputation that preserves the intended functionality is simply to reverse the first two layers of preparation gates, as illustrated in Fig.~\ref{fig:Fourier-basis}.

However, under the formulation for classical ancilla use defined in Eq.~\eqref{eq:computation-uncomputation}, the uncomputation \(V\) would be required to act as
\[
{\color{blue}|\gamma\>_\cA} W|\psi\>_\cS = \sum_{y=0}^7 {\color{blue}|y\>_\cA} \pare{\frac{e^{-\frac{i\pi y}{4}}}{\sqrt{8}}W|\psi\>_\cS} \xmapsto[\text{unitary }V]{\text{uncomputation}} {\color{blue}|0\>_\cA} \pare{\sum_{y=0}^7 \frac{e^{-\frac{i\pi y}{4}}}{\sqrt{8}}W|\psi\>_\cS } =\bzero,
\]
which is impossible for any unitary \(V\).
This exposes a fundamental limitation of prior formulations: they assume a fixed mode of ancilla use, namely storing classical information in the computational basis, thereby failing to accommodate other legitimate forms of ancilla use, such as phase kickback using Fourier-basis states in this example.

\paragraph{Limitation 2: Even when uncomputation is well-defined, it alters the intended functionality.}
A practical example illustrating this limitation is shown in Fig.~\ref{fig:controlled-T}, where a controlled-\(T\) gate is implemented on a two-qubit primary system \(\cS\) using a single ancilla qubit~\cite{controlled-T}.

According to the formulation of uncomputation for classical ancilla use in Eq.~\eqref{eq:computation-uncomputation}, for an arbitrary primary-system state \(|\psi\>_\cS=\alpha_0|00\>_\cS+\alpha_1|01\>_\cS+\alpha_2|10\>_\cS+\alpha_3|11\>_\cS\), the computation \(U\) and uncomputation \(V\) are expected to act as follows:
\vspace{-.5em}
\[
{\color{blue}|0\>_\cA}|\psi\>_\cS\xmapsto[\text{unitary }U]{\text{computation}}{\color{blue}|0\>_\cA}\pare{\begin{smallmatrix}\alpha_0|00\>_\cS+\\ \alpha_1|01\>_\cS+\\ \alpha_2|10\>_\cS\ \,\end{smallmatrix}}+{\color{blue}|1\>_\cA}\pare{e^{\frac{7\pi i}{4}}\alpha_3|11\>_\cS}\xmapsto[\text{unitary }V]{\text{uncomputation}}{\color{blue}|0\>_\cA}\pare{\begin{smallmatrix*}[r]\alpha_0|00\>_\cS+\\ \alpha_1|01\>_\cS+\\ \alpha_2|10\>_\cS+\\ e^{7\pi i/4}\alpha_3|11\>_\cS\ \,\end{smallmatrix*}}.
\]

Unlike the previous example, the uncomputation here is well-defined and can be implemented by a Toffoli gate targeting the ancilla qubit.
However, the resulting circuit implements a controlled-\(T^\dagger\) gate rather than the intended controlled-\(T\) gate.
Thus, although the formulation for classical ancilla use in Eq.~\eqref{eq:computation-uncomputation} admits a valid uncomputation, it fails to preserve the intended functionality.

This exposes another fundamental limitation of prior formulations: they do not explicitly account for phase operations on the ancilla, implicitly attributing all resulting phases to the intended functionality.
In this example, the prescribed uncomputation consists solely of a Toffoli gate, without any phase correction, thereby leaving the phase arising from operations on the ancilla incorrectly incorporated into the functionality of the overall circuit. 

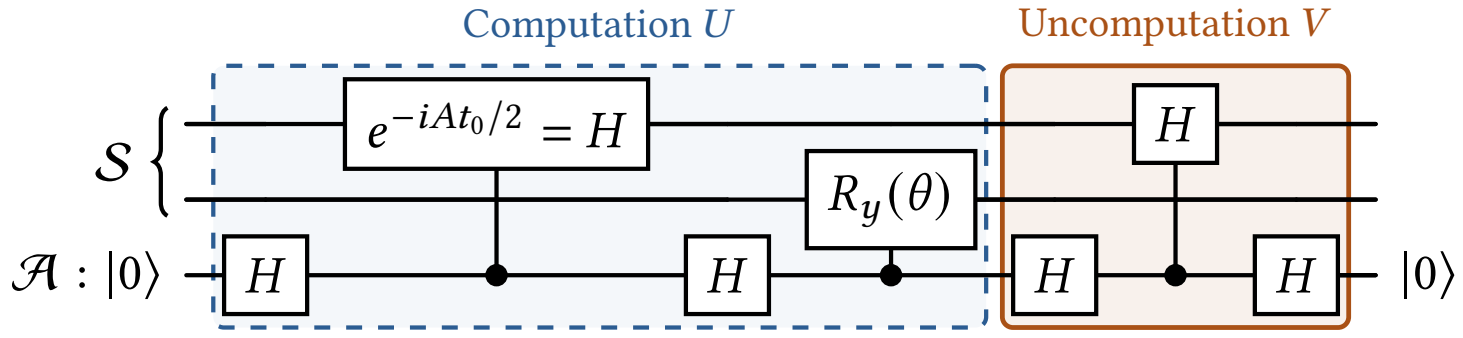
\begin{figure}[t]
    \centering
    \begin{quantikz}[row sep={0.50cm,between origins},column sep=0.25cm]
        \lstick[2]{\(\cS\)} & \qw\csgateframe{comp}{3}{4}{Computation \(U\)} & \gate{e^{-iA t_0/2}=H} & \qw & \qw & \qw\csgateframe{clean}{3}{3}{Uncomputation \(V\)} & \gate{H} & \qw & \qw \\
             & \qw & \qw & \qw & \gate{R_y(\theta)} & \qw & \qw & \qw & \qw \\
        \lstick{\(\cA:\ket{0}\)} & \gate{H} & \ctrl{-2} & \gate{H} & \ctrl{-1} & \gate{H} & \ctrl{-2} & \gate{H} & \qw\rstick{\(\ket{0}\)}
    \end{quantikz}
    \caption{The HHL circuit in the three-qubit case, adapted from~\cite[Fig.~6]{single-clock-HHL}, with the final measurement omitted.
    Parameters are set to \(A=(5I-H)/2\) following~\cite[Eq.~(22)]{single-clock-HHL}, which, together with the phase-estimation convention \(t_0=2\pi\) adopted in the source, yields \(e^{iAt_0/2}=H\) and \(\theta=-2\arccos(2/3)\).}
    \label{fig:HHL}
\end{figure}

\paragraph{Limitation 3: Even when uncomputation is well-defined and preserves the intended functionality, existing tools fail to synthesize it.}
Existing automated uncomputation tools~\cite{silq,qurts,reqomp,unqomp,RwUn} largely rely on the analysis of classical reversible functions under a formulation tailored to classical ancilla use in the computational basis.
Specifically, they analyze circuits by assuming the existence of classical Boolean functions \(f\) such that
\begin{equation}\label{eq:Bennett}
|0\>_\cA|x\>_\cS
\xmapsto[\text{unitary }U]{\text{computation}}
e^{i\theta_x}|f(x)\>_\cA|\psi_x\>_\cS,
\quad
\mbox{for all computational-basis states \(|x\>_\cS\).\footnotemark}
\end{equation}
In other words, the computation must map each computational-basis state of the ancilla system to another, without introducing superposition for any computational-basis input.
Uncomputation is then synthesized using techniques such as dependency
tracking and circuit rewriting to identify and reverse classically
reversible ancilla updates.
Consequently, these approaches remain within the scope of Bennett's trick~\cite{Bennett}, improving its efficiency without extending its applicability.
\footnotetext{
We allow input-dependent phases as a simple extension, since they can be ignored during classical analysis and incorporated into the functionality, as discussed above.
Some prior approaches do not permit such phases~\cite{reqomp,unqomp}.
}

This restriction becomes apparent when Hadamard gates on the ancilla create a coherent superposition of branches associated with different states of the primary system, for example,
\[
{\color{blue}|0\>_\cA}|x\>_\cS
\xmapsto[\text{unitary }U]{\text{computation}}
\frac{1}{\sqrt{2}}{\color{blue}|0\>_\cA}|\psi_0\>_\cS
+
\frac{1}{\sqrt{2}}{\color{blue}|1\>_\cA}|\psi_1\>_\cS.
\]
A practical example is the three-qubit HHL circuit shown in Fig.~\ref{fig:HHL}, in which the ancilla coherently controls unitary operations on the primary system.
Although its uncomputation is well-defined under the formulation for classical ancilla use in Eq.~\eqref{eq:computation-uncomputation} and realizes the desired functionality, existing tools cannot synthesize it.
The obstacle lies not in the existence or correctness of uncomputation, but in the reliance on the analysis of classical reversible functions, which cannot capture the non-classical use of ancillas as coherent quantum controls rather than classical information registers.


\section{Our Formulation: From Specification to Uncomputation}\label{sec:formulation}

To address the limitations discussed in the previous section, we introduce a new perspective on ancilla use and develop a corresponding formulation of uncomputation.
As outlined in Section~\ref{sec:intro-approach}, the central idea is to specify each gate in an ancilla-assisted quantum circuit from two complementary components: functionality implementation and ancilla-role transformation.
We first provide formal definitions of these notions.

\begin{definition}[Functionality, Ancilla Role, and Specification]\label{def:semantics}
    Consider a primary system \(\cS\) and an ancilla system \(\cA\), both assumed to be finite-dimensional qubit registers throughout this paper.
    We introduce the following notions:
    \begin{enumerate}
        \item A \emph{(logical) functionality} \(W\) is a unitary acting on the primary system \(\cS\).
        \item An \emph{ancilla role} \(R\) is an isometry from the primary system \(\cS\) to the joint system \(\cA\otimes\cS\).
        \item For a unitary \(U\) acting on the joint system \(\cA\otimes\cS\), a \emph{specification} \(U:R_1\ixRightarrow{W}R_2\) is valid if and only if the diagram in Fig.~\ref{fig:commutative-diagram} commutes, i.e., \(UR_1=R_2W\).
    \end{enumerate}
\end{definition}

In an ancilla-assisted quantum circuit, although the {\color{csRoleBlue}physical} gates \(U_1,\ldots,U_n\) act on the joint system \(\cA\otimes\cS\), their ultimate purpose is to contribute to implementing the {\color{csPhaseGreen}logical} functionality \(W\) on the primary system.
Operations on ancilla system are not ends in themselves; they are intended to facilitate the subsequent implementation of functionalities.
Even when a physical gate appears merely to prepare the ancilla without immediately implementing any logical functionality,  such preparation ultimately serves to facilitate the implementation of the intended functionality by subsequent gates.

\begin{figure}[t]
    \centering
    \hspace{1em}
    \begin{tikzpicture}[x=0.7cm, y=0.75cm, >=stealth, every node/.style={inner sep=2pt}]
        \node (logical-0) at (-5.4,0.95) {$\cS$};
        \node (physical-0) at (-5.4,-0.95) {$\cA\otimes\cS$};
        \node (logical-1) at (-2.8,0.95) {$\cS$};
        \node (physical-1) at (-2.8,-0.95) {$\cA\otimes\cS$};
        \node (logical-2) at (-0.4,0.95) {$\cS$};
        \node (physical-2) at (-0.4,-0.95) {$\cA\otimes\cS$};
        \node (logical-n-1) at (2.4,0.95) {$\cS$};
        \node (physical-n-1) at (2.4,-0.95) {$\cA\otimes\cS$};
        \node (logical-n) at (5.0,0.95) {$\cS$};
        \node (physical-n) at (5.0,-0.95) {$\cA\otimes\cS$};

        \draw[->,line width=0.6pt,shorten >=-1.8pt] (logical-0) -- node[left] (role-initial) {$R_0=|0\>_\cA\otimes I_\cS$} (physical-0);

        \draw[double=csRoleBlue!4, double distance=0.8pt, ->] (physical-0) -- node[below] (physical-gate-1) {$U_1$} (physical-1);
        \draw[double=csRoleBlue!4, double distance=0.8pt, ->] (physical-1) -- node[below] (physical-gate-2) {$U_2$} (physical-2);
        \draw[double=csRoleBlue!4, double distance=0.8pt, ->] (physical-2) -- node[below=2pt] (physical-gate-dots) {$\cdots$} (physical-n-1);
        \draw[double=csRoleBlue!4, double distance=0.8pt, ->] (physical-n-1) -- node[below] (physical-gate-n) {$U_n$} (physical-n);

        \coordinate (role-functionality-separator) at (0,0.53);
        \coordinate (role-physical-separator) at (0,-0.53);
        \begin{scope}[on background layer]
            \node[rectangle,draw=none,
                inner xsep=4pt,inner ysep=2pt,outer sep=0pt,
                fit=(physical-0)(physical-1)(physical-2)(physical-n-1)(physical-n)
                    (physical-gate-1)(physical-gate-2)(physical-gate-dots)(physical-gate-n)] (physical-view-background) {};
            \path[draw=none,fill=csRoleBlue!5]
                ([yshift=2pt]physical-view-background.south west)
                rectangle ([xshift=-3pt]physical-view-background.north east |- role-physical-separator);
            \path[draw=none,fill=csPhaseGreen!5]
                let \p1=([yshift=2pt]physical-view-background.south west),
                    \p2=([xshift=-3pt]physical-view-background.north east |- role-functionality-separator)
                in (\x1,-\y1) rectangle (\x2,\y2);
        \end{scope}

        \draw[double=csPhaseGreen!4, double distance=0.8pt, ->] (logical-0) -- node[above] {$W_1$} (logical-1);
        \draw[double=csPhaseGreen!4, double distance=0.8pt, ->] (logical-1) -- node[above] {$W_2$} (logical-2);
        \draw[double=csPhaseGreen!4, double distance=0.8pt, ->] (logical-2) -- node[above=2pt] {$\cdots$} (logical-n-1);
        \draw[double=csPhaseGreen!4, double distance=0.8pt, ->] (logical-n-1) -- node[above] {$W_n$} (logical-n);

        \draw[->,line width=0.6pt,shorten >=-1.8pt] (logical-1) -- node[left] {$R_1$} (physical-1);
        \draw[->,line width=0.6pt,shorten >=-1.8pt] (logical-2) -- node[left] (role-second) {$R_2$} (physical-2);
        \draw[->,line width=0.6pt,shorten >=-1.8pt] (logical-n-1) -- node[left] (role-penultimate) {$R_{n-1}$} (physical-n-1);
        \draw[->,line width=0.6pt,shorten >=-1.8pt] (logical-n) -- node[right] (role-final) {$R_n$} (physical-n);
        \coordinate (role-dots-center) at ($(role-second.east)!0.5!(role-penultimate.west)$);
        \node[anchor=base] at (role-dots-center |- role-second.base) {$\cdots$};

        \begin{scope}[on background layer]
            \path[draw=none,fill=csCleanOrange!5]
                ([xshift=-3pt]role-initial.west |- role-functionality-separator)
                rectangle ([xshift=3pt]role-final.east |- role-physical-separator);
        \end{scope}

        \draw[dashed,csPhaseGreen,line width=0.8pt] ([xshift=-3pt]role-initial.west |- role-functionality-separator)
            -- node[pos=0,above right=2pt,yshift=1pt,text=csPhaseGreen,font=\footnotesize\itshape,align=left,inner sep=0pt] (logical-view-label) {Logical\\functionality}
            ([xshift=3pt]role-final.east |- role-functionality-separator);
        \draw[dashed,csRoleBlue,line width=0.8pt] ([xshift=-3pt]role-initial.west |- role-physical-separator)
            -- node[pos=0,below right=2pt,yshift=-1pt,text=csRoleBlue,font=\footnotesize\itshape,align=left,inner sep=0pt] (physical-view-label) {Physical\\realization}
            ([xshift=3pt]role-final.east |- role-physical-separator);
        \definecolor{ancillaRoleBeige}{HTML}{625440}
        \node[anchor=west,align=center,text=ancillaRoleBeige,font=\footnotesize,inner sep=0pt]
            at ([xshift=8pt]role-final.east) {\emph{Ancilla roles}\\{\scriptsize(logical to physical)}};
    \end{tikzpicture}
    \caption{Physical realization of logical functionalities through ancilla-role transformations.}
    \label{fig:two-view}
\end{figure}
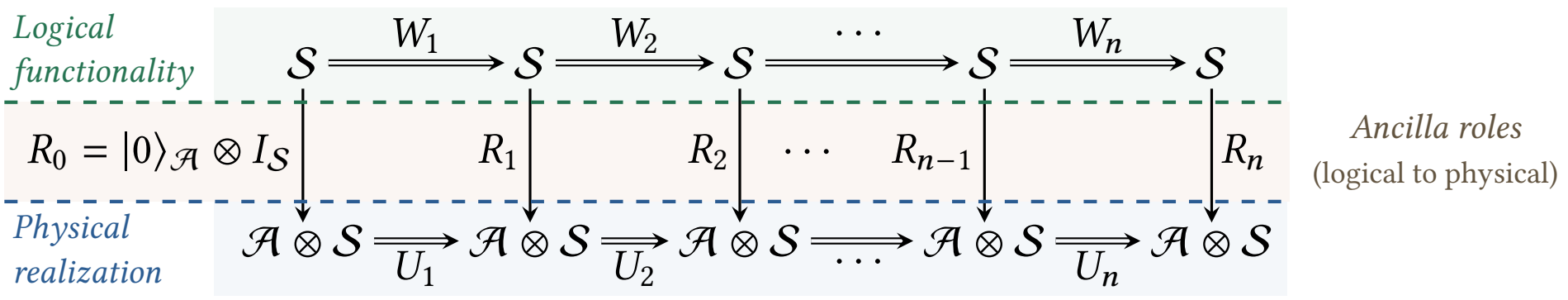

As illustrated in Fig.~\ref{fig:two-view}, our formulation extracts the underlying logical functionality \(W_i\) from each gate \(U_i\) in the computation circuit \(U\), providing two equivalent perspectives on ancilla-assisted computation: physical gates acting on the joint system \(\cA\otimes\cS\) are concrete and implementable, whereas logical functionalities acting on the primary system \(\cS\) are abstract and may not be directly implementable.

This leads us to the role of ancilla qubits: they bring abstract logical functionalities into the physical realm, enabling their realization as concrete quantum circuits.
By appropriately introducing ancilla qubits in the required forms, which is characterized by the notion of \emph{ancilla roles}, a logical computation that would otherwise be difficult or intractable to implement can become more tractable at the physical level through a concrete ancilla-assisted quantum circuit.

The specification \(U:R_1\ixRightarrow{W}R_2\) of a physical gate \(U\) can now be understood as follows.
Given a physically available operation \(U\) and the current ancilla role \(R_1\), which characterizes the current form of ancilla use, applying \(U\) implements the logical functionality \(W\) while transforming the ancilla role into \(R_2\), which then becomes  physically available for subsequent operations, as illustrated in Fig.~\ref{fig:two-view}.
In the following subsection, we further elaborate on the meaning and significance of ancilla roles through concrete examples.

\subsection{Ancilla Roles: Abstractions of Ancilla Use and Invariants of Computation}\label{sec:formulation-role}
Conceptually, an ancilla role characterizes the form of ancilla use by abstracting how ancilla qubits are introduced and employed, for instance, as registers for storing classical information.

The significance of this notion lies in two aspects.
Theoretically, it serves a formal interface for separating the operations on ancillas from the overall semantics of a computation, which is essential for formulating and synthesizing uncomputation.
Practically, it often serves as an \emph{invariant} throughout the computation, as illustrated by the following examples.
\begin{example}[Ancilla Role \(\State(\phi)\)]\label{eg:state-role}
    For a fixed pure state \(|\phi\>_\cA\) of the ancilla system, define the ancilla role \(\State(\phi)\) by
    \[
    \State(\phi)=|\phi\>_\cA\otimes I_\cS,\quad\text{i.e.,}\quad\State(\phi):|\psi\>_\cS\mapsto|\phi\>_\cA|\psi\>_\cS.
    \]
\end{example}
Intuitively, the ancilla role \(\State(\phi)\) characterizes a form of ancilla use in which the ancilla system is in a fixed pure state \(|\phi\>_\cA\), uncorrelated with the primary system.
This role is particularly useful for phase kickback, as formalized by the following proposition.

\begin{proposition}[Phase Kickback via Ancilla Role \(\State(\phi)\)]\label{prop:state-role}
    Let \(G\) be a unitary on the ancilla system \(\cA\) such that
    \[
    G|\phi\>_\cA=e^{i\theta}|\phi\>_\cA.
    \]
    Then, a \(\mathrm{controlled-}G\) gate, with a single-qubit primary system as the control and the ancilla system as the target, admits the following specification:
    \[
    \mathrm{controlled-}G:\State(\phi)\xRightarrow{P(\theta)}\State(\phi),
    \quad\text{where }
    P(\theta)=|0\>\<0|+e^{i\theta}|1\>\<1|\mbox{ is the phase gate}.
    \]
    In other words, the controlled\(-G\) gate preserves the ancilla role \(\State(\phi)\) while implementing the phase gate \(P(\theta)\) on the primary system.
\end{proposition}
\begin{proof}
    By Definition~\ref{def:semantics}, it suffices to check the commutativity: \(\text{controlled-}G \cdot \State(\phi) = \State(\phi) \cdot P(\theta)\). This can be verified by direct calculation:
    \[
    \begin{aligned}
        \text{controlled-}G\cdot \State(\phi)&=(I_\cA\otimes |0\>_\cS\<0|+G_\cA\otimes |1\>_\cS\<1|)\cdot \pare{|\phi\>_\cA\otimes I_\cS}\\
        &=|\phi\>_\cA\otimes |0\>_\cS\<0|+G|\phi\>_\cA\otimes |1\>_\cS\<1|\\
        &=|\phi\>_\cA\otimes |0\>_\cS\<0|+|\phi\>_\cA\otimes e^{i\theta}|1\>_\cS\<1|\\
        &=\State(\phi)\cdot P(\theta).
    \end{aligned}
    \]
\end{proof}
The Fourier-basis phase-kickback circuit discussed under the first limitation in Section~\ref{sec:limitation}, shown in Fig.~\ref{fig:Fourier-basis}, can be understood as a generalization of this mechanism for implementing functionality.
Specifically, the circuit first \emph{prepares} the ancilla by transforming its role while implementing trivial functionality, then \emph{uses} the ancilla to implement nontrivial functionality while preserving its role, and finally \emph{uncomputes} the ancilla by reversing the preparation:
\[
\begin{aligned}
(\mathsf{Prepare})\quad&\text{First two layers of gates}:|0\>_\cA\otimes I_\cS\ixRightarrow{I}\State(\gamma),\\
(\mathsf{Use})\quad&\operatorname{ADD}:\State(\gamma)\ixRightarrow{W}\State(\gamma),\quad \mbox{(Generalization of Proposition~\ref{prop:state-role})}\\
(\mathsf{Uncompute})\quad&\text{Reverse preparation}:\State(\gamma)\ixRightarrow{I}|0\>_\cA\otimes I_\cS.
\end{aligned}
\]

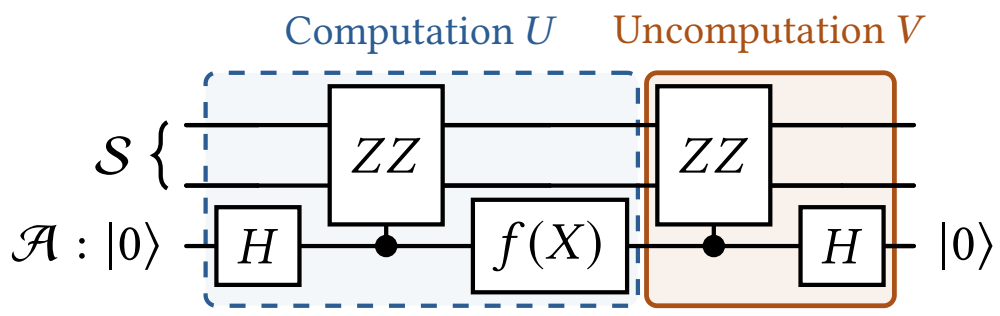
\begin{figure}[t]
    \centering
    \begin{quantikz}[row sep={0.40cm,between origins},column sep=0.2cm]
        \lstick[wires=2,braces={decoration={pre=moveto,pre length=2.5pt,post=moveto,post length=2.5pt}}]{\(\cS\)} & \qw\csgateframe{comp}{3}{3}{Computation \(U\)} & \gate[2]{ZZ} & \qw & \gate[2]{ZZ}\csgateframe{clean}{3}{2}{Uncomputation \(V\)} & \qw & \qw \\
        & \qw &  & \qw &  & \qw & \qw \\
        \lstick{\(\cA:\ket{0}\)} & \gate{H} & \ctrl{-2} & \gate{f(X)} & \ctrl{-2} & \gate{H} & \qw\rstick{\(\ket{0}\)}
    \end{quantikz}
    \caption{Circuit implementing a fractional query with functionality \(W=f(ZZ)\), where \(f(A)=\exp\pare{\tfrac{-i\pi(I-A)}{8}}\). Adapted from~\cite[Fig.~1]{fractional-query}.}
    \label{fig:fractional-query}
\end{figure}

\begin{example}[Ancilla Role \(\Choice(G)\)]\label{eg:choice-role}
    Fix a single-qubit ancilla system \(\cA\) and a unitary \(G\) on the primary system \(\cS\).
    Define the ancilla role \(\Choice(G)\) by
    \[
    \Choice(G)=\frac{|0\>_\cA\otimes I_\cS+|1\>_\cA\otimes G_\cS}{\sqrt{2}},\quad\text{i.e.,}\quad\Choice(G):|\psi\>_\cS\mapsto\frac{|0\>_\cA|\psi\>_\cS+|1\>_\cA\pare{G|\psi\>_\cS}}{\sqrt{2}}.
    \]
\end{example}
Intuitively, the ancilla role \(\Choice(G)\) characterizes a form of ancilla use in which the ancilla qubit coherently controls the application of \(G\) to the primary system, potentially resulting in entanglement between the two systems.
This role is particularly useful for reasoning about quantum computations involving coherent control and interference, as illustrated by the following proposition.
\begin{proposition}[Coherent Control via Ancilla Role \(\Choice(G)\)]\label{prop:choice-role}
    Let \(G\) be a unitary on the primary system \(\cS\) such that \(G^2=I_\cS\).
    Then an \(X\) gate on the one-qubit ancilla system \(\cA\) admits the following specification:
    \[
    X_\cA:\Choice(G)\xRightarrow{G}\Choice(G).
    \]
\end{proposition}
The proof follows by direct calculation, analogous to that of the phase-kickback proposition, and is deferred to the appendix.


This role is useful for analyzing the circuit for implementing a fractional query~\cite{fractional-query} as shown in Fig.~\ref{fig:fractional-query}.
Specifically, by instantiating \(G=ZZ\) and applying the linearity property to the proof of Proposition~\ref{prop:choice-role}, we can generalize the above specification to obtain
\[
f(X)_\cA:\Choice(ZZ)\ixRightarrow{f(ZZ)}\Choice(ZZ),
\quad \mbox{where }
f(A)=\exp\pare{\tfrac{-i\pi(I-A)}{8}},
\]
which is exactly realized by the third gate in Fig.~\ref{fig:fractional-query}.

\begin{example}[Ancilla Role \(\Record(P)\)]\label{eg:record-role}
    Fix a single-qubit ancilla system \(\cA\) and a projection operator \(P\) on the primary system \(\cS\).
    Define the ancilla role \(\Record(P)\) by
    \[
    \Record(P)=|0\>_\cA\otimes P^\perp_\cS+|1\>_\cA\otimes P_\cS,\quad\text{i.e.,}\quad\Record(P):|\psi\>_\cS\mapsto|0\>_\cA(I-P)|\psi\>_\cS+|1\>_\cA P|\psi\>_\cS.
    \]
\end{example}
Intuitively, the ancilla role \(\Record(P)\) characterizes a form of ancilla use in which the ancilla qubit coherently records information of the primary system, as illustrated by the following example.
\begin{example}
    Consider a classical Boolean function \(f:\{0,1\}^n\to\{0,1\}\) with \(\dim\cS=2^n\).
    Let \(P_f=\sum_{b\in\{0,1\}^n:f(b)=1}|b\>\<b|\) denote the projector onto the subspace spanned by computational-basis states \(|b\>\) satisfying \(f(b)=1\).
    For any primary-system state \(|\psi\>_\cS=\sum_b\alpha_b|b\>_\cS\), we have
    \[
    \begin{aligned}
       \Record\pare{P_f}:|\psi\>_\cS
       &\mapsto |0\>_\cA (I-P_f)|\psi\>_\cS+|1\>_\cA P_f|\psi\>_\cS\\
       &=|0\>_\cA\sum_{b:f(b)=0}\alpha_b|b\>_\cS
       +|1\>_\cA\sum_{b:f(b)=1}\alpha_b|b\>_\cS\\
       &=\sum_b \alpha_b|f(b)\>_\cA|b\>_\cS,
    \end{aligned}
    \]
    which coherently records the value of \(f\) in the ancilla qubit.
\end{example}


This role captures the ancilla use in the middle Toffoli gate of the \(C^3X\) circuit shown in Fig.~\ref{fig:C3X}, which admits the following specification:
\[
\operatorname{Toffoli}_{a,q_3,q_4}: \Record\pare{P_f}\xRightarrow{C^3X}\Record\pare{P_f},\quad\mbox{for }n=4,\ f(x_1,x_2,x_3,x_4)=x_1\wedge x_2.
\]

In summary, ancilla roles describe the functional role of the ancilla system in a computation, such as storing an isolated state, coherently controlling the primary system or recording the information of the primary system.
Furthermore, the notion of ancilla roles naturally distinguishes two types of operations commonly found in practical circuits: preparing ancillas by transforming their roles, and using the prepared ancillas to implement functionality while preserving their roles.

\subsection{Uncomputation: Restoring Ancilla Roles with Trivial Functionality}
With the preceding notions and formulation, we can now distinguish two aspects of a unitary's semantics, allowing us to define uncomputation in terms of the component characterizing ancilla operations, rather than the computation unitary \(U\) itself.
\begin{definition}[Uncomputation]\label{def:uncomputation}
    Fix a primary system \(\cS\) and an ancilla system \(\cA\).
    An \emph{uncomputation} \(V\) with respect to an ancilla role \(R:\cS\to\cA\otimes \cS\) is a unitary acting on the joint system \(\cA\otimes\cS\) admitting the following specification:
    \vspace{-.7em}
    \[
    V:R\ixRightarrow{I}|0\>_\cA\otimes I_\cS.
    \]
    Intuitively, uncomputation is a unitary that restores the ancilla to its initial role while implementing the identity functionality on the primary system.
\end{definition}
\begin{remark}
    More generally, uncomputation need not restore ancillas to the ground state \(|0\rangle_\cA\), but may instead restore a specified ancilla role, such as \(\State(\phi)\), corresponding to a particular ancilla state \(|\phi\rangle_\cA\). 
\end{remark}
In our formulation, uncomputation \(V\) is defined with respect to an ancilla role \(R\), rather than the entire unitary \(U\) of the preceding computation, which carries two aspects of information: (1) ancilla operations and (2) the implemented logical functionality. 
Mathematically, \(V\) is thus defined in terms of an isometry \(R:\cS\to\cA\otimes\cS\), which captures only the former aspect of \(U\), rather than the full information contained in a unitary \(U:\cA\otimes\cS\to\cA\otimes\cS\).
Such a definition is natural and intuitive because uncomputation requires only knowledge of the current role of ancillas, without needing to know the functionality implemented by the preceding computation.

By comparison, prior work defines \(V\) with respect to \(U\), incorporating additional information about functionality \(W\) and thereby imposing the burden of prescribing how \(U\) determines \(W\).
Indeed, since computation \(U\) and uncomputation \(V\) together determine functionality \(W\), requiring \(V\) to be determined by \(U\) implicitly requires \(W\) to be determined by \(U\) as well.

To this end, prior work adopts a fixed interpretation of ancilla use based on classical computation in the computational basis, effectively defining \(W=\pare{\sum_x\<x|_\cA\otimes I_\cS}U\pare{|0\>_\cA\otimes I_\cS}\), as in Eq.~\eqref{eq:computation-uncomputation}.
However, such a fixed interpretation can limit the generality of uncomputation, potentially rendering it undefined even when a valid solution exists.

Our formulation avoids this limitation by introducing the notion of ancilla role to isolate the ancilla-related information encoded in a unitary, retaining only the minimal information necessary to define uncomputation and thereby decoupling uncomputation from the implemented functionality.
This decoupling provides greater flexibility and generality in interpreting functionality and forms of ancilla use, leading to the following characterization:
\[
(\text{Computation})\quad U:|0\>_\cA\otimes I_\cS\ixRightarrow{(\text{functionality})\ W}R,\qquad
(\text{Uncomputation})\quad V:R\ixRightarrow{I}|0\>_\cA\otimes I_\cS.
\]
Prior uncomputation approaches for classical ancilla use in Eq.~\eqref{eq:computation-uncomputation} can then be understood as a special case of our formulation through the following procedure:
\begin{enumerate}
    \item[Step 1:] Given a computation unitary \(U\) and a designated ancilla system \(\cA\), compute the functionality \(W=\pare{\sum_x\<x|_\cA\otimes I_\cS}U\pare{|0\>_\cA\otimes I_\cS}\) and check whether \(W\) is unitary.

    \item[Step 2:] If \(W\) is unitary, determine the ancilla role \(R=U(|0\>_\cA\otimes I_\cS)W^\dagger\), which by construction yields a valid specification \(U:|0\>_\cA\otimes I_\cS\ixRightarrow{W}R\).

    \item[Step 3:] Synthesize a unitary \(V\) satisfying \(VR=|0\>_\cA\otimes I_\cS\), or equivalently, \(V|_{\mathrm{im}R}=(|0\>_\cA\otimes I_\cS)R^\dagger\), thereby ensuring \(V:R\ixRightarrow{I}|0\>_\cA\otimes I_\cS\).
\end{enumerate}
The functionality \(W\) and ancilla role \(R\) are uniquely determined by the first two steps whenever \(W\) is unitary.
Moreover, a suitable unitary \(V\) always exists, since the prescribed action on \(\mathrm{im}R\) can be extended to a unitary on the full Hilbert space.



\section{A Use-First Strategy for Automatic Uncomputation Synthesis}\label{sec:strategy}

In this section, building on the preceding formulation, we develop a simple yet powerful strategy for automatically synthesizing uncomputation beyond classical ancilla use, demonstrating the theoretical well-foundedness and practical significance of our formulation.

\begin{figure}[t]
    \centering
    \begin{minipage}{0.45\textwidth}
    \begin{algorithm}[H]
    \caption{\textbf{Use-first} strategy for automatic uncomputation synthesis.}
    \label{alg:use-first}
        \DontPrintSemicolon
        \SetKwFor{ForEach}{for each}{do}{end}
        \KwIn{Computation circuit \(U\), and ancilla desigination \(\cA\).}
        \KwOut{Uncomputation \(V\).}
        \(\cS\gets \dom U\setminus \cA;\quad R\gets |0\>_\cA\otimes I_\cS;\)\;
        \For{\textup{gate \(U_i\in U\) in execution order}}{
            \eIf{\(U_i(\mathrm{im}R)=\mathrm{im}R\)}{
                Mark \(U_i\) as {\color{csPhaseGreen}Use}\;
            }{
                Mark \(U_i\) as {\color{csRoleBlue}Prepare};\;
                \(R\gets U_iR\)\;
            }
        }
        \Return{\textup{\(V=\) inverses of the gates marked as {\color{csRoleBlue}Prepare}, applied in reverse order.}}\;
    \end{algorithm}
    \label{fig:use-first}
    \end{minipage}
    \hspace{2em}
    \begin{minipage}{0.45\textwidth}
    \begin{algorithm}[H]
    \caption{Algorithm for mid-circuit uncomputation generation.}
    \label{alg:mid-circuit}
        \DontPrintSemicolon
        \SetKwFor{ForEach}{for each}{do}{end}
        \KwIn{Computation circuit \(U\), and ancilla desigination \(\cA\).}
        \KwOut{A Circuit \(U'\) including uncomputations.}
        \newcommand{\algAssignLhs}[1]{\makebox[1.4em][l]{\(#1\)}}
        \(\cA'\gets \cA;\quad U'\gets U;\)\;
        \While{\(\cA'\neq\emptyset\)}{
            Pick \(a\in\cA'\) whose last gate \(U_i'\) occurrence in \(U'\) is earliest;\;
            \(V\gets
            \begin{aligned}[t]
                &\textbf{Use-first}(U_1';\ldots;U_i', \{a\})\\
                &\text{in Algorithm~\ref{alg:use-first}};
            \end{aligned}\)\;
            \(U'\gets\) insert \(V\) after \(U_i'\) in \(U'\);\;
            \(\cA'\gets\cA'\setminus \{a\}\)\;
        }
        \Return{\(U'\)}\;
    \end{algorithm}
    \label{fig:use-first-algorithm}
    \end{minipage}
\end{figure}
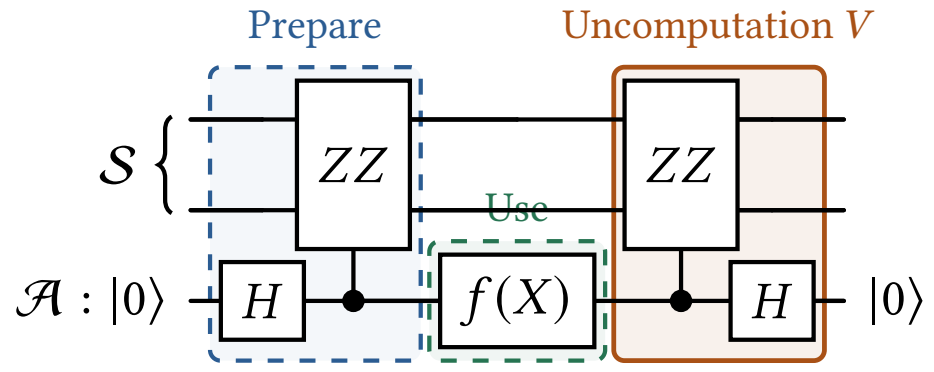
\begin{figure}[t]
    \centering
    \begin{minipage}{0.4\textwidth}
        \begin{quantikz}[row sep={0.60cm,between origins},column sep=0.2cm]
            \lstick[wires=2,braces={decoration={pre=moveto,pre length=2.5pt,post=moveto,post length=2.5pt}}]{\(\cS\)} & \qw\csgateframe{comp}{3}{2}{Prepare} &[-7pt] \gate[2]{ZZ} & \qw & \gate[2]{ZZ}\csgateframe{clean}{3}{2}{Uncomputation \(V\)} &[-7pt] \qw & \qw \\
            & \qw &  & \qw &  & \qw & \qw \\
            \lstick{\(\cA:\ket{0}\)} & \gate{H} & \ctrl{-2} & \gate{f(X)}\qw\csgateframe{use}{1}{1}{Use} & \ctrl{-2} & \gate{H} & \qw\rstick{\(\ket{0}\)}
        \end{quantikz}
    \end{minipage}
    \hfill
    \begin{minipage}{0.54\textwidth}
        \scriptsize
        \begin{tabular}{ll@{\(\,:\,\)}l@{\(\;\)}c@{\(\;\)}ll}
            \toprule
            & \multicolumn{4}{c}{\textbf{Specification}} & \multicolumn{1}{c}{\textbf{Class}} \\
            \midrule 
            1. & \(H_\cA\) & \(\State(0)\) & \(\ixRightarrow{I}\) & \(\State(+)\) & {\color{csRoleBlue}Prepare} \\
            2. & \(\mathrm{Control}ZZ\) & \(\State(+)\) & \(\ixRightarrow{I}\) & \(\Choice(ZZ)\) & {\color{csRoleBlue}Prepare} \\
            3. & \(f(X)_\cA\) & \(\Choice(ZZ)\) & \(\ixRightarrow{f(ZZ)}\) & \(\Choice(ZZ)\) & {\color{csPhaseGreen}Use} \\
            4. & \(\mathrm{Control}ZZ\) & \(\Choice(ZZ)\) & \(\ixRightarrow{I}\) & \(\State(+)\) & {\color{csCleanOrange} Uncompute} \\
            5. & \(H_\cA\) & \(\State(+)\) & \(\ixRightarrow{I}\) & \(\State(0)\) & {\color{csCleanOrange}Uncompute} \\
            \bottomrule
        \end{tabular}
    \end{minipage}
   \caption{Result of applying the Use-first strategy (Algorithm~\ref{alg:use-first}) to the circuit in Fig.~\ref{fig:fractional-query}. 
The ancilla roles $\State(\phi)$ and $\Choice(ZZ)$ are defined in Examples~\ref{eg:state-role} and \ref{eg:choice-role}, respectively. 
The specification of the third gate can be established using the argument following Proposition~\ref{prop:choice-role}.}
    \label{fig:fractional-query-result}
\end{figure}

As observed in Section~\ref{sec:formulation-role}, our formulation naturally gives rise to two forms of gate specification by separating functionality implementation from ancilla-role transformations:
\begin{enumerate}
    \item {\color{csRoleBlue}Prepare} the ancilla, formalized as \(U:R_1\ixRightarrow{I}R_2\). Intuitively, \(U\) transforms the ancilla role without implementing any nontrivial logical functionality.
    \item {\color{csPhaseGreen}Use} the ancilla, formalized as \(U:R\ixRightarrow{W}R\). Intuitively, \(U\) implements a nontrivial logical functionality \(W\) while preserving the ancilla role.
\end{enumerate}
While every gate admits a valid {\color{csRoleBlue}{Prepare}} specification, a {\color{csPhaseGreen}{Use}} specification is valid only under additional conditions, as characterized by the following lemma.

\begin{lemma}[Conditions for {Prepare} and {Use} Specifications]\label{lemma:prepare-use}
    Let \(U\) be a unitary on \(\cA\otimes\cS\) and \(R:\cS\to\cA\otimes\cS\) an ancilla role. Then:
    \begin{enumerate}
        \item There exists a unique ancilla role \(R'\) such that \(U:R\ixRightarrow{I}R'\), given by \(R'=UR\).
        \item There exists a functionality \(W\) such that \(U:R\ixRightarrow{W}R\) if and only if \(U(\operatorname{im}R)=\operatorname{im}R\). In this case, \(W\) is uniquely determined by \(W=R^\dagger UR\).
    \end{enumerate}
\end{lemma}

The proof is deferred to appendix.
This lemma reveals a fundamental asymmetry: every gate can be viewed as {\color{csRoleBlue}preparing} the ancilla, whereas only certain gates can be viewed as {\color{csPhaseGreen}using} the ancilla under a given ancilla role.
This asymmetry reflects an underlying intuition: an ancilla system can be used to implement functionality only when it has been prepared in an appropriate role.

This observation motivates a simple \emph{Use-first} strategy for automatic uncomputation synthesis.
Roughly speaking, for each physical gate, we first attempt to specify it as {\color{csPhaseGreen}using} the ancilla; if this is invalid, we instead specify it as {\color{csRoleBlue}preparing} the ancilla.
Uncomputation is then synthesized by reversing the gates specified as {\color{csRoleBlue}preparing} the ancilla.

This strategy naturally aligns with the design principle of ancilla-assisted circuits: gates should primarily be intended to implement functionality, resorting to ancilla preparation only when necessary to enable subsequent use.
The pseudocode of Use-first strategy is presented in Algorithm~\ref{alg:use-first}, where a circuit \(U\) is represented as a sequence of gates \(U_1;U_2;\ldots;U_n\) in execution order.

As a running example, we revisit the circuit in Fig.~\ref{fig:fractional-query}, which cannot be handled by prior tools because the functionality induced by the prior formulation of uncomputation for classical ancilla use is not unitary.
The result of applying the Use-first strategy to this circuit is shown in Fig.~\ref{fig:fractional-query-result}, where our twofold formulation precisely captures the intent of each gate in the ancilla-assisted circuit, thereby revealing its computation-use-uncomputation structure.
In the following subsection, we further evaluate the strategy on more practical examples to demonstrate the effectiveness of our formulation and strategy.

\subsection{Evaluation of the Use-first Strategy}
In this subsection, we evaluate our Use-first strategy against existing tools on a benchmark suite spanning five categories of practical examples.
We describe the benchmark construction, experimental methodology, and comparative results, followed by a discussion of the limitations of our strategy, to answer the following central research question:

\begin{quote}
    \noindent\(\boxed{\parbox{\dimexpr\linewidth-2\fboxsep-2\fboxrule\relax}{%
        Are our formulation and Use-first strategy effective in characterizing and synthesizing uncomputation for quantum circuits beyond classical ancilla use?
    }}\)\par
\end{quote}
The evaluation results are summarized in Table~\ref{tab:benchmark}. 
We next describe the experimental setup and discuss the results in detail.


\begin{table}[t]
\centering
\begingroup
\small
\caption{Comparison of our strategy with existing tools across five categories of practical cases. 
For existing tools, we configure Reqomp~\cite{reqomp} with ancilla reuse and relative-phase Toffoli optimization disabled, provide the gate annotations required by Unqomp~\cite{unqomp}, and try RwUn's four modes in sequence until a circuit is successfully synthesized~\cite{RwUn}.
We also decompose and rewrite circuit gates to accommodate the gate-set restrictions of certain tools, ensuring a fairer comparison and that the reported failures reflect genuine limitations of the underlying approaches rather than artifacts of gate representations or tool configurations.}
\label{tab:benchmark}
\setlength{\tabcolsep}{2.5pt}
\setlength{\aboverulesep}{0pt}
\setlength{\belowrulesep}{0pt}
\renewcommand{\arraystretch}{1}
\setbox\strutbox=\hbox{\vrule height10.5pt depth5.5pt width0pt}
\setlength{\arrayrulewidth}{0.25pt}
\newcommand{\benchrow}{\strut}
\newcommand{\benchexample}[2]{\begin{adjustbox}{max width=99.5pt}#1~\cite{#2}\end{adjustbox}}
\newcommand{\benchnotedexample}[3]{\begin{adjustbox}{max width=99.5pt}#1~\cite{#2}\textsuperscript{#3}\end{adjustbox}}
\newcommand{\benchpaddedrow}{\rule[-7.5pt]{0pt}{20pt}}
\newcommand{\benchshift}[2]{\raisebox{#1}[0pt][0pt]{#2}}
\newcommand{\benchheadrow}{\strut}
\newcommand{\benchpair}[1]{\multicolumn{2}{@{\hspace{2pt}}c@{\hspace{2pt}}}{#1}}
\newcommand{\benchheading}[2]{\raisebox{-8.15pt}[0pt][0pt]{\parbox[c]{#1}{\centering\bfseries #2}}}
\newcommand{\benchcategory}[3][16pt]{\raisebox{\dimexpr#1/2-#1*#2/2\relax}[0pt][0pt]{\parbox[c]{76.5pt}{\centering #3}}}

\begin{adjustbox}{max width=\textwidth,center}
\begin{tabular}{@{}m{76.5pt}|@{\hspace{6pt}}>{\raggedright\arraybackslash}m{99.5pt}*{4}{>{\centering\arraybackslash\hspace*{3pt}}m{19.5pt}<{\hspace*{-3pt}}@{\hspace{3pt}}>{\centering\arraybackslash\hspace*{-3pt}}m{21.5pt}<{\hspace*{3pt}}}@{}}
\toprule
\benchheading{76.5pt}{Category} & \benchheading{99.5pt}{Examples} & \benchpair{\benchheadrow\textbf{Our Strategy}} & \benchpair{\textbf{Reqomp~\cite{reqompArtifact}}} & \benchpair{\textbf{Unqomp~\cite{unqomp}}} & \benchpair{\textbf{RwUn~\cite{RwUn}}} \\
\cmidrule(lr){3-4}\cmidrule(lr){5-6}\cmidrule(lr){7-8}\cmidrule(lr){9-10}
 & & \benchpair{\benchheadrow\footnotesize\textbf{Success}\hspace{3pt}\textbf{\#Gates}} & \benchpair{\benchheadrow\footnotesize\textbf{Success}\hspace{3pt}\textbf{\#Gates}} & \benchpair{\benchheadrow\footnotesize\textbf{Success}\hspace{3pt}\textbf{\#Gates}} & \benchpair{\benchheadrow\footnotesize\textbf{Success}\hspace{3pt}\textbf{\#Gates}} \\
\midrule

%

\benchcategory{2}{I. Undefined\\functionality} & \benchrow \benchexample{Fractional query}{fractional-query} & $\checkmark$ & 5 & $\times$ & \textemdash & $\times$ & \textemdash & $\times$ & \textemdash \\
 & \benchrow \benchexample{Fourier-basis phase kickback}{Fourier-basis} & $\checkmark$ & 13 & $\times$ & \textemdash & $\times$ & \textemdash & $\times$ & \textemdash \\
\hline

%

\benchcategory{2}{II. Unintended\\functionality} & \benchrow \benchexample{Controlled-T}{controlled-T} & $\checkmark$ & 3 & $\times$ & \textemdash & $\times$ & \textemdash & $\times$ & \textemdash \\
 & \benchrow \benchexample{Relative-phase-Toffoli $C^3X$}{relative-phase-toffoli} & $\checkmark$ & 19 & $\times$ & \textemdash & $\times$ & \textemdash & $\times$ & \textemdash \\
\hline

%

\benchcategory[20pt]{2}{III. Intended, but\\unsupported by\\baselines} & \benchpaddedrow \benchshift{-2pt}{\benchexample{QSVT projector phase}{QSVT}} & \benchshift{-2pt}{$\checkmark$} & \benchshift{-2pt}{7} & \benchshift{-2pt}{$\times$} & \benchshift{-2pt}{\textemdash} & \benchshift{-2pt}{$\times$} & \benchshift{-2pt}{\textemdash} & \benchshift{-2pt}{$\times$} & \benchshift{-2pt}{\textemdash} \\
 & \benchpaddedrow \benchshift{2pt}{\benchexample{Single-clock HHL}{single-clock-HHL}} & \benchshift{2pt}{$\checkmark$} & \benchshift{2pt}{7} & \benchshift{2pt}{$\times$} & \benchshift{2pt}{\textemdash} & \benchshift{2pt}{$\times$} & \benchshift{2pt}{\textemdash} & \benchshift{2pt}{$\times$} & \benchshift{2pt}{\textemdash} \\
\hline

%
%

\benchcategory{12}{IV. Existing\\benchmarks} & \benchrow \benchnotedexample{Adder}{reqompArtifact}{1} & $\checkmark$ & 12 & $\checkmark$ & 10 & $\checkmark$ & 8 & $\checkmark$ & 11 \\
 & \benchrow \benchexample{WeightedAdder}{reqompArtifact} & $\checkmark$ & 5 & $\checkmark$ & 5 & $\checkmark$ & 5 & $\checkmark$ & 6 \\
 & \benchrow \benchexample{Multiplier}{reqompArtifact} & $\checkmark$ & 4 & $\checkmark$ & 4 & $\checkmark$ & 4 & $\checkmark$ & 4 \\
 & \benchrow \benchexample{IntegerComparator}{reqompArtifact} & $\checkmark$ & 3 & $\checkmark$ & 3 & $\checkmark$ & 3 & $\checkmark$ & 8 \\
 & \benchrow \benchexample{Clean Incrementer}{RwUn} & $\checkmark$ & 5 & $\checkmark$ & 5 & $\checkmark$ & 5 & $\checkmark$ & 8 \\
 & \benchrow \benchexample{MCX}{reqompArtifact} & $\checkmark$ & 3 & $\checkmark$ & 3 & $\checkmark$ & 3 & $\checkmark$ & 3 \\
 & \benchrow \benchnotedexample{MCRY}{reqompArtifact}{2} & $\checkmark$ & 3 & $\checkmark$ & 3 & $\checkmark$ & 3 & $\times$ & \textemdash \\
 & \benchrow \benchexample{PiecewiseLinearR}{reqompArtifact} & $\checkmark$ & 7 & $\checkmark$ & 7 & $\checkmark$ & 7 & $\times$ & \textemdash \\
 & \benchrow \benchexample{PolynomialPauliR}{reqompArtifact} & $\checkmark$ & 6 & $\checkmark$ & 6 & $\checkmark$ & 6 & $\times$ & \textemdash \\
 & \benchrow \benchnotedexample{ConditionalCleanMCS}{RwUn,ConditionalCleanMCS}{3} & $\times$ & \textemdash & $\checkmark$ & 17 & $\times$ & \textemdash & $\checkmark$ & 26 \\
 & \benchrow \benchexample{Deutsch--Jozsa}{reqompArtifact} & $\checkmark$ & 11 & $\checkmark$ & 11 & $\checkmark$ & 11 & $\checkmark$ & 9 \\
 & \benchrow \benchexample{Grover}{reqompArtifact} & $\checkmark$ & 41 & $\checkmark$ & 41 & $\checkmark$ & 41 & $\checkmark$ & 37 \\
\hline


\centering\arraybackslash V. General hard case & \benchrow \benchexample{Gate teleportation}{gate-teleportation} & $\times$ & \textemdash & $\times$ & \textemdash & $\times$ & \textemdash & $\times$ & \textemdash \\
\hline
\multicolumn{2}{l}{Verified successes} & 17/19 & \textemdash & 12/19 & \textemdash & 11/19 & \textemdash & 9/19 & \textemdash \\
\bottomrule
\end{tabular}
\end{adjustbox}
\par\smallskip
\begin{minipage}{\textwidth}
    \footnotesize\raggedright
    \textsuperscript{1} Unqomp retains the annotated \texttt{qfree} carry helper as one atomic instruction~\cite[Sec.~6.2]{unqomp}.\par
    \textsuperscript{2} This MCRY instance uses \(\theta=\pi/4\), instead of \(\pi/2\) in ~\cite{RwUn,RwUnArtifactV2} which results in a degenerate gate.\par
    \textsuperscript{3} For ConditionalCleanMCS, Reqomp uses atomic controlled-\(S\); RwUn receives its equivalent CX/\(T\)/\(T^\dagger\) decomposition.
\end{minipage}
\endgroup
\end{table}

\paragraph{Benchmark construction} We construct a benchmark suite comprising five categories of practical examples.
The first three categories correspond to the three levels of limitations of prior formulations discussed in Section~\ref{sec:limitation}, each containing two practical examples.

The fourth category is adapted from benchmarks used in prior work~\cite{reqomp,unqomp,RwUn,RwUnArtifactV2}.
We select one representative from each group of semantically equivalent and structurally similar circuits (e.g., ConditionalCleanMCS and RiseConditionalCleanMCS in~\cite{RwUn}) and exclude randomly generated examples from~\cite{RwUn} that lack practical significance or fail to reflect the general principles and intuition underlying the design of ancilla-assisted circuits.

For this category, we choose the smallest parameter values that yield nontrivial functionality and nondegenerate circuits, as our primary goal is to evaluate the effectiveness of our formulation and strategy rather than their efficiency.
We discuss efficiency as a limitation of our current approach later.
Furthermore, since existing tools are sensitive to gate sets, we adopt the input-circuit settings noted in Table~\ref{tab:benchmark}, following practices established in prior work to ensure faithful reproduction of prior results and provide favorable conditions for existing tools.

Finally, the fifth category presents a challenging general case that is difficult to handle with automatic strategies.

\paragraph{Experimental methodology} For the first three categories, we implement the Use-first strategy in Algorithm~\ref{alg:use-first} and apply it directly to the examples.
The fourth category, however, presents an additional challenge: uncomputation may occur in the middle of a circuit, at different points for different ancilla qubits.
To accommodate this, we extend the Use-first strategy with a simple algorithm for determining where to insert uncomputation.
Specifically, we prioritize the ancilla whose last participating gate occurs earliest in the circuit and eagerly synthesize its uncomputation using the Use-first strategy, as formalized in Algorithm~\ref{alg:mid-circuit}.
Here, by ``\(a\)'s last gate,'' we mean the last gate in which \(a\) participates, either as a control or as a target qubit.

To keep our approach simple and intuitive, we deliberately forgo the more elaborate dependency analysis that has been a key focus and contribution of prior work~\cite{unqomp,RwUn,reqomp}.
Our primary goal is instead to demonstrate the well-foundedness of our formulation, with the Use-first strategy in Algorithm~\ref{alg:use-first} serving as the core component.
The limitation caused by this simplification and our solution to it will be discussed later.

\paragraph{Results and analysis.}
As shown in Table~\ref{tab:benchmark}, our Use-first strategy successfully handles all cases in the first three categories, demonstrating its applicability beyond classical ancilla use, whereas the baseline tools fail on certain cases due to their inability to handle phases or Hadamard gates acting on ancilla qubits.
In the appendix, we present the complete circuit for each benchmark case, together with the per-gate specifications inferred by our strategy and the resulting uncomputation.

For benchmarks drawn from prior work, our strategy also demonstrates broad applicability and produces relatively efficient circuits.
This is expected, as the Use-first strategy essentially identifies which gates in the computation circuit need to be reversed for uncomputation, thereby typically avoiding unnecessary operations in the synthesized circuit.
RwUn~\cite{RwUn} also produces circuits with fewer gates for Deutsch--Jozsa and Grover, benefiting from its native rewriting and cancellation optimizations.
These gate counts are reported according to each tool's own instruction granularity.

An application of Algorithm~\ref{alg:mid-circuit} to WeightedAdder from Category IV is illustrated in Fig.~\ref{fig:weightedadder-result}.
We first identify the third gate as the last gate acting on ancilla \(a\), and apply our Use-first strategy (Algorithm~\ref{alg:use-first}) to the first three gates, thereby avoiding interference from subsequent gates.
The first gate is naturally classified as {\color{csPhaseGreen}Use}, since it acts only on the data qubits, implementing functionality without changing the ancilla role.
The next two gates are classified in the same manner, and reversing the second gate, classified as {\color{csRoleBlue}Prepare}, successfully uncomputes \(a\) while preserving the implemented functionality.

In summary, our Use-first strategy demonstrates the effectiveness of our formulation in supporting the synthesis of uncomputation beyond classical ancilla use.

\begin{figure}[t]
    \centering
    \begin{minipage}[t]{0.4\textwidth}
        \vspace{0pt}
        \centering
        \begin{quantikz}[row sep={0.60cm,between origins},column sep=0.3cm,execute at end picture={\node[anchor=west,inner sep=1pt,font=\scriptsize,text=black] (last-gate-a-label) at ([xshift=0.45cm,yshift=0.45cm]ggroup-1-4.north east) {last gate of \(a\)}; \draw[-{Latex[length=1.4mm,width=0.9mm]},thick,black] ([xshift=-2pt]ggroup-1-4.north east) to[out=70,in=180] (last-gate-a-label.west);}]
            \lstick{\(x\)} & \ctrl{2}\csgateframe{use}{3}{1}{Use} & \ctrl{3}\csgateframed{comp}{4}{1}{Prepare} & \ctrl{2}\csgateframe{use}{4}{1}{Use} & \ctrl{3}\csgateframed{clean}{4}{1}{Uncompute} & \ctrl{1} & \rstick{\(x\)} \\
            \lstick{\(s_0\)} & \qw & \ctrl{2} & \qw & \ctrl{2} & \targ{} & \rstick{\(s_0'\)} \\
            \lstick{\(s_1\)} & \targ{} & \qw & \targ{} & \qw & \qw & \rstick{\(s_1'\)} \\
            \lstick{\(a:|0\>\)} & \qw & \targ{} & \ctrl{-1} & \targ{} & \qw &\rstick{\(\ket0\)}
        \end{quantikz}
        \caption{Result of our strategy on the WeightedAdder case, where \(\cS=\{x,s_0,s_1\}\), \(\cA=\{a\}\) and funcitonality \(W:|x\>|s\>\mapsto |x\>|s+3x\!\!\mod 4\>\)}
        \label{fig:weightedadder-result}
    \end{minipage}%
    \hfill%
    \begin{minipage}[t]{0.55\textwidth}
        \vspace{0pt}
        \centering
        \begin{quantikz}[row sep={0.40cm,between origins},column sep=0.23cm]
            \lstick{\(x_0\)} & \ctrl{2}\csgateframe{use}{6}{1}{Use \(a_1\)} &[4pt] \qw\gategroup[wires=6,steps=2,background,
                style={cscomp,rounded corners=2pt,inner xsep=1.5pt,inner ysep=1pt},
                label style={cscomplabel,label position=above,anchor=south,yshift=-6pt}]{\footnotesize Prepare \(a_1\)} & \ctrl{2}\csgateframed{clean}{6}{1}{Uncompute \(a_0\)} &[4pt] \qw\csgateframe{use}{6}{1}{Use \(a_1\)} & \ctrl{2}\csgateframed{clean}{6}{2}{Uncompute \(a_1\)} & \qw & \qw \\
            \lstick{\(x_1\)} & \qw & \gate[3]{\mathrm{OR}} & \qw & \qw & \qw & \gate[3]{\mathrm{OR}} & \qw \\
            \lstick{\(a_0:\ket0\)} & \targ{} & & \targ{} & \qw & \targ{} & & \qw\rstick{\(\color{purple}\ket{x_0}\)} \\
            \lstick{\(a_1:\ket0\)} & \qw & & \qw & \gate[3]{\mathrm{OR}} & \qw & & \qw\rstick{\(\ket0\)} \\
            \lstick{\(x_2\)} & \qw & \qw & \qw & & \qw & \qw & \qw \\
            \lstick{\(b\)} & \qw & \qw & \qw & & \qw & \qw & \qw
        \end{quantikz}
        \vspace{-.7em}
        \caption{Failure of Algorithm~\ref{alg:mid-circuit} on the three-bit IntegerComparator: uncomputation for \(a_1\) disrupts the previously synthesized uncomputation for \(a_0\). Here, \(\mathrm{OR}:|u,v,t\>\mapsto |u,v,t\oplus(u\lor v)\>\), \(\cS=\{x_0,x_1,x_2,b\}\), and \(\cA=\{a_0,a_1\}\).}
        \label{fig:failure}
    \end{minipage}
\end{figure}

\paragraph{Discussion of limitations} 
Our current formulation and strategy admit three primary limitations.
The first concerns \emph{efficiency}.
Since our core contribution focuses on fundamental conceptual advances, we adopt an extensional perspective in defining the notions of ancilla roles, functionality, and specifications in Definition~\ref{def:semantics}.
Each is treated as a mathematical object, such as a unitary or an isometry, without developing a concrete syntax or representation.
Consequently, checking the condition \(U(\mathrm{im}R)=\mathrm{im}R\) in Lemma~\ref{lemma:prepare-use} and Algorithm~\ref{alg:use-first} requires a naive and potentially expensive computation, limiting the scalability of our current implementation.

However, as described at the beginning of this subsection, our primary purpose in conducting the implementation and evaluation is to demonstrate that our formulation is well-motivated and useful, rather than to benchmark the efficiency of our strategy.
We therefore keep our strategy as simple as possible.
Addressing this efficiency limitation would require more intensional representations of ancilla roles and functionality, such as \(\State(\phi)\), \(\Choice(G)\), and \(\Record(P)\) discussed in Examples~\ref{eg:state-role}, \ref{eg:choice-role}, and \ref{eg:record-role}, respectively.
Such representations can enable symbolic and efficient reasoning about gate specifications and checking the condition for {\color{csPhaseGreen}Use} semantics in Lemma~\ref{lemma:prepare-use}.

The second limitation concerns \emph{dependency analysis}, a crucial mechanism for ensuring correct uncomputation in circuits involving classical ancilla use.
Extensively studied in prior work~\cite{reqomp,unqomp,RwUn}, this mechanism tracks information flow between qubits to guide the ordering and construction of uncomputation operations.
However, Algorithm~\ref{alg:mid-circuit} does not explore this aspect in detail, instead adopting a naive approach to determining the order and placement of uncomputation.
This limitation causes the failure on ConditionalCleanMCS in Table~\ref{tab:benchmark}, which, despite being a single benchmark instance, represents a broader class of problems.

One such failure occurs in the three-bit IntegerComparator shown in Fig.~\ref{fig:failure}.
Algorithm~\ref{alg:mid-circuit} determines the order of uncomputation based on the last gate involving each ancilla qubit, leading it to synthesize uncomputation for \(a_0\) before \(a_1\).
However, the subsequent uncomputation for \(a_1\) disrupts the previously synthesized uncomputation for \(a_0\), resulting in incorrect uncomputation.

Nevertheless, this limitation does not compromise our main conclusion regarding the effectiveness of our formulation and Use-first strategy, in the sense that we separate the synthesis of uncomputation (Algorithm~\ref{alg:use-first}) from the determination of its ordering and placement (Algorithm~\ref{alg:mid-circuit}), treating them as two distinct problems.

To address this limitation, we extend Algorithm~\ref{alg:mid-circuit} into a more sophisticated algorithm incorporating preliminary dependency analysis, while retaining repeated applications of the Use-first strategy in Algorithm~\ref{alg:use-first} as the core mechanism for synthesizing uncomputation.
Our extended strategy successfully handles 59 out of 60 circuits in an expanded benchmark covering a wider range of parameters, larger scales, and more complex scenarios.
We defer the extension of Algorithm~\ref{alg:mid-circuit} and the corresponding experimental results to appendix.

\begin{figure}[t]
    \begin{minipage}{0.35\textwidth}
        \begin{quantikz}[row sep={0.50cm,between origins},column sep=0.23cm,
            /tikz/remember picture,
            /tikz/execute at end picture={
                \node[fit={(ggroup-1-2)},inner sep=0pt,outer sep=0pt]
                    (fig12-first-cnot-frame) {};
            }]
            \lstick{\(\cS\)}    & \ctrl{1}\gategroup[wires=2,steps=1,background,
                style={draw=black,fill=none,solid,line width=0.6pt,inner xsep=-1.5pt,inner ysep=-2pt,yshift=2pt}]{}
            & \qw      & \ctrl{1} & \qw      &  \\
            \lstick{\(a:|Y\>\)} & \targ{}  & \gate{H} & \targ{}  & \gate{H} & \rstick{\(|Y\>\)}
        \end{quantikz}
    \end{minipage}
    \hspace{.5em}
    {\huge \(=\)}
    \hspace{.5em}
    \begin{minipage}{0.55\textwidth}
        \begin{quantikz}[row sep={0.50cm,between origins},column sep=0.23cm,
            /tikz/remember picture,
            /tikz/execute at end picture={
                \node[draw=black,fill=none,line width=0.6pt,inner sep=1.5pt,
                    fit={(ggroup-1-2)(ggroup-1-3)
                         (fig13-use-label.west)(fig13-use-label.east)
                         (fig13-prepare-label.west)(fig13-prepare-label.east)
                         ([yshift=-4pt]fig13-use-label.north)([yshift=-4pt]fig13-prepare-label.north)}]
                    (fig13-cnot-frame) {};
                \node[anchor=north,yshift=1pt,font=\footnotesize]
                    at (fig13-cnot-frame.south) {CNOT gate};
                \draw[overlay,black,line width=0.8pt,
                    -{Latex[length=1.6mm,width=1.1mm]}]
                    (fig12-first-cnot-frame.north)
                    |- ([yshift=-3pt]fig13-cnot-frame.north west);
            }]
            \lstick{\(\cS\)} & \ctrl{1}\gategroup[wires=2,steps=1,background,
                style={csuse,rounded corners=2pt,inner xsep=-1.5pt,inner ysep=-1pt,yshift=0pt},
                label style={csuselabel,label position=above,anchor=south,yshift=-8pt,name=fig13-use-label}]{\footnotesize\strut Use}
            & \qw\gategroup[wires=2,steps=3,background,
                style={cscomp,rounded corners=2pt,inner xsep=-1.5pt,inner ysep=-1pt,yshift=0pt},
                label style={cscomplabel,label position=above,anchor=south,yshift=-8pt,name=fig13-prepare-label}]{\footnotesize\strut Prepare}
            & \ctrl{1} & \qw &[2pt] \qw\csgateframe{clean}{2}{3}{Uncomputation} & \ctrl{1} & \qw & \\
            \lstick{\(a:|Y\>\)} & \gate{iY} & \gate{H} & \targ{}  & \gate{H} & \gate{H} & \targ{}  & \gate{H} & \rstick{\(|Y\>\)}
        \end{quantikz}
    \end{minipage}
        \caption{
            Circuit for teleporting an \(S\) gate, adapted from Fig.~29 of~\cite{gate-teleportation}.
            The ancilla register is \(\cA=\{a\}\), initialized in the magic state \(|Y\>=S|+\>\), and the intended functionality is \(W=S\).
            The semantics of the first CNOT gate involves both using and preparing the ancilla simultaneously.
        }
        \label{fig:gate-teleportation}
\end{figure}
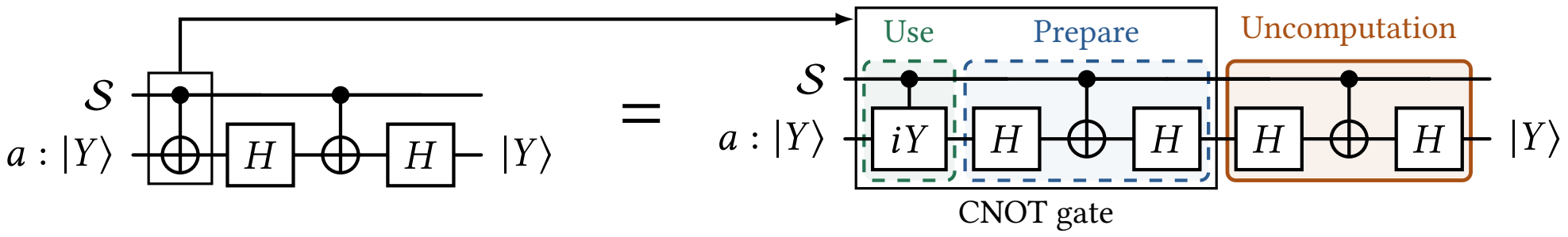

The final limitation concerns the inability to handle more general and challenging cases, as exemplified by the gate teleportation circuit illustrated in Fig.~\ref{fig:gate-teleportation} and listed in Category V of Table~\ref{tab:benchmark}.
In this case, neither a {\color{csRoleBlue}Prepare} nor a {\color{csPhaseGreen}Use} specification alone captures the intended behavior of the first CNOT gate, since it simultaneously implements nontrivial functionality and transforms the ancilla role.

Although a controlled-\(iY\) gate alone suffices to implement the desired functionality, a CNOT gate is used instead, as it is more standard in fault-tolerant quantum computing schemes~\cite{gate-teleportation}.
This CNOT gate can be decomposed into a controlled-\(iY\) gate and a controlled-\(Z\) gate, with the latter further decomposed into two Hadamard gates and one CNOT gate.
Then, uncomputation only needs to reverse the controlled-\(Z\) component, yielding the circuit shown in Fig.~\ref{fig:gate-teleportation}.

Nevertheless, this limitation does not undermine the effectiveness of our formulation, but rather highlights the challenges of automatic uncomputation synthesis.
To address this limitation, in the next section, we develop a complete reasoning system for gate specifications and systematic uncomputation synthesis to handle such general challenging cases.

\begin{table}[t]
    \centering
    \caption{Guarantees and intended use cases of automatic uncomputation synthesis for different inputs.}
    \label{tab:automatic-uncomputation-synthesis}
    \renewcommand{\arraystretch}{1.2}
    \renewcommand{\tabularxcolumn}[1]{p{#1}}
    \footnotesize
    \begin{tabularx}{\linewidth}{@{}>{\raggedright\arraybackslash}p{0.18\linewidth}>{\raggedright\arraybackslash}X>{\raggedright\arraybackslash}X@{}}
        \toprule
        \centering\arraybackslash\makecell[c]{\textbf{Input}\\\textbf{Information}}
        & \centering\arraybackslash\makecell[c]{\textbf{Guarantees of automatically}\\\textbf{synthesized uncomputation}}
        & \centering\arraybackslash\makecell[c]{\textbf{Intended use cases}} \\
        \midrule
        Computation circuit \(U\) alone & Neither successful synthesis nor preservation of the intended functionality is guaranteed, due to inherent ambiguity. & Greater automation and smaller uncompu\-tation circuits are prioritized over general correctness guarantees.\\
        \addlinespace
        Ancilla role \(R\) alone & A valid \(V:R\ixRightarrow{I}|0\>_\cA\otimes I_\cS\) can always be synthesized, with no circuit-size guarantee. & Correctness is prioritized over uncomputation circuit size. \\
        \addlinespace
        Derivation \(\pi\) of the circuit specification \(U:|0\>_\cA\otimes I_\cS\ixRightarrow{W}R\) & A valid uncomputation \(V\) can always be synthesized by structural induction on \(\pi\), with circuit size linear in the length of \(\pi\). & Both correctness and circuit-size efficiency are guaranteed, at the cost of reduced automation due to additional programmer input. \\
        \bottomrule
\end{tabularx}
\end{table}

\section{Complete Reasoning and Systematic Synthesis of Uncomputation}\label{sec:reasoning}
In this section, we develop a more systematic framework for uncomputation synthesis to address cases beyond the capabilities of the preceding automatic strategy.
This development begins with identifying an inherent ambiguity in uncomputation, as formalized in the following theorem.

\begin{theorem}[Inherent Ambiguity of Functionality]\label{thm:ambiguity}
    For any unitary \(U\) acting on \(\cA\otimes\cS\) and any unitary \(W\) acting on \(\cS\), there exist an ancilla role \(R\) and a unitary \(V\) acting on \(\cA\otimes\cS\) such that
    \[
        U:|0\>_\cA\otimes I_\cS\xRightarrow{W}R
        \mbox{ and }
        V:R\xRightarrow{I}|0\>_\cA\otimes I_\cS\mbox{ are both valid specifications.}
    \]
\end{theorem}
\begin{proof}
    It suffices to choose \(R=U\pare{|0\>_\cA\otimes W_\cS^\dagger}\) and \(V=(I_\cA\otimes W)U^\dagger\).
\end{proof}
This theorem implies that, in principle, the computation circuit \(U\) alone conveys no more information about the desired functionality \(W\) than no information at all, since an uncomputation \(V\) realizing any desired \(W\) can always be constructed extensionally by first reversing \(U\) and then applying \(W\) to the primary system \(\cS\).
Therefore, an automatic uncomputation synthesis strategy must adopt some predetermined assumptions under which the functionality \(W\) is uniquely determined by the computation unitary \(U\), and then synthesize the corresponding uncomputation \(V\) from the input circuit.

As discussed before, prior formulations for classical ancilla use adopt a fixed interpretation of functionality, effectively fixing \(W=\pare{\sum_x\<x|\otimes I}U(|0\>\otimes I)\).
Our Use-first strategy introduced in the previous section does not impose predetermined assumptions on the form of functionality or the mode of ancilla use, thereby accommodating a broader range of cases.
However, it still presupposes two types of gate specifications: preparing the ancilla or using the ancilla.
Circuits may go beyond these two cases, with individual gates simultaneously implementing nontrivial functionality and transforming the ancilla role, as exemplified by the gate teleportation circuit in Figure~\ref{fig:gate-teleportation}, included as a general hard case in Table~\ref{tab:benchmark}.

In such cases, the boundary between computation and uncomputation becomes blurred, making it difficult to intuitively identify which parts of the circuit constitute uncomputation.
Nevertheless, we seek to accommodate such general ancilla-assisted circuits within our formulation through an inference system that allows circuit designers to formally express their design intentions.
As long as the circuit designer provides a derivation of the circuit's specifications, which can be regarded as a formal annotation that captures the designer's intended functionality and ancilla use, we can automatically synthesize an efficient uncomputation for the circuit.

In summary, depending on the intended use cases, automatic uncomputation synthesis can be categorized into the scenarios summarized in Table~\ref{tab:automatic-uncomputation-synthesis}.
Having discussed the first two scenarios in previous sections, we now focus on the third by developing an inference system that allows programmers to formally express their design intentions for ancilla-assisted circuits.

\subsection{A Complete Inference System for Specifications}
Specifically, we write \(U:R_1\ixrightarrow{W}R_2\) for the derivability relation induced by the inference rules in Fig.~\ref{fig:inference-rules}, distinguishing it from the specification \(U:R_1\ixRightarrow{W}R_2\) defined earlier.
As formalized in the following theorem, the inference system is sound and complete with respect to specifications.

\begin{theorem}[Soundness and Completeness]
    Fix primary and ancillary systems \(\cS\) and \(\cA\).
    For any unitary \(U\) acting on \(\cA\otimes\cS\), ancilla roles \(R_1,R_2\), and functionality \(W\), we have
    \[
    U:R_1\ixRightarrow{W}R_2\mbox{ is valid} \iff U:R_1\ixrightarrow{W}R_2\mbox{ is derivable in the inference system.}
    \]
\end{theorem}
\begin{figure}[t]
    \centering
    \small
    \[
    \begin{array}{c}
        \multicolumn{1}{l}{\boxed{\textbf{Axioms}}}\\[.5em]
        \inferrule[\textsc{Ax-X}]{q\in\cS}{X_q:I_q\xrightarrow{X_q}I_q}
        \hspace{2em}
        \inferrule[\textsc{Ax-H}]{q\in\cS}{H_q:I_q\xrightarrow{H_q}I_q}
        \hspace{2em}
        \inferrule[\textsc{Ax-X-Anc}]{a\in \cA\\ b\in\{0,1\}}{X_a:|b\>_a\xrightarrow{1}|b\oplus 1\>_a}
        \hspace{2em}
        \inferrule[\textsc{Ax-H-Anc}]{a\in\cA\\ b\in\{0,1\}}{H_a:|b\>_a\xrightarrow{1}\tfrac{1}{\sqrt{2}}\pare{|0\>_a+(-1)^b|1\>_a}}\\[2em]
        \multicolumn{1}{l}{\boxed{\textbf{Structural rules}}}\\[.5em]
        \inferrule[\textsc{Seq}]{U_1:R_1\xrightarrow{W_1}R_2\\ U_2:R_2\xrightarrow{W_2}R_3}{U_1;U_2:R_1\xrightarrow{W_1;W_2}R_3}
        \hspace{3em}
        \inferrule[\textsc{Sum}]{U_j:R\xrightarrow{W_j}S\\ c_j\in\bC\\ j=1,...,\ell}{\sum_{j=1}^\ell c_jU_j:R\xrightarrow{\sum_{j=1}^\ell c_jW_j}S}
        \\[2em]
        \inferrule[\textsc{Tensor}]{U_1:R_1\xrightarrow{W_1}S_1\\ U_2:R_2\xrightarrow{W_2}S_2}{U_1\otimes U_2:R_1\otimes R_2\xrightarrow{W_1\otimes W_2}S_1\otimes S_2}
        \hspace{3em}
        \inferrule[\textsc{Glue}]{U:R_j\xrightarrow{W_j}S_j\\ A_j\in \cL(\cS)\\ j=1,...,\ell}{U:\sum_{j=1}^\ell R_jA_j\xrightarrow{I}\sum_{j=1}^\ell S_jW_jA_j}
    \end{array}
    \]
    \caption{Inference rules for the relation \(U:R_1\ixrightarrow{W}R_2\).
    Here, \(\cL(\cS)\) denotes the space of linear operators on the primary system \(\cS\).
    We implicitly require all expressions for gates, functionalities, and ancilla roles to be well-defined and denote valid objects of their respective types.}
    \label{fig:inference-rules}
\end{figure}

The inference system essentially provides a formal description of the intended behavior of each gate in an ancilla-assisted circuit. We briefly discuss the intuition behind each inference rule below.

\paragraph{Axioms} For the axioms, we distinguish between gates acting on the ancilla system and those acting on the primary system.
A gate acting solely on the ancilla system naturally implements trivial functionality while transforming the ancilla role, as captured by Rules~\textsc{Ax-X-Anc} and \textsc{Ax-H-Anc}.
In contrast, a gate acting solely on the primary system implements functionality while preserving the ancilla role, as captured by Rules~\textsc{Ax-X} and \textsc{Ax-H}.
These axioms capture gate specifications locally using only minimal ancilla roles and two gates, \(X\) and \(H\), which can then be composed through Rules~\textsc{Tensor}, \textsc{Seq}, and \textsc{Sum} to derive the specifications of more general gates over larger ancilla roles.

\paragraph{Structural Rules}
Rules~\textsc{Seq}, \textsc{Tensor}, and \textsc{Sum} are relatively straightforward: \textsc{Seq} composes gate specifications sequentially, \textsc{Tensor} extends local gate specifications to larger systems, and \textsc{Sum} derives the specifications of more general gates from those of simpler ones.
Rule~\textsc{Glue} is less intuitive: it transforms the ancilla role in a gate's specification while implementing trivial functionality, allowing us to obtain a desired ancilla role and
thereby bridge mismatched intermediate ancilla roles when composing gates sequentially.

\subsection{Derivation-Directed Automatic Uncomputation Synthesis}
The inference system in Fig.~\ref{fig:inference-rules} can also be viewed as a formal language for programmers to annotate their ancilla-assisted circuits, expressing the intended operation of each gate.
Given a derivation \(\pi\) of \(U:|0\>_\cA\otimes I_\cS\ixrightarrow{W}R\), we can then automatically synthesize an uncomputation \(V:R\ixRightarrow{I}|0\>_\cA\otimes I_\cS\) by structural induction on \(\pi\), as summarized in Table~\ref{tab:uncomputation-construction}.
The resulting uncomputation is guaranteed to be correct, with circuit size linear in the length of \(\pi\), as formalized in the following theorem.

\begin{table}[t]
    \centering
    \caption{Recursive construction of the uncomputation circuit \(V(\pi)\) according to the final inference rule of the derivation \(\pi\).}
    \label{tab:uncomputation-construction}
    \begin{tabular}{lll}
        \toprule
        \textbf{Last rule in the proof \(\pi\)} & \textbf{Conclusion} & \textbf{Extracted \(V(\pi)\)} \\
        \midrule
        \textsc{Ax-X}     &\(X_q:I_q\ixrightarrow{X_q}I_q\) & \(V(\pi)=I_q\) \\
        \textsc{Ax-H}     &\(H_q:I_q\ixrightarrow{H_q}I_q\) & \(V(\pi)=I_q\) \\
        \textsc{Ax-X-Anc} &\(X_a:|b\>_a\ixrightarrow{1}|b\oplus 1\>_a\) & \(V(\pi)=X_a\) \\
        \textsc{Ax-H-Anc} & \(H_a:|b\>_a\ixrightarrow{1}\tfrac{1}{\sqrt{2}}\pare{|0\>_a+(-1)^b|1\>_a}\) & \(V(\pi)=H_a\) \\
        \textsc{Seq}      & \(Seq(\pi_1,\pi_2)\) & \(V(\pi)=V(\pi_2);V(\pi_1)\) \\
        \textsc{Tensor}   & \(Tensor(\pi_1,\pi_2)\) & \(V(\pi)=V(\pi_1) \otimes V(\pi_2)\) \\
        \textsc{Sum}      & \(Sum(\pi_1,\pi_2,...,\pi_\ell)\) & \(V(\pi)=V(\pi_i)\) for arbitrary \(i\) \\
        \textsc{Glue}     & \(U:R\ixrightarrow{I}S\) & \(V(\pi)=U^\dagger\)\\
        \bottomrule
    \end{tabular}
\end{table}

\begin{theorem}[Soundness and Efficiency Guarantee]
    Given a derivation \(\pi\) of \(U:|0\>_\cA\otimes I_\cS\ixrightarrow{W}R\), the uncomputation \(V(\pi)\) constructed according to Table~\ref{tab:uncomputation-construction} satisfies the following properties:
    \begin{enumerate}
        \item \emph{Soundness}: \(V(\pi)\) admits the specification \(V(\pi):R\ixRightarrow{I}|0\>_\cA\otimes I_\cS\), ensuring that \(V(\pi)\) is a valid uncomputation.
        \item \emph{Efficiency}: The circuit size of \(V(\pi)\), measured by the number of sequential gate applications, is linear in the length of \(\pi\), i.e., \(|V(\pi)|=O(|\pi|)\).
    \end{enumerate}
\end{theorem}
The proof of this theorem is deferred to appendix.

\subsection{Application to General Hard Case}\label{sec:case}

We now apply the recursive construction to the challenging case of the gate teleportation circuit shown in Fig.~\ref{fig:gate-teleportation}.
The first CNOT gate simultaneously implements nontrivial functionality and transforms the ancilla role, making it difficult to handle automatically.
However, our inference system allows programmers to explicitly specify these two effects.
Specifically, by applying rule \textsc{Seq}, the final step in the derivation for the CNOT gate is given by
\[
\inferrule{
    \mathrm{CiY}:\State(Y)\ixrightarrow{S}\State(Y)
    \\
    H_a;\mathrm{CNOT};H_a:\State(Y)\ixrightarrow{I}
    \Choice(iZ)
}{
    \mathrm{CNOT}= \mathrm{CiY};H_a;\mathrm{CNOT};H_a:\State(Y)\xrightarrow{S}
    \Choice(iZ)
}{\textsc{(Seq)}}
\]
where both premises can be further derived using the inference rules.
The complete derivation and corresponding uncomputation extraction are deferred to the appendix.

This derivation captures the programmer's intent to decompose the CNOT gate into two semantic components: the first {\color{csPhaseGreen}{Use}}s the ancilla to implement the intended functionality, while the second {\color{csRoleBlue}{Prepare}}s it by transforming its role.
Although the second component may appear redundant, since the intended functionality is already realized by the first, the original CNOT gate is more natural and standard in practical fault-tolerant quantum circuits.
Thus, this semantic decomposition can be explicitly annotated within the inference system without modifying the underlying circuit.

The uncomputation can then be constructed either by reversing the {\color{csRoleBlue}{Prepare}} gates, as in the preceding strategy, or by structural induction on the derivation according to Table~\ref{tab:uncomputation-construction}.
Both approaches yield the complete circuit shown in Fig.~\ref{fig:gate-teleportation}.

\section{Related Work}\label{sec:rw} 
In this section, we organize our discussion of related work into three aspects, focusing on the use and benefits of ancilla qubits in quantum computing, as well as the synthesis and verification of their uncomputation.

\paragraph{Ancilla-assisted computation and circuit constructions} 
Ancilla qubits are used in two forms. 
First, \emph{clean ancillas} are initialized to specific states and restored after computation, enabling resource trade-offs in circuits such as arithmetic~\cite{ancilla-arithmetic,cuccaro-adder,ancilla-arithmetic-1,ancilla-arithmetic-2}, state preparation~\cite{state-preparation-1,state-preparation-2}, and quantum Fourier transforms~\cite{QFT-1,QFT-2}. 
Second, \emph{dirty ancillas} may start in unknown states but must be restored afterward, allowing more flexible reuse of available qubits~\cite{dirty-1,dirty-2,dirty-3}.

Beyond case-by-case design, ancilla use has also been studied from more general perspectives.
For example, \citet{controlled-T} characterizes trade-offs between clean ancillas and $T$-depth; 
\citet{catalystic} use reusable catalysts for gate-set embeddings, shifting approximation costs to state preparation; and
\citet{Khattar2025riseofconditionally} study conditionally clean ancillas and their resource benefits across multiple circuit families.

\paragraph{Uncomputation synthesis and ancilla resource management} 

\citet{Bennett} establishes compute-copy-uncompute as a foundation for reversible cleanup, which \citet{pebbling} extend through reversible pebbling to synthesize space-constrained cleanup schedules. 
SQUARE~\cite{SQUARE} balances uncomputation costs against the benefits of ancilla reuse. 
Unqomp~\cite{unqomp} synthesizes uncomputation through dependency analysis with additional annotations, while Reqomp~\cite{reqomp} incorporates recomputation and recycling under qubit budgets. 
Extending these approaches, Qrisp~\cite{qrisp} introduces composite-gate analysis and permeability, while \citet{modular} develop an intermediate representation for modular synthesis of uncomputation and adjoints. 

To handle dependency cycles, \citet{partial} combine subset selection with partial reversal to recover a subset of ancillas.
SPARE~\cite{SPARE} further optimizes compute--uncompute patterns using ancilla-state information.
For dirty ancillas, BONA~\cite{BONA} optimizes borrowing schedules under restoration guarantees, while RwUn~\cite{RwUn} combines static reasoning and rewriting to synthesize uncomputation for both clean and dirty ancillas.

\paragraph{Language abstractions and verifications of ancillas}
Quipper~\cite{Quipper} supports scoped ancillas and automatic cleanup when compiling classical functions into reversible oracles.
ReQWIRE~\cite{ReQWIRE} supports proofs of ancilla termination assertions, while Silq~\cite{silq} uses type annotations for safe implicit uncomputation of temporary values. Extending type-based reasoning, Twist~\cite{Twist} combines purity types with static and runtime checks to verify ancilla disentanglement, while Qurts~\cite{qurts} employs lifetime-parameterized affine types to ensure safe uncomputation and discarding under different cleanup strategies. 
At the circuit level, \citet{proto-quipper} extend Proto-Quipper with modalities for reversibility and controllability, formalizing reversal, control, and the \texttt{with-computed} pattern. 

Focusing on restoration guarantees, \citet{ancilla-safety} reduce clean and dirty ancilla safety to commutativity checks, enabling error diagnosis and local repair of certain separable faults.
For automatic verification, AutoQ~\cite{AutoQ} uses symbolic tree automata to check input--output specifications, including ancilla restoration, while \citet{autoq-spec} extend the set-based specification language and improve its translation to compact automata.


\section*{Disclosure of Use of AI in Research}
We used ChatGPT-6 Astra Ultra to assist with experimental implementation and benchmarking under explicit instructions to follow our pseudocode and methodology, including configuring existing tools. 
We also used it to assess the fairness of comparisons with prior work, verify factual statements, examine theorem correctness, and check the coverage of related work.

\section*{Acknowledgments}
The authors thank Professor Charles Yuan for his mentorship and guidance throughout this work, and Adela Deng, Bhargav Subramanian, Daksh Shami, and Gunagya Singh Mamak for insightful discussions and valuable feedback.

\bibliography{ref}
\bibliographystyle{ACM-Reference-Format}

\clearpage
\appendix

\section{Deferred Proofs}\label{app:proofs}
In this section, we provide the omitted proofs in the main text.

\subsection{Proof of Proposition~\ref{prop:choice-role}}
\begin{proposition}[Coherent Control via Ancilla Role \(\Choice(G)\)]
    Let \(G\) be a unitary on the primary system \(\cS\) such that \(G^2=I_\cS\).
    Then an \(X\) gate on the one-qubit ancilla system \(\cA\) admits the following specification:
    \[
    X_\cA:\Choice(G)\xRightarrow{G}\Choice(G).
    \]
\end{proposition}
\begin{proof}
    According to Definition~\ref{def:semantics}, it suffices to check the commutativity \(X_\cA \cdot \Choice(G) = \Choice(G) \cdot G\). This can be verified by direct calculation:
    \[
    \begin{aligned}
        X_\cA\cdot \Choice(G)&=X_\cA\cdot \tfrac{1}{\sqrt{2}}\pare{|0\>_\cA\otimes I_\cS+|1\>_\cA\otimes G_\cS}\\
        &=\tfrac{1}{\sqrt{2}}\pare{|0\>_\cA\otimes G_\cS+|1\>_\cA\otimes I_\cS}\\
    &=\tfrac{1}{\sqrt{2}}\pare{|0\>_\cA\otimes I_\cS+|1\>_\cA\otimes G_\cS}\cdot G_\cS\quad (\mbox{because }G^2=I)\\
    &=\Choice(G)\cdot G_\cS.
    \end{aligned}
    \]
\end{proof}

\subsection{Proof of Lemma~\ref{lemma:prepare-use}}
\begin{lemma}[Conditions for {Prepare} and {Use} Specifications]
    Let \(U\) be a unitary on \(\cA\otimes\cS\) and \(R:\cS\to\cA\otimes\cS\) an ancilla role. Then:
    \begin{enumerate}
        \item There exists a unique ancilla role \(R'\) such that \(U:R\ixRightarrow{I}R'\), given by \(R'=UR\).
        \item There exists a functionality \(W\) such that \(U:R\ixRightarrow{W}R\) if and only if \(U(\operatorname{im}R)=\operatorname{im}R\). In this case, \(W\) is uniquely determined by \(W=R^\dagger UR\).
    \end{enumerate}
\end{lemma}
\begin{proof}
    Recall that \(R^\dagger R=I_\cS\), since an ancilla role is an isometry, and that \(RR^\dagger\) is the orthogonal projector onto \(\operatorname{im}R\).

    \paragraph{Prepare specifications.}
    By Definition~\ref{def:semantics}, the specification \(U:R\ixRightarrow{I}R'\) is valid exactly when \(UR=R'\).
    Thus the only possible choice is \(R'=UR\).
    This choice is indeed an ancilla role, because
    \[
        (UR)^\dagger(UR)=R^\dagger U^\dagger UR=R^\dagger R=I_\cS.
    \]
    This proves existence and uniqueness.

    \paragraph{Use specifications: necessity.}
    Suppose that \(U:R\ixRightarrow{W}R\) is valid, so \(UR=RW\).
    Since \(W\) is unitary, it maps \(\cS\) onto itself. Hence
    \[
        U(\operatorname{im}R)
        =\operatorname{im}(UR)
        =\operatorname{im}(RW)
        =\operatorname{im}R.
    \]
    Multiplying \(UR=RW\) on the left by \(R^\dagger\) also gives \(W=R^\dagger UR\), proving uniqueness.

    \paragraph{Use specifications: sufficiency.}
    Conversely, suppose that \(U(\operatorname{im}R)=\operatorname{im}R\), and set \(W=R^\dagger UR\).
    The image of \(UR\) lies in \(\operatorname{im}R\), so the projector \(RR^\dagger\) acts as the identity on \(UR\). Therefore,
    \[
        RW=RR^\dagger UR=UR,
        \qquad
        W^\dagger W
        =R^\dagger U^\dagger RR^\dagger UR
        =R^\dagger U^\dagger UR
        =I_\cS.
    \]
    Since \(W\) acts on the finite-dimensional space \(\cS\), the equality \(W^\dagger W=I_\cS\) implies that \(W\) is unitary.
    Thus \(W\) is a functionality, and \(UR=RW\) establishes the required specification.
\end{proof}

\subsection{Proof of the Soundness and Completeness Theorem}
\begin{theorem}[Soundness and Completeness]
    Fix primary and ancillary systems \(\cS\) and \(\cA\).
    For any unitary \(U\) acting on \(\cA\otimes\cS\), ancilla roles \(R_1,R_2\), and functionality \(W\), we have
    \[
        U:R_1\ixRightarrow{W}R_2\mbox{ is valid}
        \iff U:R_1\ixrightarrow{W}R_2\mbox{ is derivable in the inference system.}
    \]
\end{theorem}
\begin{proof}
    We first check that each inference rule preserves validity.
    We then prove completeness by constructing a derivation for an arbitrary valid specification.
    Throughout, we use the typing conditions in Fig.~\ref{fig:inference-rules}: gates and functionalities must be unitary, and roles must be isometries.
    For empty registers, we use the usual empty-tensor convention: the empty tensor product of judgments is the scalar identity judgment \(1:1\ixrightarrow{1}1\).

    \paragraph{Soundness.}
    Each axiom satisfies the defining equality \(UR=SW\), by the action of \(X\) and \(H\) on computational-basis states.
    The structural rules preserve this equality as follows.
    \begin{itemize}
        \item \emph{\textsc{Seq}.}
        Recall that \(U_1;U_2\) applies \(U_1\) first, so its matrix is \(U_2U_1\).
        The premise equalities imply
        \[
            (U_2U_1)R_1=U_2(U_1R_1)=U_2(R_2W_1)=(U_2R_2)W_1=(R_3W_2)W_1=R_3(W_2W_1),
        \]
        which is the equality required by the conclusion.

        \item \emph{\textsc{Tensor}.}
        Tensoring the two premise equalities gives
        \[
            (U_1\otimes U_2)(R_1\otimes R_2)=
            (U_1R_1)\otimes (U_2R_2)=
            (S_1W_1)\otimes (S_2W_2)
            =(S_1\otimes S_2)(W_1\otimes W_2),
        \]
        with the natural identification of the register orderings.

        \item \emph{\textsc{Sum}.}
        Since all premises have the same roles \(R,S\),
        \[
            \pare{\sum_{j=1}^\ell c_jU_j}R
            =\sum_{j=1}^\ell c_jU_jR
            =\sum_{j=1}^\ell c_jSW_j
            =S\pare{\sum_{j=1}^\ell c_jW_j}.
        \]
        The typing conditions ensure that both sums are unitaries.

        \item \emph{\textsc{Glue}.}
        Using \(UR_j=S_jW_j\) for each premise, we obtain
        \[
            U\pare{\sum_{j=1}^{\ell}R_jA_j}
            =\sum_{j=1}^{\ell}UR_jA_j
            =\sum_{j=1}^{\ell}S_jW_jA_j.
        \]
        This is exactly the conclusion with identity functionality; its two roles are isometries by the typing conditions.
    \end{itemize}
    Induction on a derivation proves soundness.

    \paragraph{Completeness.}
    The idea is to derive arbitrary changes of role with identity functionality, and then insert the desired functionality between two such changes.
    We establish the required derivations in five steps.

    \noindent\emph{Step 1: elementary gates on arbitrary roles.}
    Let \(m\) and \(n\) be the numbers of qubits in \(\cA\) and \(\cS\), respectively.
    For each computational-basis string \(b\in\{0,1\}^m\) of the ancilla register, write
    \[
        R_b=|b\>_\cA\otimes I_\cS.
    \]
    Identity judgments follow from \(X;X=I\), so \textsc{Tensor} can extend the local axioms to all the registers.
    For an elementary gate \(U=X_q\) or \(H_q\), these axioms give
    \[
        \begin{cases}
            U:R_b\ixrightarrow{I}UR_b, & q\in\cA,\\
            U:R_b\ixrightarrow{U}R_b, & q\in\cS.
        \end{cases}
    \]
    In the second case, the functionality is understood as the corresponding gate on \(\cS\), and \(UR_b=R_bU\).

    Now let \(R\) be any role and set \(A_b=R_b^\dagger R\).
    The basis decomposition of the ancilla gives
    \[
        \sum_{b\in\{0,1\}^m}R_bR_b^\dagger=I_{\cA\otimes\cS},
        \qquad R=\sum_{b\in\{0,1\}^m}R_bA_b.
    \]
    In either case above, \textsc{Glue} therefore derives
    \[
        U:R\ixrightarrow{I}UR.
    \]
    Its output role is an isometry, since \(U\) is unitary.
    This also allows \textsc{Seq} to compose elementary gates when the intermediate role is a superposition.

    \noindent\emph{Step 2: diagonal gates on arbitrary roles.}
    First we derive a useful specification for \(Z\) on an ancilla qubit.
    Step~1 and \(Z=HXH\) give
    \[
        Z_a:|b\>_a\ixrightarrow{1}z_b|b\>_a,
        \qquad z_b=(-1)^b,\quad b\in\{0,1\}.
    \]
    We need the phase \(z_b\) in the functionality rather than in the output role.
    This transfer follows from three legal rule applications:
    \[
        \begin{aligned}
            z_bZ_a&:|b\>_a\ixrightarrow{z_b}z_b|b\>_a
                &&\text{by a one-term \textsc{Sum}},\\
            z_bZ_a&:|b\>_a\ixrightarrow{1}|b\>_a
                &&\text{by \textsc{Glue} with }A_1=1,\\
            Z_a&:|b\>_a\ixrightarrow{z_b}|b\>_a
                &&\text{by a one-term \textsc{Sum}}.
        \end{aligned}
    \]
    Both uses of \textsc{Sum} have coefficient \(z_b\); the middle step uses \(z_b^2=1\).
    All gates and scalar functionalities here are unitary, and all displayed states are normalized.

    On a primary qubit, the axioms and \textsc{Seq} directly derive \(Z_q:I_q\ixrightarrow{Z_q}I_q\).
    Tensoring these judgments, including identities on the remaining qubits, gives
    \[
        Z_\cA^r\otimes Z_\cS^s:
        R_b\ixrightarrow{(-1)^{r\cdot b}Z_\cS^s}R_b,
    \]
    where \(b,r\in\{0,1\}^m\), \(s\in\{0,1\}^n\), and
    \[
        Z_\cA^r=\bigotimes_{j=1}^{m}Z^{r_j},
        \qquad Z_\cS^s=\bigotimes_{j=1}^{n}Z^{s_j}.
    \]

    Let \(U\) be any diagonal unitary on the joint register.
    The Pauli \(Z\) strings form a basis for diagonal operators, so write
    \[
        U=\sum_{r\in\{0,1\}^m}\sum_{s\in\{0,1\}^n}
        c_{r,s}Z_\cA^r\otimes Z_\cS^s.
    \]
    For each \(b\), \textsc{Sum} yields
    \[
        U:R_b\ixrightarrow{W_b}R_b,
        \qquad W_b=R_b^\dagger U R_b.
    \]
    Here \(W_b\) is the diagonal block obtained by fixing the ancilla to \(b\).
    Its diagonal entries have modulus one, so \(W_b\) is unitary and this application of \textsc{Sum} is well typed.
    For an arbitrary role \(R\), use \(A_b=R_b^\dagger R\) again.
    Since \(UR_b=R_bW_b\), \textsc{Glue} derives
    \[
        U:R\ixrightarrow{I}\sum_{b\in\{0,1\}^m}R_bW_bA_b=UR.
    \]
    Thus every diagonal unitary admits a derivable Prepare specification on every role.

    \noindent\emph{Step 3: arbitrary physical unitaries.}
    Every unitary on a finite qubit register has an exact finite decomposition into diagonal unitaries and individual Hadamard gates.
    We explain this fact to make clear that no approximation is needed here.
    Unitary Gaussian elimination expresses a unitary as a product of two-level unitaries, each acting only on the span of two computational-basis states, and a diagonal remainder~\cite{qcqi}.
    By conjugating with swaps of basis states along a path whose successive strings differ in one bit, the two states can be brought to a pair that differs in just one bit.
    The resulting two-level unitary is a one-qubit gate controlled on a fixed pattern of all other bits; the swaps are controlled \(X\) gates of the same form.
    A one-qubit gate has an Euler decomposition~\cite{qcqi}:
    \[
        e^{i\alpha}R_z(\beta)R_x(\gamma)R_z(\delta),
        \qquad R_x(\gamma)=H R_z(\gamma)H,
    \]
    where \(R_z(\theta)=e^{-i\theta Z/2}\) and \(R_x(\theta)=e^{-i\theta X/2}\).
    Adding the fixed controls makes the phase and \(Z\)-rotation factors diagonal on the full register.
    The controlled \(X\)-rotation is a controlled \(Z\)-rotation conjugated by an ordinary \(H\) on the target: outside the selected control pattern, the two Hadamards cancel.
    Likewise, a controlled \(X\) equals a controlled \(Z\) conjugated by these Hadamards.
    This establishes the asserted decomposition.

    Steps~1 and~2 handle every factor of that decomposition.
    Applying \textsc{Seq} to the factors therefore gives the derived rule
    \begin{equation}\label{eq:app-arbitrary-prepare}
        U:R\ixrightarrow{I}UR
        \qquad\text{for every joint unitary }U\text{ and every role }R.
    \end{equation}
    Every intermediate role is an isometry, since it is obtained by applying unitaries to \(R\).

    \noindent\emph{Step 4: arbitrary functionality at the initial role.}
    On the primary system alone, we can derive
    \[
        W:I_\cS\ixrightarrow{W}I_\cS
        \qquad\text{for every primary unitary }W.
    \]
    Indeed, \textsc{Seq} gives \(I=X^2\) and \(Z=HXH\), and a one-term \textsc{Sum} supplies the phase in \(Y=iXZ\).
    Hence the primary axioms derive the corresponding judgment for each of \(I,X,Y,Z\).
    Rule \textsc{Tensor} gives the corresponding judgments for all primary Pauli strings \(P_1,\ldots,P_{4^n}\).
    Expand \(W\) in this basis as
    \[
        W=\sum_{j=1}^{4^n}c_jP_j.
    \]
    Applying \textsc{Sum} proves the claim: both sums in the conclusion equal the unitary \(W\).
    Tensoring with the ancilla identity judgment yields
    \[
        I_\cA\otimes W:R_{\mathrm{init}}\ixrightarrow{W}R_{\mathrm{init}},
        \qquad R_{\mathrm{init}}=R_{0^m}=|0\>_\cA\otimes I_\cS.
    \]

    \noindent\emph{Step 5: the desired specification.}
    Consider an arbitrary valid specification \(U:R_{\mathrm{in}}\ixRightarrow{W}R_{\mathrm{out}}\), so \(UR_{\mathrm{in}}=R_{\mathrm{out}}W\).
    Choose a joint unitary \(U_{\mathrm{in}}\) satisfying \(U_{\mathrm{in}}R_{\mathrm{init}}=R_{\mathrm{in}}\).
    Such a unitary exists: the columns of \(R_{\mathrm{init}}\) and \(R_{\mathrm{in}}\) are orthonormal families of the same size, so we can extend both to orthonormal bases and map the first basis to the second.
    Set
    \[
        U_W=I_\cA\otimes W,
        \qquad U_{\mathrm{out}}=UU_{\mathrm{in}}U_W^\dagger.
    \]
    These are unitaries, and
    \[
        U_{\mathrm{out}}R_{\mathrm{init}}
        =UU_{\mathrm{in}}R_{\mathrm{init}}W^\dagger
        =UR_{\mathrm{in}}W^\dagger
        =R_{\mathrm{out}}.
    \]
    Equation~\eqref{eq:app-arbitrary-prepare} and Step~4 give
    \[
        \begin{aligned}
            U_{\mathrm{in}}^\dagger&:R_{\mathrm{in}}\ixrightarrow{I}R_{\mathrm{init}},\\
            U_W&:R_{\mathrm{init}}\ixrightarrow{W}R_{\mathrm{init}},\\
            U_{\mathrm{out}}&:R_{\mathrm{init}}\ixrightarrow{I}R_{\mathrm{out}}.
        \end{aligned}
    \]
    Applying \textsc{Seq} twice derives \(U_{\mathrm{in}}^\dagger;U_W;U_{\mathrm{out}}:R_{\mathrm{in}}\ixrightarrow{W}R_{\mathrm{out}}\).
    Its physical unitary is
    \[
        U_{\mathrm{out}}U_WU_{\mathrm{in}}^\dagger
        =UU_{\mathrm{in}}U_W^\dagger U_WU_{\mathrm{in}}^\dagger
        =U,
    \]
    so this is the required derivation of \(U:R_{\mathrm{in}}\ixrightarrow{W}R_{\mathrm{out}}\).
\end{proof}

\subsection{Proof of the Soundness and Efficiency Guarantee Theorem}
\begin{theorem}[Soundness and Efficiency Guarantee]
    Given a derivation \(\pi\) of \(U:|0\>_\cA\otimes I_\cS\ixrightarrow{W}R\), the uncomputation \(V(\pi)\) constructed according to Table~\ref{tab:uncomputation-construction} satisfies the following properties:
    \begin{enumerate}
        \item \emph{Soundness}: The specification \(V(\pi):R\ixRightarrow{I}|0\>_\cA\otimes I_\cS\) is valid, ensuring that \(V(\pi)\) is a valid uncomputation.
        \item \emph{Efficiency}: The circuit size of \(V(\pi)\), measured by the number of sequential gate applications, is linear in the length of \(\pi\), i.e., \(|V(\pi)|=O(|\pi|)\).
    \end{enumerate}
\end{theorem}
\begin{proof}
    We establish soundness first and then bound the size of the extracted circuit.

    \paragraph{Soundness.}
    We prove a stronger statement that also covers intermediate steps of a derivation: whenever \(\pi\) derives \(U:R_{\mathrm{in}}\ixrightarrow{W}R_{\mathrm{out}}\), the extracted circuit \(V(\pi)\) is unitary and satisfies
    \[
        V(\pi)R_{\mathrm{out}}=R_{\mathrm{in}}.
    \]
    In other words, it restores the input role without applying any logical functionality.
    We proceed by induction on the last rule of \(\pi\).
    \begin{itemize}
        \item \emph{Axioms.}
        For \textsc{Ax-X} and \textsc{Ax-H}, both roles are \(I_q\), and the extracted circuit is \(I_q\), so the claim is immediate.
        For \textsc{Ax-X-Anc}, it follows from \(X|b\oplus1\>=|b\>\).
        For \textsc{Ax-H-Anc}, the output role is \(H|b\>\), which the extracted \(H\) maps back to \(|b\>\), since \(H^2=I\).
        All four extracted circuits are unitary.

        \item \emph{\textsc{Seq}.}
        Suppose that the two premises have roles \(R_1\to R_2\) and \(R_2\to R_3\), respectively, and write \(V_i=V(\pi_i)\).
        By induction, \(V_1R_2=R_1\) and \(V_2R_3=R_2\).
        The extracted circuit applies \(V_2\) first and then \(V_1\), giving
        \[
            V(\pi)R_3=V_1V_2R_3=V_1R_2=R_1.
        \]
        It is unitary as a composition of unitaries.

        \item \emph{\textsc{Tensor}.}
        If the premises have roles \(R_i\to S_i\), the induction hypotheses give \(V(\pi_i)S_i=R_i\). Hence
        \[
            \bigl(V(\pi_1)\otimes V(\pi_2)\bigr)(S_1\otimes S_2)
            =R_1\otimes R_2.
        \]
        The tensor product of the two extracted unitaries is unitary.

        \item \emph{\textsc{Sum}.}
        All \(\ell\) premises have the same input role \(R\) and output role \(S\).
        Thus \(V(\pi_i)S=R\) for every \(i=1,\ldots,\ell\), by induction.
        Choosing any one of these circuits gives a unitary with the required action; there is no need to combine the extracted circuits.

        \item \emph{\textsc{Glue}.}
        For a rule instance with \(\ell\) premises, write the input and output roles of the conclusion as
        \[
            R=\sum_{j=1}^{\ell}R_jA_j,
            \qquad
            S=\sum_{j=1}^{\ell}S_jW_jA_j.
        \]
        Soundness of the inference system gives \(UR_j=S_jW_j\) for every premise. Consequently,
        \[
            UR=\sum_{j=1}^{\ell}UR_jA_j=\sum_{j=1}^{\ell}S_jW_jA_j=S.
        \]
        The extracted circuit \(U^\dagger\) is unitary and satisfies \(U^\dagger S=R\).
    \end{itemize}
    This completes the induction.
    Applying it to \(R_{\mathrm{in}}=|0\>_\cA\otimes I_\cS\) and \(R_{\mathrm{out}}=R\) proves soundness.

    \paragraph{Efficiency.}
    We count gate applications at the granularity of the circuit descriptions supplied in the derivation, and include those descriptions in its length \(|\pi|\).
    In particular, if a \textsc{Glue} step records a circuit \(U=U_1;\cdots;U_k\), that description contributes at least \(k\) symbols to \(|\pi|\), and its inverse is
    \[
        U^\dagger=U_k^\dagger;\cdots;U_1^\dagger,
    \]
    with the same number of gate applications.

    During extraction, a \textsc{Seq} or \textsc{Tensor} step processes both subderivations, a \textsc{Sum} step processes only the selected subderivation, and a \textsc{Glue} step emits the inverse circuit without descending into its premises.
    Each visited axiom contributes at most one gate.
    The terminal pieces visited in this way are disjoint parts of the derivation tree, and each contributes at most its description length to the extracted circuit.
    Sequential and tensor composition add these gate counts, so
    \[
        |V(\pi)|=O(|\pi|).
    \]
    This bound uses the supplied gate granularity; it does not include the cost of subsequently compiling an abstract gate into a different gate set.
\end{proof}

\begingroup
\raggedbottom
\section{Visualized Results of Our Strategy to the Benchmark}
\label{app:benchmark-examples}
We give the input circuit and the output of our strategy for every case in Table~\ref{tab:benchmark}, in the same order as the table.
Each diagram shows the complete output: deleting the orange frames recovers the input.
Blue dashed frames mark \emph{Prepare}, green dashed frames mark \emph{Use}, and orange solid frames mark the inserted \emph{Uncomputation}.
An orange frame records what the algorithm inserts, including in the two cases whose outputs fail verification.
Unframed gates form a continuation beyond the scanned prefix and are retained unchanged.
Neutral gates encountered by the algorithm are included in Use.



\renewcommand{\csgateframe}[4]{%
 \gategroup[wires=#2,steps=#3,background,
 style={cs#1,rounded corners=2pt,inner xsep=-1.5pt,inner ysep=-1pt}]{}%
 \gategroup[wires=#2,steps=#3,
 style={draw=none,fill=none,inner xsep=-1.5pt,inner ysep=-1pt},
 label style={cs#1label,label position=above,anchor=south,yshift=-6pt,
 fill=white,inner sep=0.5pt}]{\footnotesize #4}}
\newcommand{\exlabelspace}[1]{\hphantomgate{\text{\footnotesize #1}}}

\newcommand{\exroundframe}[3]{%
 \gategroup[wires=#1,steps=#2,background,
 style={draw=csRoleBlue,dashed,fill=none,rounded corners=3pt,
 inner xsep=3pt,inner ysep=5pt}]{}%
 \gategroup[wires=#1,steps=#2,
 style={draw=none,fill=none,inner xsep=3pt,inner ysep=5pt},
 label style={text=csRoleBlue,label position=below,anchor=north,yshift=-2pt,
 fill=white,inner sep=1pt}]
 {\footnotesize #3}}

\subsection{Fractional Query}
\label{app:example-fractional-query}
\emph{Source.} The two-query construction in Fig.~1 of Cleve et al.~\cite{fractional-query}, instantiated with $Q_x=Z_{q_0}Z_{q_1}$ and $\theta'=\pi/4$.
The source ancilla starts in $\ket{+}$; we include initial and final Hadamards to use a zero-initialized scratch wire.

For the fractional-query construction, let $G=Z_{q_0}Z_{q_1}$ and $r(\pi/4)=\exp[-i\pi(I-X)/8]$.
The algorithm classifies $H_a;\mathrm{controlled-}G$ as Prepare and $r_a(\pi/4)$ as Use.
Reversing the two Prepare gates restores the coherent-choice ancilla and implements $\exp[-i\pi(I-G)/8]$ on the logical system.
The controlled-$ZZ$ query is one atomic gate.

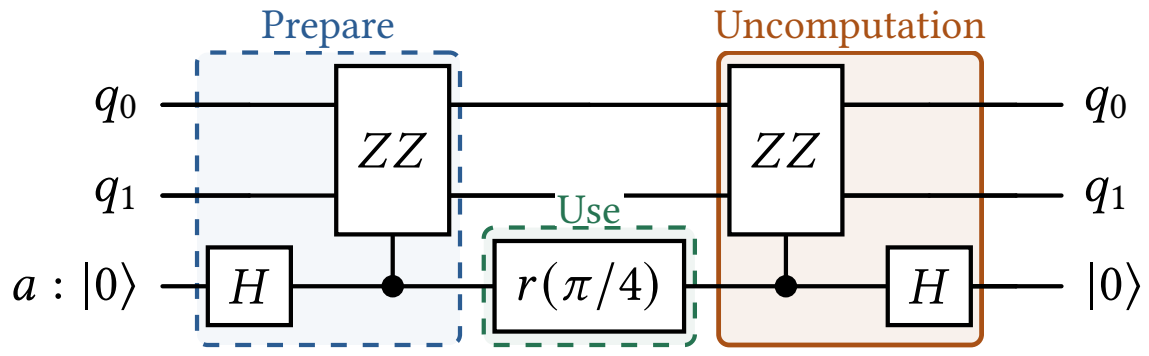
\begin{figure}[H]
\centering
\begin{adjustbox}{max width=\linewidth}
\begin{quantikz}[row sep={0.60cm,between origins},column sep=0.30cm]
\lstick{\(q_0\)} & \qw\csgateframe{comp}{3}{2}{$\mathrm{Prepare}$} & \gate[2]{ZZ} & \qw & \gate[2]{ZZ}\csgateframe{clean}{3}{2}{$\mathrm{Uncomputation}$} & \qw & \qw & \rstick{\(q_0\)} \\
\lstick{\(q_1\)} & \qw &  & \qw &  & \qw & \qw & \rstick{\(q_1\)} \\
\lstick{\(a:\ket{0}\)} & \gate{H} & \ctrl{-2} & \gate{r(\pi/4)}\csgateframe{use}{1}{1}{$\mathrm{Use}$} & \ctrl{-2} & \gate{H} & \qw & \rstick{\(\ket{0}\)}
\end{quantikz}
\end{adjustbox}
\caption{Fractional query: two Prepare gates, one Use gate, and two inserted gates. The five-gate output passes verification.}
\Description{Fractional Query: algorithm output with Prepare, Use, and inserted uncomputation frames.}
\label{fig:example-fractional-query}
\end{figure}

\subsection{Fourier-Basis Phase Kickback}
\label{app:example-fourier-basis-phase-kickback}
\emph{Source.} The phase-kickback construction in Sec.~2.1, Eqs.~(2)--(3), and Fig.~2 of Jones et al.~\cite{Fourier-basis} (arXiv:1204.0567).
Our adaptation takes $n=3$ and $k=1$, replaces the fixed increment by a coherent three-qubit addend $x$, and explicitly prepares and unprepares the Fourier ancilla.

This instance uses three data qubits and three Fourier-state ancillas.
The six Prepare gates produce
\[
 \ket{\gamma}=\frac{1}{\sqrt{8}}\sum_{y=0}^{7}e^{-i\pi y/4}\ket{y}.
\]
The atomic six-wire gate $\mathrm{ADD}_8:\ket{x,y}\mapsto\ket{x,(y+x)\bmod8}$ preserves this ancilla state and is classified as Use.
The algorithm then reverses the six preparation gates.
Its output implements $\ket{x}\mapsto e^{i\pi x/4}\ket{x}$ and returns all three ancillas to zero.
Numbered ports retain the operand order of $\mathrm{ADD}_8$.

\begin{figure}[H]
\centering
\begin{adjustbox}{max width=\linewidth}
\begin{quantikz}[row sep={0.60cm,between origins},column sep=0.30cm]
\lstick{\(x_0\)} & \qw & \qw & \gate[wires=6]{\mathrm{ADD}_8}\gateinput{\(\scriptstyle 0\)}\csgateframe{use}{6}{1}{$\mathrm{Use}$} & \qw & \qw & \qw & \rstick{\(x_0\)} \\
\lstick{\(x_1\)} & \qw & \qw & \gateinput{\(\scriptstyle 1\)} & \qw & \qw & \qw & \rstick{\(x_1\)} \\
\lstick{\(x_2\)} & \qw & \qw & \gateinput{\(\scriptstyle 2\)} & \qw & \qw & \qw & \rstick{\(x_2\)} \\
\lstick{\(a_0:\ket{0}\)} & \gate{H}\csgateframe{comp}{3}{2}{$\mathrm{Prepare}$} & \gate{P(-\pi/4)} & \gateinput{\(\scriptstyle 3\)} & \gate{P(\pi/4)}\csgateframe{clean}{3}{2}{$\mathrm{Uncomputation}$} & \gate{H} & \qw & \rstick{\(\ket{0}\)} \\
\lstick{\(a_1:\ket{0}\)} & \gate{H} & \gate{P(-\pi/2)} & \gateinput{\(\scriptstyle 4\)} & \gate{P(\pi/2)} & \gate{H} & \qw & \rstick{\(\ket{0}\)} \\
\lstick{\(a_2:\ket{0}\)} & \gate{H} & \gate{P(-\pi)} & \gateinput{\(\scriptstyle 5\)} & \gate{P(\pi)} & \gate{H} & \qw & \rstick{\(\ket{0}\)}
\end{quantikz}
\end{adjustbox}
\caption{Fourier-basis phase kickback. Each preparation or cleanup column contains three gates; the full output has 13 gates and passes verification.}
\Description{Fourier-Basis Phase Kickback: algorithm output with Prepare, Use, and inserted uncomputation frames.}
\label{fig:example-fourier-basis-phase-kickback}
\end{figure}
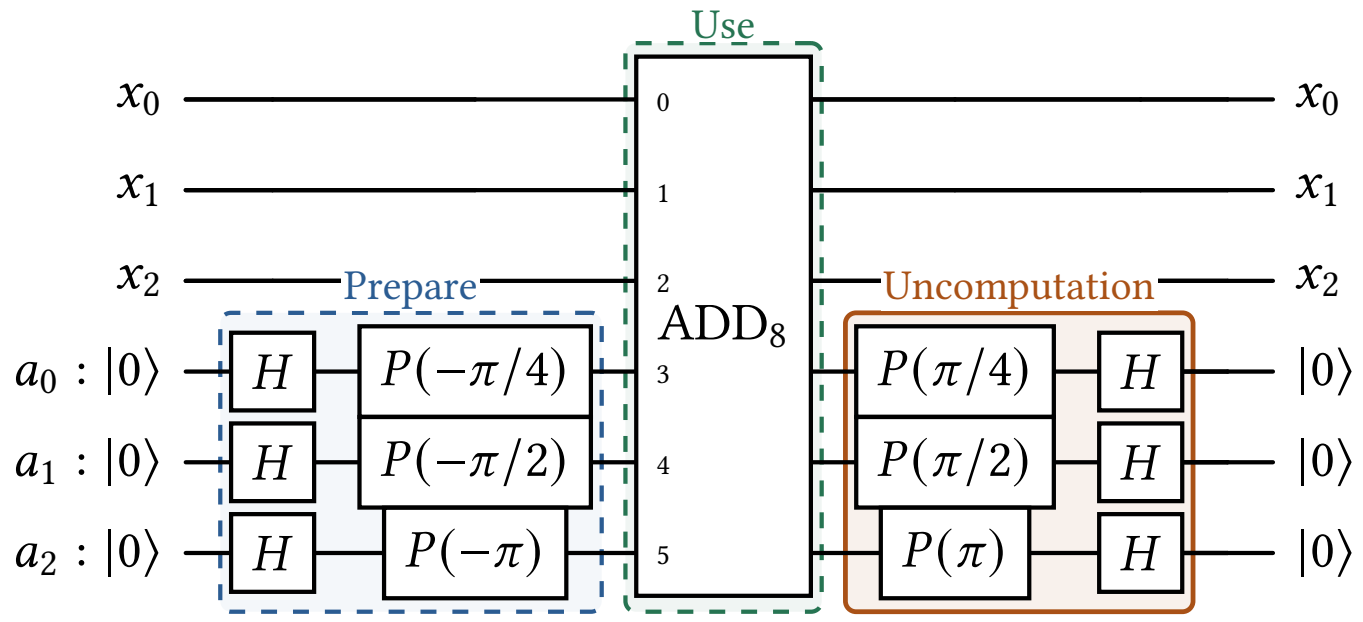

\subsection{Controlled-\texorpdfstring{$T$}{T}}
\label{app:example-controlled-t}
\emph{Source.} Selinger~\cite{controlled-T}, Remark~3.2, Eq.~(14).
We retain $\mathrm{CC}(-iX)$ and its inverse as atomic gates.

The input is the conditional encoder $\mathrm{CC}(-iX)_{q_0,q_1\to a}$ followed by $T_a$.
These are Prepare and Use, respectively.
The inserted inverse is $\mathrm{CC}(iX)$, retaining the encoder's conditional phase.
The resulting logical operation is $\operatorname{diag}(1,1,1,e^{i\pi/4})$, and the ancilla is restored.

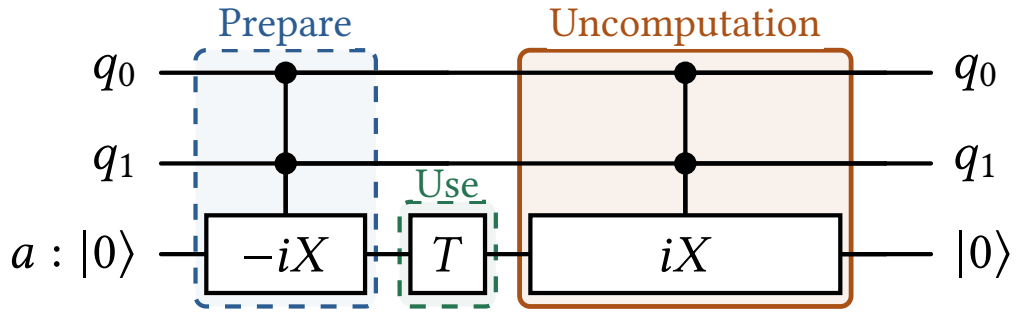
\begin{figure}[H]
\centering
\begin{adjustbox}{max width=\linewidth}
\begin{quantikz}[row sep={0.60cm,between origins},column sep=0.30cm]
\lstick{\(q_0\)} & \ctrl{2}\csgateframe{comp}{3}{1}{$\mathrm{Prepare}$} & \qw & \ctrl{2}\csgateframe{clean}{3}{1}{$\mathrm{Uncomputation}$} & \qw & \rstick{\(q_0\)} \\
\lstick{\(q_1\)} & \ctrl{1} & \qw & \ctrl{1} & \qw & \rstick{\(q_1\)} \\
\lstick{\(a:\ket{0}\)} & \gate[][1.05cm]{-iX} & \gate{T}\csgateframe{use}{1}{1}{$\mathrm{Use}$} & \gate[][2.05cm]{iX} & \qw & \rstick{\(\ket{0}\)}
\end{quantikz}
\end{adjustbox}
\caption{Controlled-$T$: the three-gate output passes verification. The $iX$ in the orange frame is the inverse of the conditional $-iX$ encoder.}
\Description{Controlled-T: algorithm output with Prepare, Use, and inserted uncomputation frames.}
\label{fig:example-controlled-t}
\end{figure}

\subsection{Relative-Phase-Toffoli \texorpdfstring{$C^3X$}{C3X}}
\label{app:example-relative-phase-c3x}
\emph{Source.} The clean-ancilla $C^3X$ construction in Eq.~(4) of Maslov~\cite{relative-phase-toffoli}.
Each relative-phase Toffoli is expanded into the nine gates numbered 2--10 in the dashed part of Fig.~3.

We use the nine elementary gates of the relative-phase Toffoli encoder.
Each is classified separately as Prepare.
The central $\mathrm{CCX}_{a,q_2\to q_3}$ is Use, and the algorithm appends all nine inverse gates in reverse order.
The inverse happens to have the same displayed sequence as the encoder.
The output restores $a$ and implements $q_3\mapsto q_3\oplus q_0q_1q_2$, including cancellation of the encoder's conditional phase.

\begin{figure}[H]
\centering
\textit{Gates 1--10.}\par\smallskip
\begin{adjustbox}{max width=\linewidth}
\begin{quantikz}[row sep={0.60cm,between origins},column sep=0.30cm]
\lstick{\(q_0\)} & \qw\csgateframe{comp}{5}{9}{$\mathrm{Prepare}$} & \qw & \qw & \qw & \ctrl{4} & \qw & \qw & \qw & \qw & \qw & \qw & \rstick{\(q_0\)} \\
\lstick{\(q_1\)} & \qw & \qw & \ctrl{3} & \qw & \qw & \qw & \ctrl{3} & \qw & \qw & \qw & \qw & \rstick{\(q_1\)} \\
\lstick{\(q_2\)} & \qw & \qw & \qw & \qw & \qw & \qw & \qw & \qw & \qw & \ctrl{1}\csgateframe{use}{3}{1}{$\mathrm{Use}$} & \qw & \rstick{\(q_2\)} \\
\lstick{\(q_3\)} & \qw & \qw & \qw & \qw & \qw & \qw & \qw & \qw & \qw & \targ{} & \qw & \rstick{\(q_3\)} \\
\lstick{\(a:\ket{0}\)} & \gate{H} & \gate{T} & \targ{} & \gate{T^\dagger} & \targ{} & \gate{T} & \targ{} & \gate{T^\dagger} & \gate{H} & \ctrl{-1} & \qw & \rstick{\(a\)}
\end{quantikz}
\end{adjustbox}
\par\medskip
\textit{Gates 11--19 (continued).}\par\smallskip
\begin{adjustbox}{max width=\linewidth}
\begin{quantikz}[row sep={0.60cm,between origins},column sep=0.30cm]
\lstick{\(q_0\)} & \qw\csgateframe{clean}{5}{9}{$\mathrm{Uncomputation}$} & \qw & \qw & \qw & \ctrl{4} & \qw & \qw & \qw & \qw & \qw & \rstick{\(q_0\)} \\
\lstick{\(q_1\)} & \qw & \qw & \ctrl{3} & \qw & \qw & \qw & \ctrl{3} & \qw & \qw & \qw & \rstick{\(q_1\)} \\
\lstick{\(q_2\)} & \qw & \qw & \qw & \qw & \qw & \qw & \qw & \qw & \qw & \qw & \rstick{\(q_2\)} \\
\lstick{\(q_3\)} & \qw & \qw & \qw & \qw & \qw & \qw & \qw & \qw & \qw & \qw & \rstick{\(q_3\)} \\
\lstick{\(a\)} & \gate{H} & \gate{T} & \targ{} & \gate{T^\dagger} & \targ{} & \gate{T} & \targ{} & \gate{T^\dagger} & \gate{H} & \qw & \rstick{\(\ket{0}\)}
\end{quantikz}
\end{adjustbox}
\caption{Relative-phase-Toffoli construction of $C^3X$. Read the upper panel and then the lower panel. All 19 elementary gates are shown; the output passes verification.}
\Description{Relative-Phase-Toffoli C^3X: algorithm output with Prepare, Use, and inserted uncomputation frames.}
\label{fig:example-relative-phase-c3x}
\end{figure}
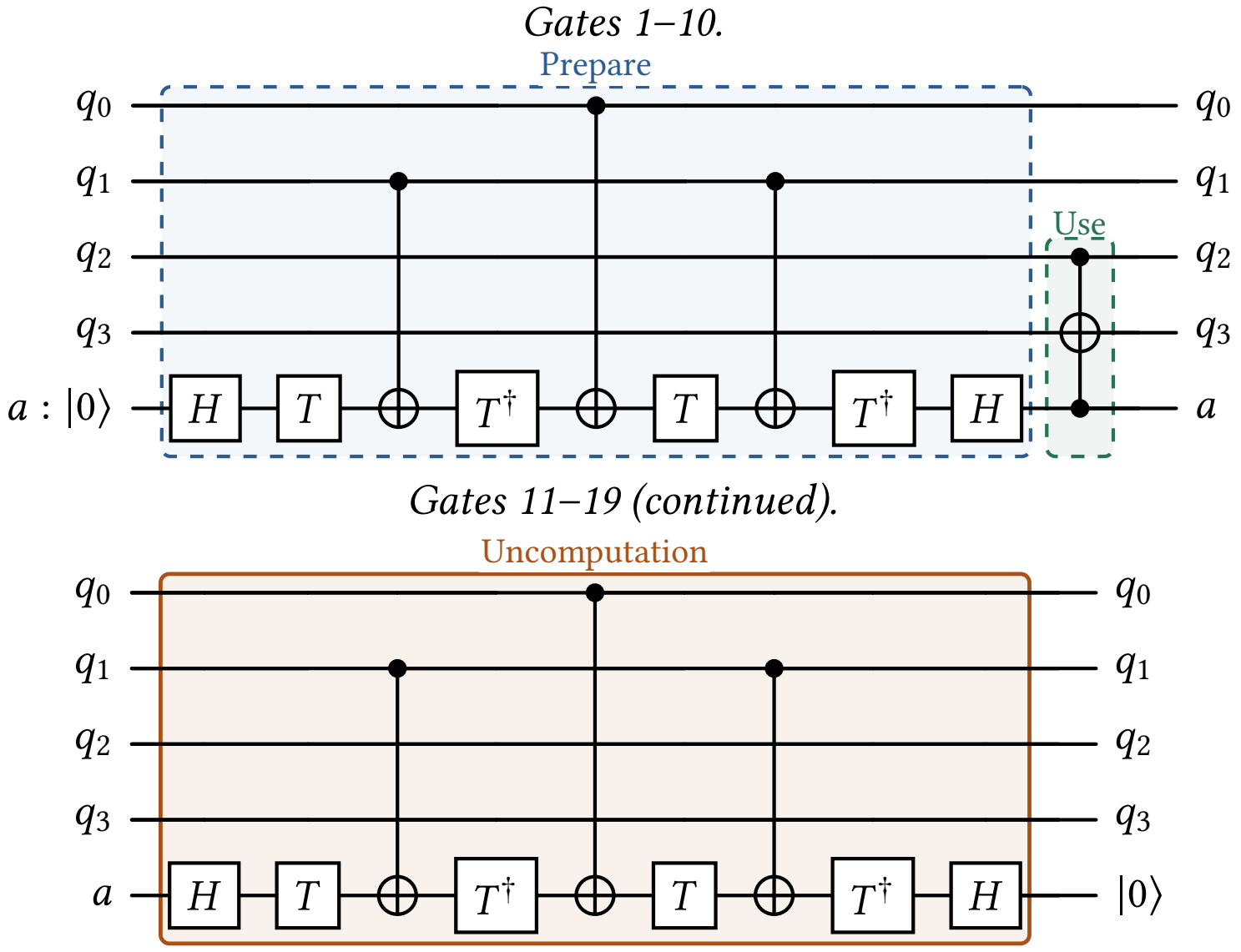

\subsection{QSVT Projector Phase}
\label{app:example-qsvt-projector-phase}
\emph{Source.} Lemma~19 and Fig.~1(b) in the full version of Gily\'en et al.~\cite{QSVT}.
We instantiate the projector-controlled phase gadget with $\Pi=(I-H_q)/2$, expand $\mathrm{C}_{\Pi}\mathrm{NOT}$ as $H_a;\mathrm{CH}_{a\to q};H_a$, and use $S_a$ for the ancillary phase.
This corresponds to $\phi=\pi/4$ in the source gadget, with the global phase convention $S=e^{i\pi/4}e^{-i\pi Z/4}$.

For the projector-phase gadget, define $\Pi_\pm=(I\pm H_q)/2$.
The encoder $H_a;\mathrm{CH}_{a\to q};H_a$ consists of three Prepare gates.
The gate $S_a$ is Use with functionality $\Pi_++i\Pi_-$.
The algorithm appends the inverse encoder, restoring the ancilla while preserving this non-Pauli projector phase.

\begin{figure}[H]
\centering
\begin{adjustbox}{max width=\linewidth}
\begin{quantikz}[row sep={0.60cm,between origins},column sep=0.30cm]
\lstick{\(q\)} & \qw\csgateframe{comp}{2}{3}{$\mathrm{Prepare}$} & \gate{H} & \qw & \qw & \qw\csgateframe{clean}{2}{3}{$\mathrm{Uncomputation}$} & \gate{H} & \qw & \qw & \rstick{\(q\)} \\
\lstick{\(a:\ket{0}\)} & \gate{H} & \ctrl{-1} & \gate{H} & \gate{S}\csgateframe{use}{1}{1}{$\mathrm{Use}$} & \gate{H} & \ctrl{-1} & \gate{H} & \qw & \rstick{\(\ket{0}\)}
\end{quantikz}
\end{adjustbox}
\caption{QSVT projector phase: the seven-gate output passes verification.}
\Description{QSVT Projector Phase: algorithm output with Prepare, Use, and inserted uncomputation frames.}
\label{fig:example-qsvt-projector-phase}
\end{figure}
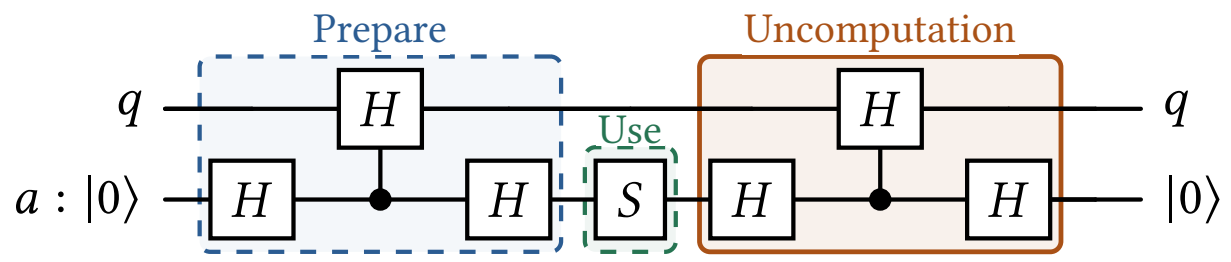

\subsection{Single-Clock HHL}
\label{app:example-hhl-single-clock}
\emph{Source.} Feng et al.~\cite{single-clock-HHL}, Sec.~IV.B, Fig.~6 (arXiv:2202.02983).
We use the exact matrix underlying the rounded entries in Eq.~(22), retain the clock Hadamards, and omit state preparation and measurement.

The coherent HHL core uses $A=(5I-H_q)/2$ and $\theta=-2\arccos(2/3)$.
Only the clock $c$ is scratch; the flag $f$ remains part of the logical output.
The three clock-encoding gates are Prepare, and the controlled $R_y(\theta)$ is Use.
Reversing the encoder restores $c$ and implements
\[
 \Pi_+\otimes I_f+\Pi_-\otimes R_y(\theta)_f,
 \qquad \Pi_\pm=(I\pm H_q)/2.
\]
The physical HHL instance initializes $f$ to $\ket{1}$; the benchmark checks the displayed unitary on arbitrary joint inputs of $q$ and $f$.

\begin{figure}[H]
\centering
\begin{adjustbox}{max width=\linewidth}
\begin{quantikz}[row sep={0.60cm,between origins},column sep=0.30cm]
\lstick{\(q\)} & \qw\csgateframe{comp}{3}{3}{$\mathrm{Prepare}$} & \gate{H} & \qw & \qw & \qw\csgateframe{clean}{3}{3}{$\mathrm{Uncomputation}$} & \gate{H} & \qw & \qw & \rstick{\(q\)} \\
\lstick{\(f\)} & \qw & \qw & \qw & \gate{R_y(\theta)}\csgateframe{use}{2}{1}{$\mathrm{Use}$} & \qw & \qw & \qw & \qw & \rstick{\(f\)} \\
\lstick{\(c:\ket{0}\)} & \gate{H} & \ctrl{-2} & \gate{H} & \ctrl{-1} & \gate{H} & \ctrl{-2} & \gate{H} & \qw & \rstick{\(\ket{0}\)}
\end{quantikz}
\end{adjustbox}
\caption{Single-clock HHL: the seven-gate output passes verification. The flag is retained, while the clock is reset.}
\Description{Single-Clock HHL: algorithm output with Prepare, Use, and inserted uncomputation frames.}
\label{fig:example-hhl-single-clock}
\end{figure}
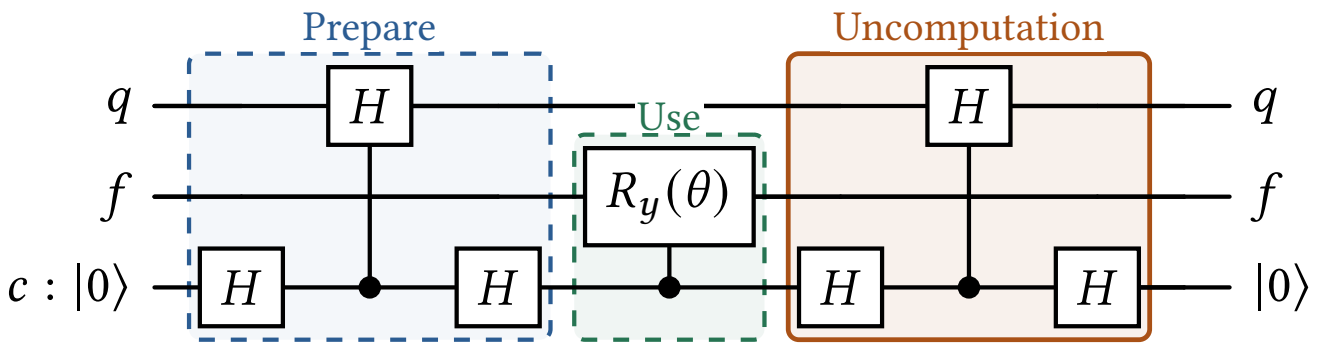

\subsection{Adder}
\label{app:example-adder}
\emph{Source.} \texttt{makesAdder} in \texttt{adder.py} of the Reqomp artifact~\cite{reqompArtifact}, with two-bit operands.

The two-bit ripple-carry instance has two clean ancillas $u_0,u_1$ and implements $b\mapsto b+a\pmod4$.
The carry helper is expanded into its three gates before synthesis.
Round $1$ selects $u_1$: four gates are Prepare, including the two updates of $b_0$, and their inverses are inserted after the last use of $u_1$.
Round $2$ selects the zero carry-in $u_0$ and classifies all twelve gates of the current circuit as Use, inserting nothing.
The first-round frames are retained below; the last two gates are first scanned in round $2$.

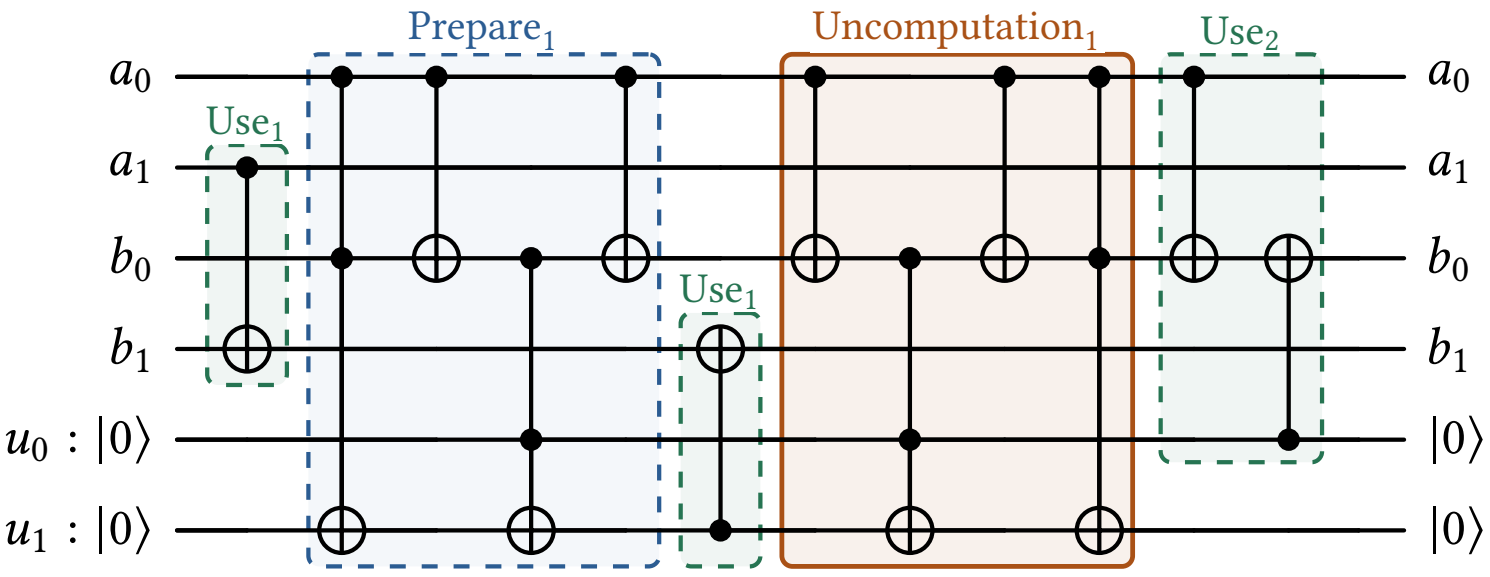
\begin{figure}[H]
\centering
\begin{adjustbox}{max width=\linewidth}
\begin{quantikz}[row sep={0.60cm,between origins},column sep=0.30cm]
\lstick{\(a_0\)} & \qw & \ctrl{5}\csgateframe{comp}{6}{4}{$\mathrm{Prepare}_{1}$} & \ctrl{2} & \qw & \ctrl{2} & \qw & \ctrl{2}\csgateframe{clean}{6}{4}{$\mathrm{Uncomputation}_{1}$} & \qw & \ctrl{2} & \ctrl{5} & \ctrl{2}\csgateframe{use}{5}{2}{$\mathrm{Use}_{2}$} & \qw & \qw & \rstick{\(a_0\)} \\
\lstick{\(a_1\)} & \ctrl{2}\csgateframe{use}{3}{1}{$\mathrm{Use}_{1}$} & \qw & \qw & \qw & \qw & \qw & \qw & \qw & \qw & \qw & \qw & \qw & \qw & \rstick{\(a_1\)} \\
\lstick{\(b_0\)} & \qw & \ctrl{3} & \targ{} & \ctrl{3} & \targ{} & \qw & \targ{} & \ctrl{3} & \targ{} & \ctrl{3} & \targ{} & \targ{} & \qw & \rstick{\(b_0\)} \\
\lstick{\(b_1\)} & \targ{} & \qw & \qw & \qw & \qw & \targ{}\csgateframe{use}{3}{1}{$\mathrm{Use}_{1}$} & \qw & \qw & \qw & \qw & \qw & \qw & \qw & \rstick{\(b_1\)} \\
\lstick{\(u_0:\ket{0}\)} & \qw & \qw & \qw & \ctrl{1} & \qw & \qw & \qw & \ctrl{1} & \qw & \qw & \qw & \ctrl{-2} & \qw & \rstick{\(\ket{0}\)} \\
\lstick{\(u_1:\ket{0}\)} & \qw & \targ{} & \qw & \targ{} & \qw & \ctrl{-2} & \qw & \targ{} & \qw & \targ{} & \qw & \qw & \qw & \rstick{\(\ket{0}\)}
\end{quantikz}
\end{adjustbox}
\caption{Two-bit adder. The four-gate orange block is inserted in round $1$. Both ancillas are restored, and the 12-gate output passes verification.}
\Description{Adder: algorithm output with Prepare, Use, and inserted uncomputation frames.}
\label{fig:example-adder}
\end{figure}

\subsection{WeightedAdder}
\label{app:example-weighted-adder-reqomp}
\emph{Source.} \texttt{makeWeightedAdder} in \texttt{weightedadder.py} of the Reqomp artifact~\cite{reqompArtifact}, with one state qubit and weight vector $(3)$.

The weighted-adder instance has one state qubit $x$, weight $3$, a two-bit sum $s$, and one carry ancilla $a$.
The first CNOT is Use; the Toffoli that computes the carry is Prepare; and the next Toffoli is Use.
The algorithm inserts the carry's inverse before the final update of $s_0$.
This last gate lies beyond the scanned prefix.
The complete output implements $s\mapsto s+3x\pmod4$.

\begin{figure}[H]
\centering
\begin{adjustbox}{max width=\linewidth}
\begin{quantikz}[row sep={0.60cm,between origins},column sep=0.30cm]
\lstick{\(x\)} & \ctrl{2}\csgateframe{use}{3}{1}{$\mathrm{Use}$} & \ctrl{3}\csgateframe{comp}{4}{1}{$\mathrm{Prepare}$} & \ctrl{2}\csgateframe{use}{4}{1}{$\mathrm{Use}$} & \ctrl{3}\csgateframe{clean}{4}{1}{$\mathrm{Uncomputation}$} & \ctrl{1} & \qw & \rstick{\(x\)} \\
\lstick{\(s_0\)} & \qw & \ctrl{2} & \qw & \ctrl{2} & \targ{} & \qw & \rstick{\(s_0\)} \\
\lstick{\(s_1\)} & \targ{} & \exlabelspace{$\mathrm{Prepare}$} & \targ{} & \exlabelspace{$\mathrm{Uncomputation}$} & \qw & \qw & \rstick{\(s_1\)} \\
\lstick{\(a:\ket{0}\)} & \qw & \targ{} & \ctrl{-1} & \targ{} & \qw & \qw & \rstick{\(\ket{0}\)}
\end{quantikz}
\end{adjustbox}
\caption{WeightedAdder: the carry is reset before the unframed continuation. The five-gate output passes verification.}
\Description{WeightedAdder: algorithm output with Prepare, Use, and inserted uncomputation frames.}
\label{fig:example-weighted-adder-reqomp}
\end{figure}
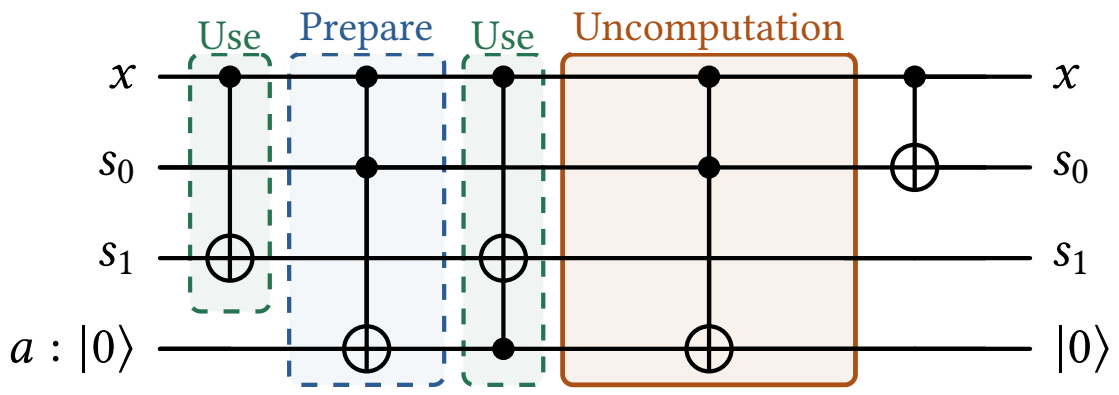

\subsection{Multiplier}
\label{app:example-multiplier}
\emph{Source.} \texttt{makesMult} in \texttt{adder.py} of the Reqomp artifact~\cite{reqompArtifact}, with one bit in each operand and in the output register.

The one-bit multiplier computes $xy$ in $a_0$, uses it to update $b$, and retains a second zero carry-in $a_1$.
Round $1$ selects $a_0$ and inserts the inverse Toffoli immediately after its use.
Round $2$ selects $a_1$: all four gates are Use, including the neutral final CNOT, so no further gate is inserted.
The output implements $b\mapsto b\oplus xy$ and restores both ancillas.

\begin{figure}[H]
\centering
\begin{adjustbox}{max width=\linewidth}
\begin{quantikz}[row sep={0.60cm,between origins},column sep=0.30cm]
\lstick{\(b\)} & \exlabelspace{$\mathrm{Prepare}_{1}$} & \targ{}\csgateframe{use}{4}{1}{$\mathrm{Use}_{1}$} & \exlabelspace{$\mathrm{Uncomputation}_{1}$} & \targ{}\csgateframe{use}{5}{1}{$\mathrm{Use}_{2}$} & \qw & \rstick{\(b\)} \\
\lstick{\(y\)} & \ctrl{2}\csgateframe{comp}{3}{1}{$\mathrm{Prepare}_{1}$} & \qw & \ctrl{2}\csgateframe{clean}{3}{1}{$\mathrm{Uncomputation}_{1}$} & \qw & \qw & \rstick{\(y\)} \\
\lstick{\(x\)} & \ctrl{1} & \qw & \ctrl{1} & \qw & \qw & \rstick{\(x\)} \\
\lstick{\(a_0:\ket{0}\)} & \targ{} & \ctrl{-3} & \targ{} & \qw & \qw & \rstick{\(\ket{0}\)} \\
\lstick{\(a_1:\ket{0}\)} & \qw & \qw & \qw & \ctrl{-4} & \qw & \rstick{\(\ket{0}\)}
\end{quantikz}
\end{adjustbox}
\caption{One-bit multiplier. Subscripts distinguish the two rounds; the four-gate output passes verification.}
\Description{Multiplier: algorithm output with Prepare, Use, and inserted uncomputation frames.}
\label{fig:example-multiplier}
\end{figure}
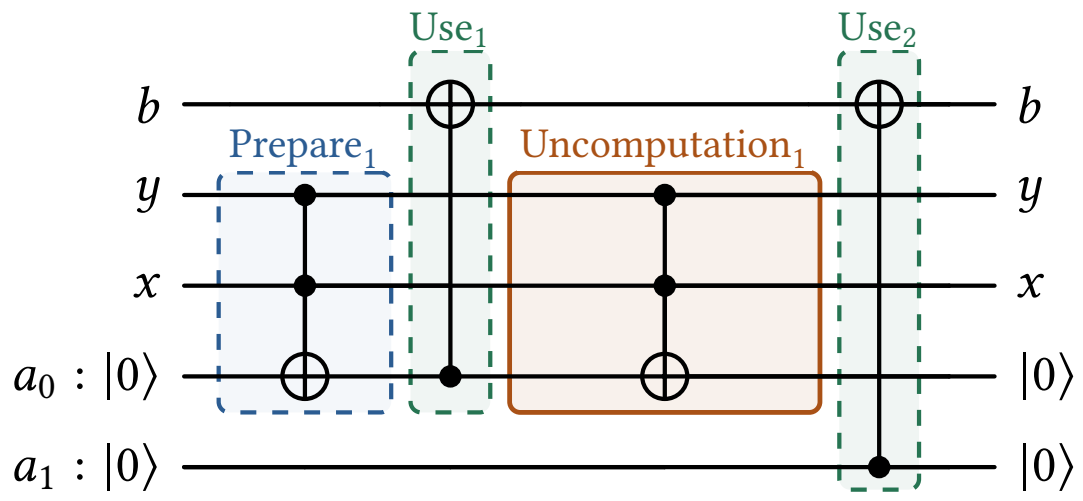

\subsection{IntegerComparator}
\label{app:example-integer-comparator}
\emph{Source.} \texttt{makeIntegerComparator} in \texttt{intergercomparator.py} of the Reqomp artifact~\cite{reqompArtifact}, with two state qubits, threshold $1$, and \texttt{geq=True}.
The filename retains the spelling used in that artifact.

The comparator uses a two-bit integer $x=x_0+2x_1$ and threshold $1$.
The CNOT copying $x_0$ into $a$ is Prepare; the atomic $\mathrm{OR}(x_1,a;f)$ is Use, with action $f\mapsto f\oplus(x_1\lor a)$.
Reversing the copy restores $a$.
The output implements $f\mapsto f\oplus[x\geq1]$.
The OR box's numbered ports are $0=x_1$, $1=a$, and $2=f$.

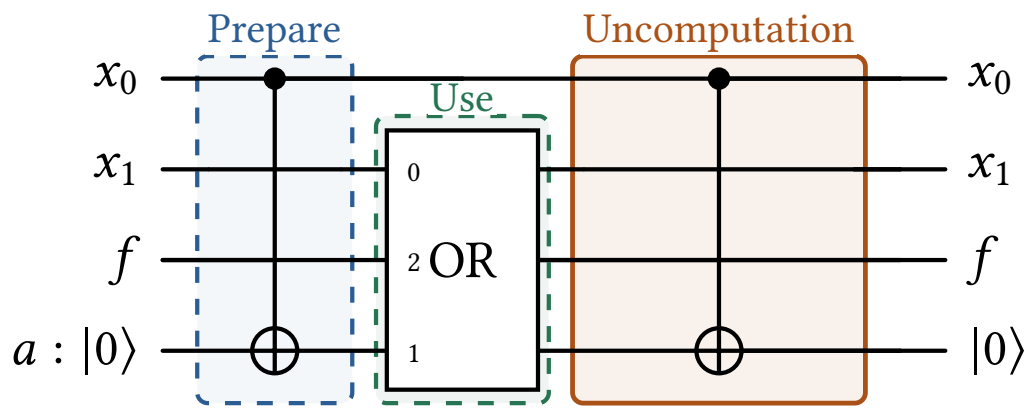
\begin{figure}[H]
\centering
\begin{adjustbox}{max width=\linewidth}
\begin{quantikz}[row sep={0.60cm,between origins},column sep=0.30cm]
\lstick{\(x_0\)} & \ctrl{3}\csgateframe{comp}{4}{1}{$\mathrm{Prepare}$} & \qw & \ctrl{3}\csgateframe{clean}{4}{1}{$\mathrm{Uncomputation}$} & \qw & \rstick{\(x_0\)} \\
\lstick{\(x_1\)} & \exlabelspace{$\mathrm{Prepare}$} & \gate[wires=3][1.0cm]{\mathrm{OR}}\gateinput{\(\scriptstyle 0\)}\csgateframe{use}{3}{1}{$\mathrm{Use}$} & \exlabelspace{$\mathrm{Uncomputation}$} & \qw & \rstick{\(x_1\)} \\
\lstick{\(f\)} & \qw & \gateinput{\(\scriptstyle 2\)} & \qw & \qw & \rstick{\(f\)} \\
\lstick{\(a:\ket{0}\)} & \targ{} & \gateinput{\(\scriptstyle 1\)} & \targ{} & \qw & \rstick{\(\ket{0}\)}
\end{quantikz}
\end{adjustbox}
\caption{IntegerComparator. The OR helper retains its atomic gate boundary, so the successful output has three gates.}
\Description{IntegerComparator: algorithm output with Prepare, Use, and inserted uncomputation frames.}
\label{fig:example-integer-comparator}
\end{figure}

\subsection{Clean Incrementer}
\label{app:example-incrementer}
\emph{Source.} The Clean Incrementer family in Sec.~7.1 and Table~1 of~\cite{RwUn}.
The three-bit instance matches \texttt{makesIncrementer(3, dirty=False)} in \texttt{RwUn/examples/incrementer.py} of the artifact~\cite{RwUnArtifactV2}.

The three-bit clean incrementer computes the carry $x_0x_1$ into $a$ and uses it to update $x_2$.
The algorithm reverses this Prepare gate before either lower bit is changed.
The last two gates are the unscanned continuation.
The output implements $x\mapsto x+1\pmod8$ and restores $a$.

\begin{figure}[H]
\centering
\begin{adjustbox}{max width=\linewidth}
\begin{quantikz}[row sep={0.60cm,between origins},column sep=0.30cm]
\lstick{\(x_0\)} & \ctrl{3}\csgateframe{comp}{4}{1}{$\mathrm{Prepare}$} & \qw & \ctrl{3}\csgateframe{clean}{4}{1}{$\mathrm{Uncomputation}$} & \ctrl{1} & \gate{X} & \qw & \rstick{\(x_0\)} \\
\lstick{\(x_1\)} & \ctrl{2} & \qw & \ctrl{2} & \targ{} & \qw & \qw & \rstick{\(x_1\)} \\
\lstick{\(x_2\)} & \exlabelspace{$\mathrm{Prepare}$} & \targ{}\csgateframe{use}{2}{1}{$\mathrm{Use}$} & \exlabelspace{$\mathrm{Uncomputation}$} & \qw & \qw & \qw & \rstick{\(x_2\)} \\
\lstick{\(a:\ket{0}\)} & \targ{} & \ctrl{-1} & \targ{} & \qw & \qw & \qw & \rstick{\(\ket{0}\)}
\end{quantikz}
\end{adjustbox}
\caption{Clean Incrementer: cleanup precedes the lower-bit updates. The five-gate output passes verification.}
\Description{Clean Incrementer: algorithm output with Prepare, Use, and inserted uncomputation frames.}
\label{fig:example-incrementer}
\end{figure}
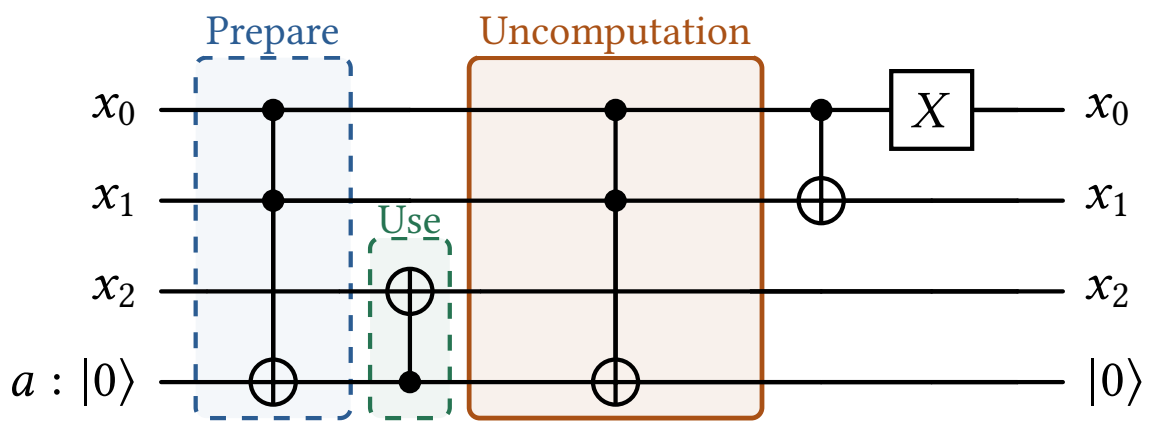

\subsection{MCX}
\label{app:example-mcx}
\emph{Source.} \texttt{makeMCX} in \texttt{mcx.py} of the Reqomp artifact~\cite{reqompArtifact}, instantiated with three controls.

The three-control NOT instance prepares $a=c_0c_1$, uses $a$ together with $c_2$ to flip $t$, and then uncomputes $a$.
The two input Toffolis are classified as Prepare and Use, respectively.
The output implements $t\mapsto t\oplus c_0c_1c_2$.

\begin{figure}[H]
\centering
\begin{adjustbox}{max width=\linewidth}
\begin{quantikz}[row sep={0.60cm,between origins},column sep=0.30cm]
\lstick{\(c_0\)} & \ctrl{4}\csgateframe{comp}{5}{1}{$\mathrm{Prepare}$} & \qw & \ctrl{4}\csgateframe{clean}{5}{1}{$\mathrm{Uncomputation}$} & \qw & \rstick{\(c_0\)} \\
\lstick{\(c_1\)} & \ctrl{3} & \qw & \ctrl{3} & \qw & \rstick{\(c_1\)} \\
\lstick{\(c_2\)} & \exlabelspace{$\mathrm{Prepare}$} & \ctrl{1}\csgateframe{use}{3}{1}{$\mathrm{Use}$} & \exlabelspace{$\mathrm{Uncomputation}$} & \qw & \rstick{\(c_2\)} \\
\lstick{\(t\)} & \qw & \targ{} & \qw & \qw & \rstick{\(t\)} \\
\lstick{\(a:\ket{0}\)} & \targ{} & \ctrl{-1} & \targ{} & \qw & \rstick{\(\ket{0}\)}
\end{quantikz}
\end{adjustbox}
\caption{MCX: one Prepare gate, one Use gate, and one inverse. The three-gate output passes verification.}
\Description{MCX: algorithm output with Prepare, Use, and inserted uncomputation frames.}
\label{fig:example-mcx}
\end{figure}
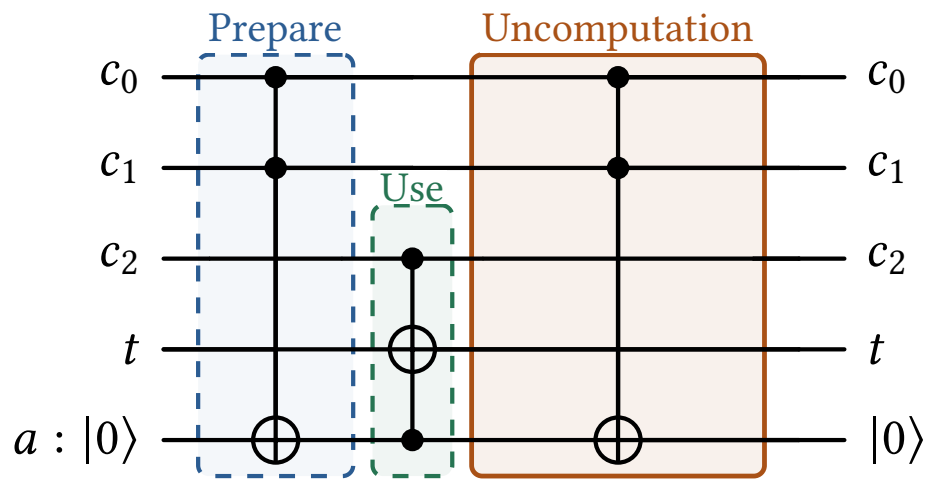

\subsection{MCRY}
\label{app:example-mcry}
\emph{Source.} \texttt{makeMCRY} in \texttt{mcx.py} of the Reqomp artifact~\cite{reqompArtifact}, instantiated with angle $\pi/4$ and one control.

The retained controlled-rotation instance has one control $c$ and angle $\pi/4$.
The CNOT copying $c$ into $a$ is Prepare; the controlled rotation is Use.
The inserted CNOT restores $a$, leaving $\mathrm{C}R_y(\pi/4)_{c\to t}$ on the logical system.

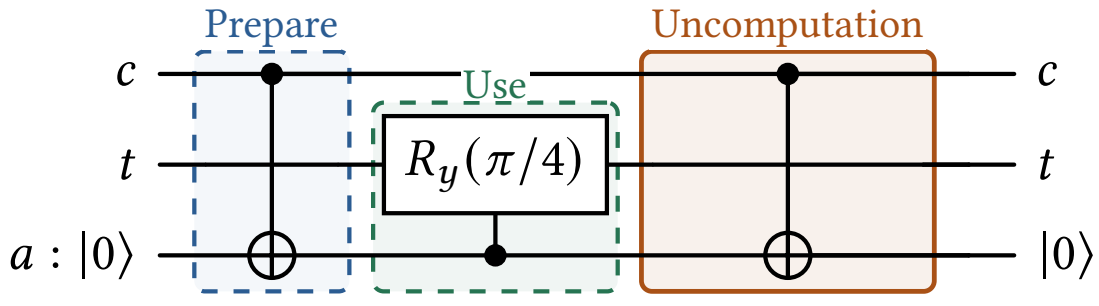
\begin{figure}[H]
\centering
\begin{adjustbox}{max width=\linewidth}
\begin{quantikz}[row sep={0.60cm,between origins},column sep=0.30cm]
\lstick{\(c\)} & \ctrl{2}\csgateframe{comp}{3}{1}{$\mathrm{Prepare}$} & \qw & \ctrl{2}\csgateframe{clean}{3}{1}{$\mathrm{Uncomputation}$} & \qw & \rstick{\(c\)} \\
\lstick{\(t\)} & \exlabelspace{$\mathrm{Prepare}$} & \gate{R_y(\pi/4)}\csgateframe{use}{2}{1}{$\mathrm{Use}$} & \exlabelspace{$\mathrm{Uncomputation}$} & \qw & \rstick{\(t\)} \\
\lstick{\(a:\ket{0}\)} & \targ{} & \ctrl{-1} & \targ{} & \qw & \rstick{\(\ket{0}\)}
\end{quantikz}
\end{adjustbox}
\caption{MCRY: the three-gate output passes verification.}
\Description{MCRY: algorithm output with Prepare, Use, and inserted uncomputation frames.}
\label{fig:example-mcry}
\end{figure}

\subsection{PiecewiseLinearR}
\label{app:example-piecewise-linear-pauli-rotation}
\emph{Source.} \texttt{makesPLR} in \texttt{piecewiselinrot.py} of the Reqomp artifact~\cite{reqompArtifact}, with one state qubit, breakpoint list $(1)$, slope list $(\pi/4)$, and offset list $(\pi/4)$.

The one-bit piecewise-linear rotation has breakpoint $1$, slope $\pi/4$, and offset $\pi/4$ on the nonzero branch.
The first CNOT prepares the branch flag $a=x$.
All five following gates are Use, including the zero-angle rotation.
The algorithm inserts the inverse CNOT, restoring $a$ and leaving $R_y(\pi x/4)$ on $t$.
Each emitted $\mathrm{CU}(\theta,0,0,0)$ is drawn as the exactly equal $\mathrm{C}R_y(\theta)$.

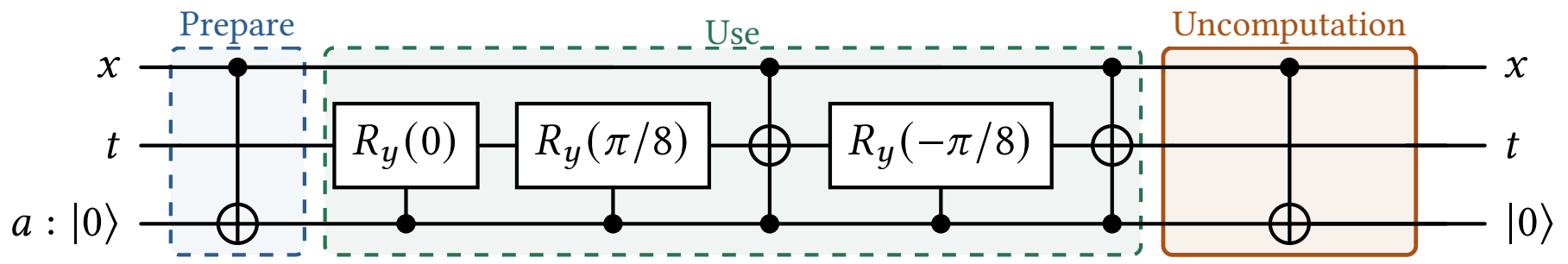
\begin{figure}[H]
\centering
\begin{adjustbox}{max width=\linewidth}
\begin{quantikz}[row sep={0.60cm,between origins},column sep=0.30cm]
\lstick{\(x\)} & \ctrl{2}\csgateframe{comp}{3}{1}{$\mathrm{Prepare}$} & \qw\csgateframe{use}{3}{5}{$\mathrm{Use}$} & \qw & \ctrl{1} & \qw & \ctrl{1} & \ctrl{2}\csgateframe{clean}{3}{1}{$\mathrm{Uncomputation}$} & \qw & \rstick{\(x\)} \\
\lstick{\(t\)} & \exlabelspace{$\mathrm{Prepare}$} & \gate{R_y(0)} & \gate{R_y(\pi/8)} & \targ{} & \gate{R_y(-\pi/8)} & \targ{} & \exlabelspace{$\mathrm{Uncomputation}$} & \qw & \rstick{\(t\)} \\
\lstick{\(a:\ket{0}\)} & \targ{} & \ctrl{-1} & \ctrl{-1} & \ctrl{-1} & \ctrl{-1} & \ctrl{-1} & \targ{} & \qw & \rstick{\(\ket{0}\)}
\end{quantikz}
\end{adjustbox}
\caption{PiecewiseLinearR. The zero-angle instruction is retained; the seven-gate output passes verification.}
\Description{PiecewiseLinearR: algorithm output with Prepare, Use, and inserted uncomputation frames.}
\label{fig:example-piecewise-linear-pauli-rotation}
\end{figure}

\subsection{PolynomialPauliR}
\label{app:example-polynomial-pauli-rotation}
\emph{Source.} \texttt{makesPolyPauliRot} in \texttt{polynomialpaulirot.py} of the Reqomp artifact~\cite{reqompArtifact}, with two state qubits and coefficient vector $(\pi/4,\pi/4,\pi/4)$.

The two-bit polynomial-rotation instance implements $R_y(p(x))$ on $t$, where $x=x_0+2x_1$ and $p(x)=\pi(1+x+x^2)/4$.
The first two rotations are Use.
The Toffoli is Prepare, storing $x_0x_1$ in $a$; the remaining two rotations are Use.
The algorithm then appends the inverse Toffoli and restores $a$.

\begin{figure}[H]
\centering
\begin{adjustbox}{max width=\linewidth}
\begin{quantikz}[row sep={0.60cm,between origins},column sep=0.30cm]
\lstick{\(x_0\)} & \qw & \qw & \ctrl{3}\csgateframe{comp}{4}{1}{$\mathrm{Prepare}$} & \ctrl{2}\csgateframe{use}{4}{2}{$\mathrm{Use}$} & \qw & \ctrl{3}\csgateframe{clean}{4}{1}{$\mathrm{Uncomputation}$} & \qw & \rstick{\(x_0\)} \\
\lstick{\(x_1\)} & \qw\csgateframe{use}{2}{2}{$\mathrm{Use}$} & \ctrl{1} & \ctrl{2} & \qw & \qw & \ctrl{2} & \qw & \rstick{\(x_1\)} \\
\lstick{\(t\)} & \gate{R_y(\pi/4)} & \gate{R_y(3\pi/2)} & \exlabelspace{$\mathrm{Prepare}$} & \gate{R_y(\pi/2)} & \gate{R_y(\pi)} & \exlabelspace{$\mathrm{Uncomputation}$} & \qw & \rstick{\(t\)} \\
\lstick{\(a:\ket{0}\)} & \qw & \qw & \targ{} & \qw & \ctrl{-1} & \targ{} & \qw & \rstick{\(\ket{0}\)}
\end{quantikz}
\end{adjustbox}
\caption{PolynomialPauliR: the six-gate output passes verification.}
\Description{PolynomialPauliR: algorithm output with Prepare, Use, and inserted uncomputation frames.}
\label{fig:example-polynomial-pauli-rotation}
\end{figure}
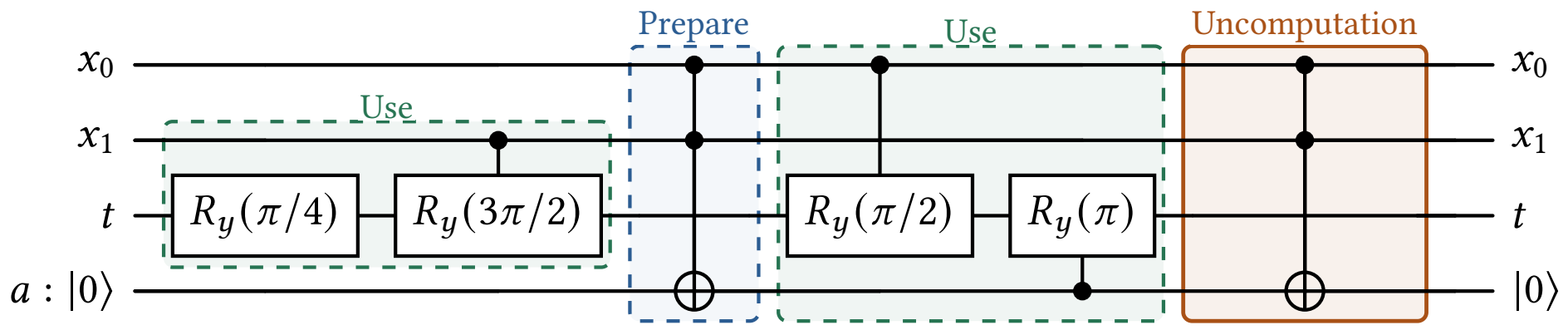

\subsection{ConditionalCleanMCS}
\label{app:example-conditional-clean-mcs}
\emph{Source.} The conditionally clean construction in Sec.~3, Fig.~3, and Corollary~2 of Nie, Zi, and Sun~\cite{ConditionalCleanMCS} (arXiv:2402.05053).
We use its two-ancilla controlled-$S$ benchmark adaptation, listed as ConditionalCleanMCS in Table~1 of~\cite{RwUn}, with six controls and atomic MCX gates.
The instance comes from \texttt{makeConditionalMCS(6, dirty=False)} in \texttt{RwUn/examples/mcx.py} of the artifact~\cite{RwUnArtifactV2}, with its controlled-$S$ decomposition represented as one atomic gate and its instruction order fixed by our shared benchmark input.

This six-control $S$ instance uses two clean ancillas $a_0,a_1$.
Round $1$ selects $a_0$.
The first five gates are Prepare ($\mathrm{Prepare}_1$); the Toffoli with three controls is Use ($\mathrm{Use}_1$).
The algorithm inserts five inverse gates ($\mathrm{Uncomputation}_1$), restoring $a_0$ at this point.
Round $2$ selects $a_1$.
The original $\mathrm{Prepare}_1$ gates are now Use, but the preceding $\mathrm{Use}_1$ gate and the entire $\mathrm{Uncomputation}_1$ block are now Prepare, as the outer $\mathrm{Prepare}_2$ frame shows.
The controlled-$S$ gate is Use ($\mathrm{Use}_2$), followed by six inserted inverse gates ($\mathrm{Uncomputation}_2$).
The remaining eight input gates are an unscanned continuation.

The second cleanup restores $a_1$ but makes $a_0$ dirty again.
On computational-basis inputs, the complete candidate acts as
\[
 \ket{c,t,0,0}\longmapsto
 i^{\,t\prod_{j=0}^{5}c_j}\ket{c,t,c_0c_1c_2c_3,0}.
\]
Thus the controlled phase is present, but the scratch register is not restored: for example, all six controls equal to $1$ leave $a_0=1$.
This is the failure reported in Table~\ref{tab:benchmark}.

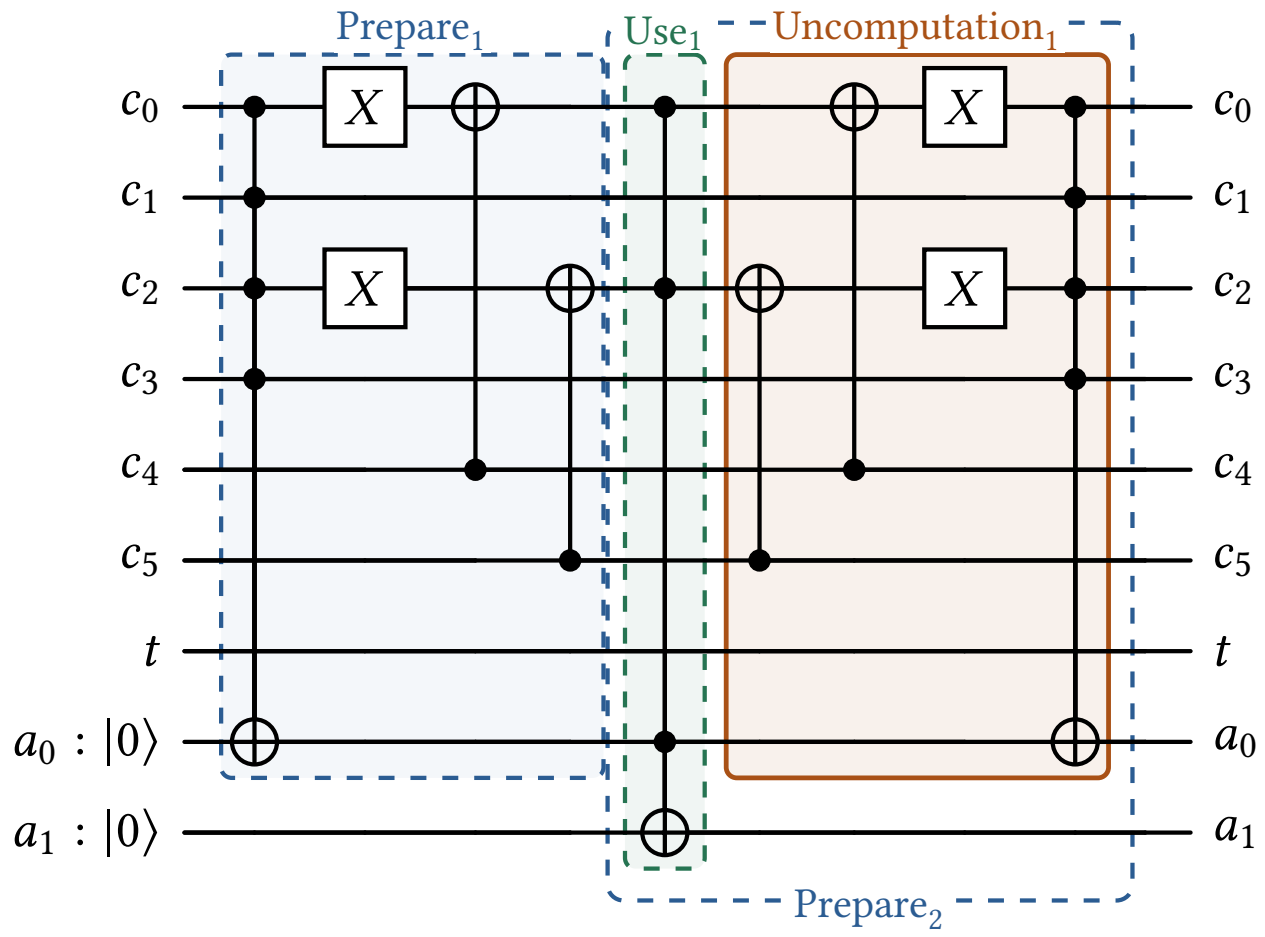
\begin{figure}[H]
\centering
\textit{Gates 1--11.}\par\smallskip
\begin{adjustbox}{max width=\linewidth}
\begin{quantikz}[row sep={0.60cm,between origins},column sep=0.30cm]
\lstick{\(c_0\)} & \ctrl{7}\csgateframe{comp}{8}{4}{$\mathrm{Prepare}_{1}$} & \gate{X} & \targ{} & \qw & \ctrl{8}\csgateframe{use}{9}{1}{$\mathrm{Use}_{1}$}\exroundframe{9}{5}{$\mathrm{Prepare}_2$} & \qw\csgateframe{clean}{8}{4}{$\mathrm{Uncomputation}_{1}$} & \targ{} & \gate{X} & \ctrl{7} & \qw & \rstick{\(c_0\)} \\
\lstick{\(c_1\)} & \ctrl{6} & \qw & \qw & \qw & \qw & \qw & \qw & \qw & \ctrl{6} & \qw & \rstick{\(c_1\)} \\
\lstick{\(c_2\)} & \ctrl{5} & \gate{X} & \qw & \targ{} & \ctrl{6} & \targ{} & \qw & \gate{X} & \ctrl{5} & \qw & \rstick{\(c_2\)} \\
\lstick{\(c_3\)} & \ctrl{4} & \qw & \qw & \qw & \qw & \qw & \qw & \qw & \ctrl{4} & \qw & \rstick{\(c_3\)} \\
\lstick{\(c_4\)} & \qw & \qw & \ctrl{-4} & \qw & \qw & \qw & \ctrl{-4} & \qw & \qw & \qw & \rstick{\(c_4\)} \\
\lstick{\(c_5\)} & \qw & \qw & \qw & \ctrl{-3} & \qw & \ctrl{-3} & \qw & \qw & \qw & \qw & \rstick{\(c_5\)} \\
\lstick{\(t\)} & \qw & \qw & \qw & \qw & \qw & \qw & \qw & \qw & \qw & \qw & \rstick{\(t\)} \\
\lstick{\(a_0:\ket{0}\)} & \targ{} & \qw & \qw & \qw & \ctrl{1} & \qw & \qw & \qw & \targ{} & \qw & \rstick{\(a_0\)} \\
\lstick{\(a_1:\ket{0}\)} & \qw & \qw & \qw & \qw & \targ{} & \qw & \qw & \qw & \qw & \qw & \rstick{\(a_1\)}
\end{quantikz}
\end{adjustbox}
\caption{ConditionalCleanMCS, gates 1--11. The orange $\mathrm{Uncomputation}_1$ block resets $a_0$ in round $1$. The outer blue frame shows that the preceding Use gate and this cleanup become Prepare in round $2$.}
\Description{First part of the ConditionalCleanMCS candidate, with the nested second-round Prepare block.}
\label{fig:example-conditional-clean-mcs-first}
\end{figure}
\begin{figure}[H]
\ContinuedFloat
\centering
\par\medskip
\textit{Gates 12--26 (continued).}\par\smallskip
\begin{adjustbox}{max width=\linewidth}
\begin{quantikz}[row sep={0.60cm,between origins},column sep=0.30cm]
\lstick{\(c_0\)} & \qw & \ctrl{7}\csgateframe{clean}{9}{5}{$\mathrm{Uncomputation}_{2}$} & \gate{X} & \targ{} & \qw & \ctrl{8} & \targ{} & \gate{X} & \qw & \qw & \qw & \qw & \rstick{\(c_0\)} \\
\lstick{\(c_1\)} & \qw & \ctrl{6} & \qw & \qw & \qw & \qw & \qw & \qw & \qw & \gate{X} & \gate{X} & \qw & \rstick{\(c_1\)} \\
\lstick{\(c_2\)} & \qw & \ctrl{5} & \gate{X} & \qw & \targ{} & \ctrl{6} & \qw & \qw & \targ{} & \gate{X} & \qw & \qw & \rstick{\(c_2\)} \\
\lstick{\(c_3\)} & \qw & \ctrl{4} & \qw & \qw & \qw & \qw & \qw & \qw & \qw & \gate{X} & \gate{X} & \qw & \rstick{\(c_3\)} \\
\lstick{\(c_4\)} & \qw & \qw & \qw & \ctrl{-4} & \qw & \qw & \ctrl{-4} & \qw & \qw & \qw & \qw & \qw & \rstick{\(c_4\)} \\
\lstick{\(c_5\)} & \qw & \qw & \qw & \qw & \ctrl{-3} & \qw & \qw & \qw & \ctrl{-3} & \qw & \qw & \qw & \rstick{\(c_5\)} \\
\lstick{\(t\)} & \gate{S}\csgateframe{use}{3}{1}{$\mathrm{Use}_{2}$} & \qw & \qw & \qw & \qw & \qw & \qw & \qw & \qw & \qw & \qw & \qw & \rstick{\(t\)} \\
\lstick{\(a_0\)} & \qw & \targ{} & \qw & \qw & \qw & \ctrl{1} & \qw & \qw & \qw & \qw & \qw & \qw & \rstick{\(a_0'\)} \\
\lstick{\(a_1\)} & \ctrl{-2} & \qw & \qw & \qw & \qw & \targ{} & \qw & \qw & \qw & \qw & \qw & \qw & \rstick{\(\ket{0}\)}
\end{quantikz}
\end{adjustbox}
\caption{ConditionalCleanMCS, gates 12--26 (continued). The second-round cleanup restores $a_1$ but makes $a_0$ dirty again. The complete 26-gate candidate therefore fails verification.}
\Description{ConditionalCleanMCS: algorithm output with Prepare, Use, and inserted uncomputation frames.}
\label{fig:example-conditional-clean-mcs}
\end{figure}
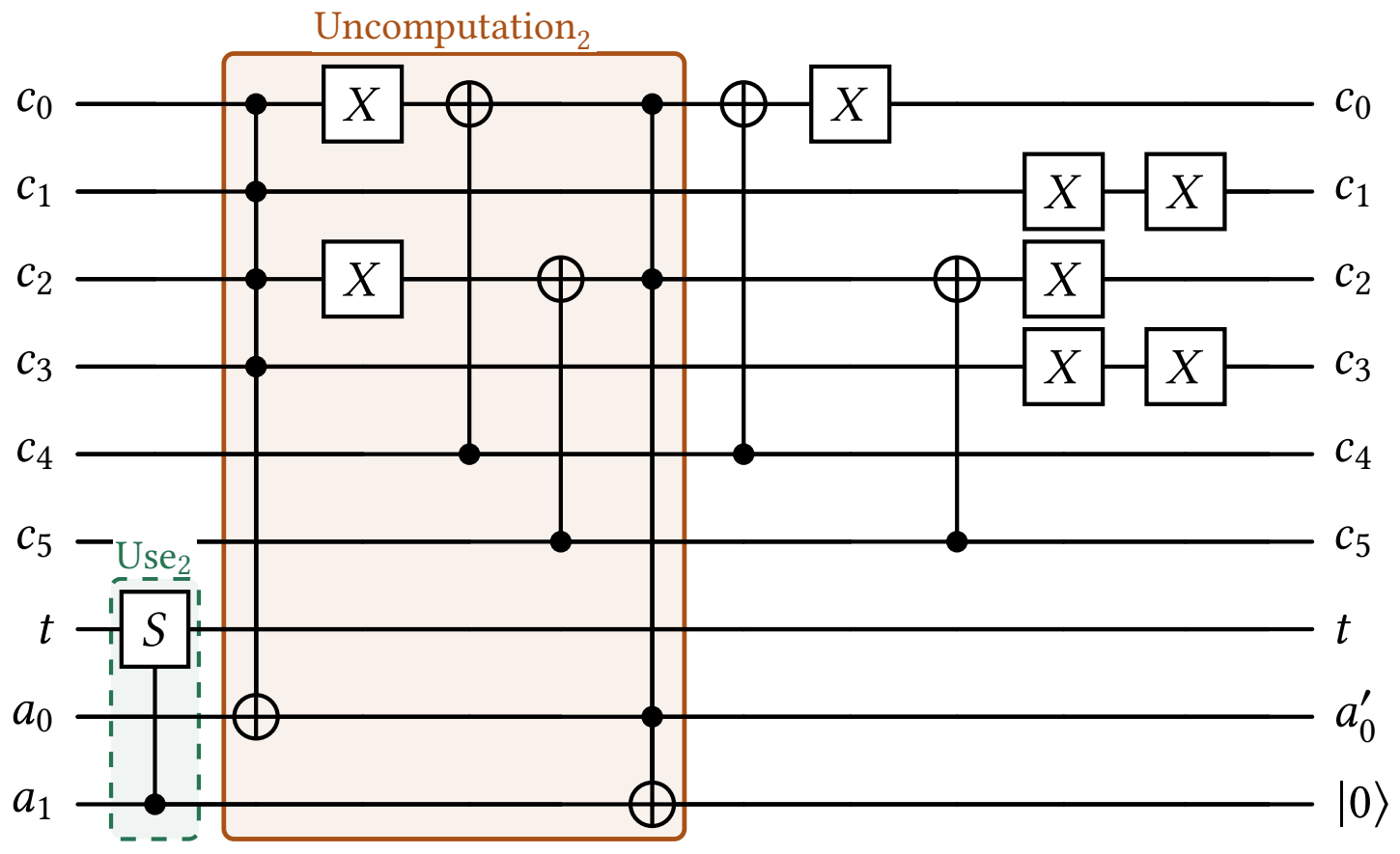

\subsection{Deutsch--Jozsa}
\label{app:example-deutsch-jozsa}
\emph{Source.} \texttt{makesDJ} in \texttt{deutschjozsa.py} of the Reqomp artifact~\cite{reqompArtifact}, with three input qubits and its default all-ones oracle (\texttt{oracle\_gate=None}).

The three-input Deutsch--Jozsa-shaped benchmark uses an oracle $y\mapsto y\oplus x_0x_1x_2$.
The initial five gates are Use.
The algorithm then prepares $x_0x_1$ in $a$, uses it in the oracle, and immediately inserts the inverse Toffoli.
The final three Hadamards are outside the scanned prefix.
For the usual initialization $\ket{000}_x\ket{0}_y$, the initial $X_y;H_y$ prepares the phase-kickback qubit in $\ket{-}$.
Here the oracle is the artifact's fixed benchmark instance; it does not satisfy the balanced-or-constant promise for three input bits.

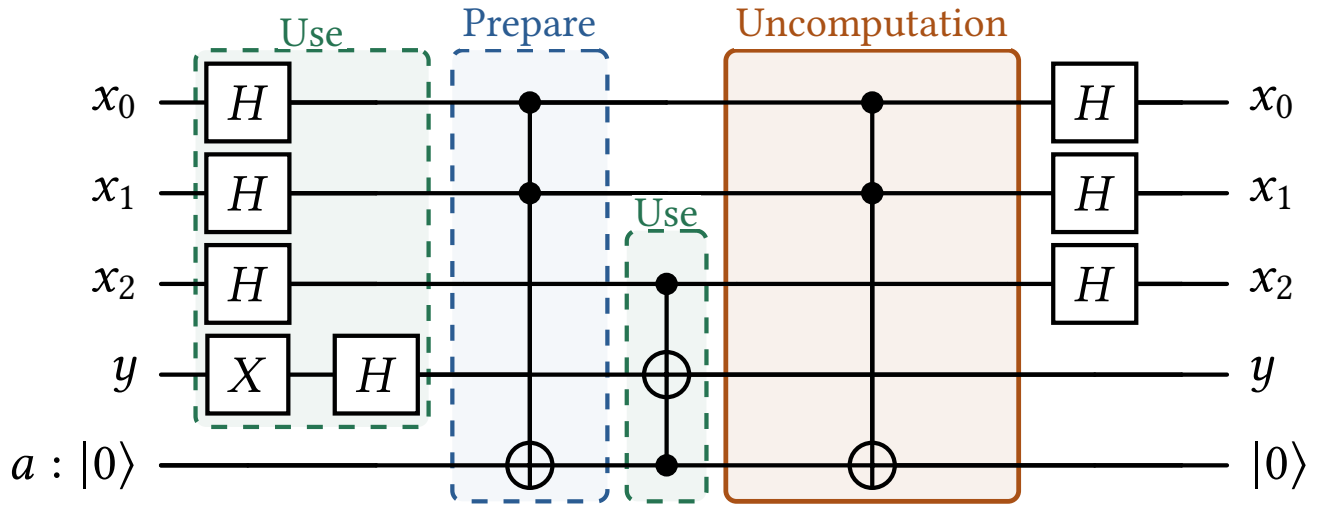
\begin{figure}[H]
\centering
\begin{adjustbox}{max width=\linewidth}
\begin{quantikz}[row sep={0.60cm,between origins},column sep=0.30cm]
\lstick{\(x_0\)} & \gate{H}\csgateframe{use}{4}{2}{$\mathrm{Use}$} & \qw & \ctrl{4}\csgateframe{comp}{5}{1}{$\mathrm{Prepare}$} & \qw & \ctrl{4}\csgateframe{clean}{5}{1}{$\mathrm{Uncomputation}$} & \gate{H} & \qw & \rstick{\(x_0\)} \\
\lstick{\(x_1\)} & \gate{H} & \qw & \ctrl{3} & \qw & \ctrl{3} & \gate{H} & \qw & \rstick{\(x_1\)} \\
\lstick{\(x_2\)} & \gate{H} & \qw & \exlabelspace{$\mathrm{Prepare}$} & \ctrl{1}\csgateframe{use}{3}{1}{$\mathrm{Use}$} & \exlabelspace{$\mathrm{Uncomputation}$} & \gate{H} & \qw & \rstick{\(x_2\)} \\
\lstick{\(y\)} & \gate{X} & \gate{H} & \qw & \targ{} & \qw & \qw & \qw & \rstick{\(y\)} \\
\lstick{\(a:\ket{0}\)} & \qw & \qw & \targ{} & \ctrl{-1} & \targ{} & \qw & \qw & \rstick{\(\ket{0}\)}
\end{quantikz}
\end{adjustbox}
\caption{Deutsch--Jozsa-shaped circuit. The oracle ancilla is reset before the final Hadamards. The 11-gate output passes verification.}
\Description{Deutsch--Jozsa: algorithm output with Prepare, Use, and inserted uncomputation frames.}
\label{fig:example-deutsch-jozsa}
\end{figure}

\subsection{Grover}
\label{app:example-grover}
\emph{Source.} \texttt{makesGroverCircuit} in \texttt{grover.py} of the Reqomp artifact~\cite{reqompArtifact}, with three search qubits, the default all-ones oracle, and \texttt{nb\_sols=1}.
These parameters give two Grover iterations.

The three-qubit Grover instance contains two oracle calls, with separate scratch wires $a_0,a_1$, and two diffusion blocks.
Each oracle computes $x_0x_1$ into its scratch wire and uses it with $x_2$ to flip the phase-kickback wire $p$.
Round $1$ selects $a_0$ and inserts one inverse Toffoli; round $2$ selects $a_1$ and does the same.
On the second scan, the entire first oracle, its cleanup, and the first diffusion block are Use relative to $a_1$.
Their earlier labels are retained where applicable.
The last diffusion block is an unscanned continuation.
With logical inputs initialized to zero, the first five gates prepare the uniform search state and $\ket{-}_p$; verification also covers arbitrary logical inputs.

\begin{figure}[H]
\centering
\textit{Gates 1--23.}\par\smallskip
\begin{adjustbox}{max width=\linewidth}
\begin{quantikz}[row sep={0.60cm,between origins},column sep=0.30cm]
\lstick{\(x_0\)} & \gate{H}\csgateframe{use}{4}{2}{$\mathrm{Use}_{1}$} & \qw & \ctrl{4}\csgateframe{comp}{5}{1}{$\mathrm{Prepare}_{1}$} & \qw & \ctrl{4}\csgateframe{clean}{5}{1}{$\mathrm{Uncomputation}_{1}$} & \gate{H}\csgateframe{use}{3}{7}{$\mathrm{Use}_{2}$} & \gate{X} & \qw & \ctrl{2} & \gate{X} & \gate{H} & \qw & \qw & \rstick{\(x_0\)} \\
\lstick{\(x_1\)} & \gate{H} & \qw & \ctrl{3} & \qw & \ctrl{3} & \gate{H} & \gate{X} & \qw & \ctrl{1} & \gate{X} & \gate{H} & \qw & \qw & \rstick{\(x_1\)} \\
\lstick{\(x_2\)} & \gate{H} & \qw & \exlabelspace{$\mathrm{Prepare}_{1}$} & \ctrl{1}\csgateframe{use}{3}{1}{$\mathrm{Use}_{1}$} & \exlabelspace{$\mathrm{Uncomputation}_{1}$} & \gate{H} & \gate{X} & \gate{H} & \targ{} & \gate{H} & \gate{X} & \gate{H} & \qw & \rstick{\(x_2\)} \\
\lstick{\(p\)} & \gate{X} & \gate{H} & \qw & \targ{} & \qw & \qw & \qw & \qw & \qw & \qw & \qw & \qw & \qw & \rstick{\(p\)} \\
\lstick{\(a_0:\ket{0}\)} & \qw & \qw & \targ{} & \ctrl{-1} & \targ{} & \qw & \qw & \qw & \qw & \qw & \qw & \qw & \qw & \rstick{\(a_0\)} \\
\lstick{\(a_1:\ket{0}\)} & \qw & \qw & \qw & \qw & \qw & \qw & \qw & \qw & \qw & \qw & \qw & \qw & \qw & \rstick{\(a_1\)}
\end{quantikz}
\end{adjustbox}
\par\medskip
\textit{Gates 24--41 (continued).}\par\smallskip
\begin{adjustbox}{max width=\linewidth}
\begin{quantikz}[row sep={0.60cm,between origins},column sep=0.30cm]
\lstick{\(x_0\)} & \ctrl{5}\csgateframe{comp}{6}{1}{$\mathrm{Prepare}_{2}$} & \qw & \ctrl{5}\csgateframe{clean}{6}{1}{$\mathrm{Uncomputation}_{2}$} & \gate{H} & \gate{X} & \qw & \ctrl{2} & \gate{X} & \gate{H} & \qw & \qw & \rstick{\(x_0\)} \\
\lstick{\(x_1\)} & \ctrl{4} & \qw & \ctrl{4} & \gate{H} & \gate{X} & \qw & \ctrl{1} & \gate{X} & \gate{H} & \qw & \qw & \rstick{\(x_1\)} \\
\lstick{\(x_2\)} & \exlabelspace{$\mathrm{Prepare}_{2}$} & \ctrl{1}\csgateframe{use}{4}{1}{$\mathrm{Use}_{2}$} & \exlabelspace{$\mathrm{Uncomputation}_{2}$} & \gate{H} & \gate{X} & \gate{H} & \targ{} & \gate{H} & \gate{X} & \gate{H} & \qw & \rstick{\(x_2\)} \\
\lstick{\(p\)} & \qw & \targ{} & \qw & \qw & \qw & \qw & \qw & \qw & \qw & \qw & \qw & \rstick{\(p\)} \\
\lstick{\(a_0\)} & \qw & \qw & \qw & \qw & \qw & \qw & \qw & \qw & \qw & \qw & \qw & \rstick{\(\ket{0}\)} \\
\lstick{\(a_1\)} & \targ{} & \ctrl{-2} & \targ{} & \qw & \qw & \qw & \qw & \qw & \qw & \qw & \qw & \rstick{\(\ket{0}\)}
\end{quantikz}
\end{adjustbox}
\caption{Grover with two oracle calls. Each diffusion block retains all 15 emitted gates. The two inserted Toffolis restore both ancillas; the complete 41-gate output passes verification.}
\Description{Grover: algorithm output with Prepare, Use, and inserted uncomputation frames.}
\label{fig:example-grover}
\end{figure}
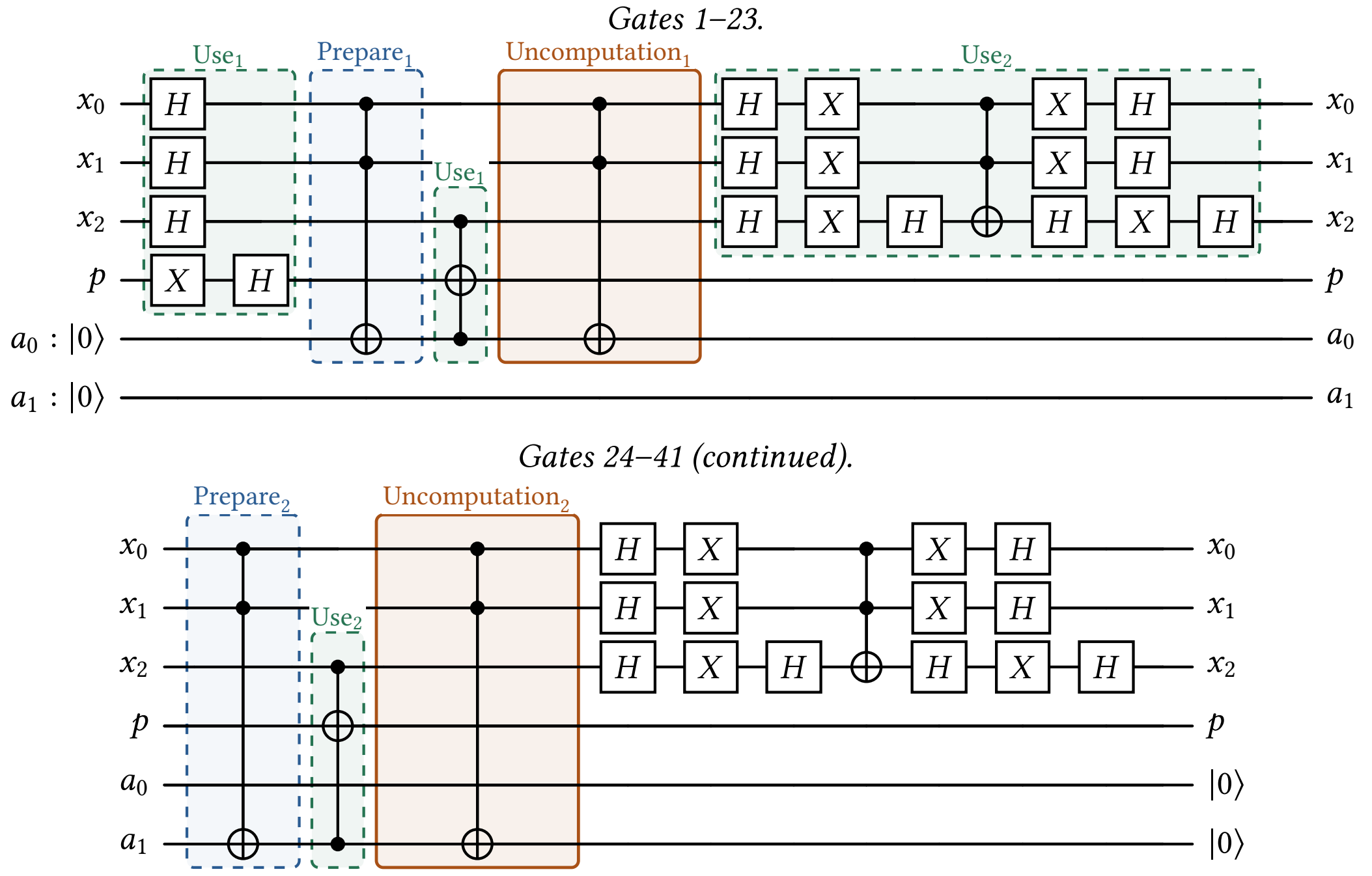

\subsection{Gate Teleportation}
\label{app:example-gate-teleportation}
\emph{Source.} The deterministic $S$-gate construction in Fig.~29 of Fowler et al.~\cite{gate-teleportation} (arXiv:1208.0928).
We make the $\ket{Y}$ preparation explicit and give the algorithm the prefix ending at the first CNOT.

The gate-teleportation example starts by preparing $\ket{Y}=(\ket{0}+i\ket{1})/\sqrt{2}$ with $H_a;S_a$, then applies $\mathrm{CX}_{q\to a}$.
These are the three input gates.
Algorithm~\ref{alg:use-first} classifies all three as Prepare; there is no Use gate.
It therefore appends their inverse, giving the identity on both wires.
The ancilla is restored, but the intended $S_q$ functionality is lost, so the six-gate candidate fails verification.
The successful catalytic restoration discussed in the main text is not the output of this automatic run.

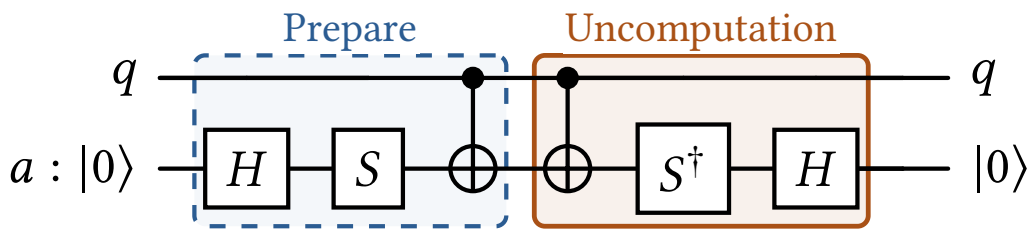
\begin{figure}[H]
\centering
\begin{adjustbox}{max width=\linewidth}
\begin{quantikz}[row sep={0.60cm,between origins},column sep=0.30cm]
\lstick{\(q\)} & \qw\csgateframe{comp}{2}{3}{$\mathrm{Prepare}$} & \qw & \ctrl{1} & \ctrl{1}\csgateframe{clean}{2}{3}{$\mathrm{Uncomputation}$} & \qw & \qw & \qw & \rstick{\(q\)} \\
\lstick{\(a:\ket{0}\)} & \gate{H} & \gate{S} & \targ{} & \targ{} & \gate{S^\dagger} & \gate{H} & \qw & \rstick{\(\ket{0}\)}
\end{quantikz}
\end{adjustbox}
\caption{Gate teleportation: the actual output of Algorithm~\ref{alg:use-first}. The Use block is empty; reversing every input gate yields identity rather than the required $S_q$.}
\Description{Gate Teleportation: algorithm output with Prepare, Use, and inserted uncomputation frames.}
\label{fig:example-gate-teleportation}
\end{figure}

\endgroup

\begingroup
\newcommand{\caseproofmark}[1]{\;\cdots\;#1}

\section{Detailed Derivation for the Case Study in Section~\ref{sec:case}}
\label{app:case}
We give the derivation of the gate-teleportation specification in Section~\ref{sec:case} and extract its uncomputation using Table~\ref{tab:uncomputation-construction}.
As in Fig.~\ref{fig:gate-teleportation}, the primary system is \(\cS=\{q\}\), the ancilla is \(a\), and every controlled gate below has control \(q\) and target \(a\).
The ancilla starts in \(|Y\>_a\), and the synthesized uncomputation restores this same state while preserving the intended functionality \(S_q\).
We write tensor products in the order \(a\otimes q\), and \(U_1;U_2\) means that \(U_1\) is applied first, so its matrix is \(U_2U_1\).

\subsection{Derivation}
\label{app:case-derivation}
We expand the final \textsc{Seq} step in the main text into a derivation tree, naming each subtree so that the synthesis can follow the same structure.
Recall the two roles
\[
    \State(Y)=\frac{|0\>_a+i|1\>_a}{\sqrt{2}}\otimes I_\cS,
    \qquad
    \Choice(iZ)=\frac{|0\>_a\otimes I_\cS+i|1\>_a\otimes Z_q}{\sqrt{2}}.
\]

\paragraph{The final composition: \(\pi\).}
The root of the derivation is the \textsc{Seq} rule already given in Section~\ref{sec:case}:
\begin{equation}
    \inferrule*[right=\textsc{(Seq)},RIGHT={\caseproofmark{\pi}.}]{
        \mathrm{CiY}:\State(Y)\ixrightarrow{S_q}\State(Y)\caseproofmark{\pi_1}
        \\
        H_a;\mathrm{CNOT};H_a:\State(Y)\ixrightarrow{I_\cS}\Choice(iZ)\caseproofmark{\pi_2}
    }{
        \mathrm{CNOT}:\State(Y)\ixrightarrow{S_q}\Choice(iZ)
    }
    \label{eq:case-complete-derivation}
\end{equation}
Here \(\mathrm{CiY};H_a;\mathrm{CNOT};H_a=\mathrm{CNOT}\): on the control branch \(|1\>_q\), the target undergoes \(H_aX_aH_a(iY_a)=X_a\), while on \(|0\>_q\) the two Hadamards cancel.
Thus it remains to construct the two premise derivations \(\pi_1\) and \(\pi_2\).

\paragraph{The Use premise: \(\pi_1\).}
To derive the first premise, expand the controlled gate into four unitary tensor products.
For \(j=1,\ldots,4\), define
\[
\begin{aligned}
    (K_1,D_1,\lambda_1)&=(I_a,I_\cS,1),
    & (K_2,D_2,\lambda_2)&=(I_a,Z_q,1),\\
    (K_3,D_3,\lambda_3)&=(iY_a,I_\cS,i),
    & (K_4,D_4,\lambda_4)&=(iY_a,Z_q,i),
\end{aligned}
\qquad
    (c_1,c_2,c_3,c_4)=\tfrac12(1,1,1,-1).
\]
Then the next rule is \textsc{Sum}:
\begin{equation}
    \inferrule*[right=\textsc{Sum},RIGHT={\caseproofmark{\pi_1}.}]{
        K_j\otimes D_j:\State(Y)\ixrightarrow{\lambda_jD_j}\State(Y)\caseproofmark{\pi_{1,j}}
        \\ j=1,\ldots,4
    }{
        \mathrm{CiY}:\State(Y)\ixrightarrow{S_q}\State(Y)
    }
    \label{eq:case-use}
\end{equation}
Indeed, the physical and logical sums are
\[
\begin{aligned}
    \sum_{j=1}^{4}c_j(K_j\otimes D_j)
        &=\tfrac12\bigl(I_a\otimes I_\cS+I_a\otimes Z_q+iY_a\otimes I_\cS-iY_a\otimes Z_q\bigr)=\mathrm{CiY},\\
    \sum_{j=1}^{4}c_j\lambda_jD_j
        &=\tfrac12\bigl(I_\cS+Z_q+iI_\cS-iZ_q\bigr)=S_q.
\end{aligned}
\]
The physical gates and functionalities in all four premises, as well as both resulting sums, are unitary, so this application of \textsc{Sum} is well typed.
Each premise is in turn obtained by \textsc{Tensor}:
\[
    \inferrule*[right=\textsc{Tensor},RIGHT={\caseproofmark{\pi_{1,j}}.}]{
        K_j:|Y\>_a\ixrightarrow{\lambda_j}|Y\>_a
        \\
        D_j:I_\cS\ixrightarrow{D_j}I_\cS
    }{
        K_j\otimes D_j:\State(Y)\ixrightarrow{\lambda_jD_j}\State(Y)
    }
\]
The ancilla premise will be derived by \(\pi_{3,Y}\) for \(j=1,2\) and by \(\pi_{4,Y}\) for \(j=3,4\); the primary premise will be derived by \(\pi_5(D_j)\).
These local derivations are expanded below after the second main premise.

\paragraph{The Prepare premise: \(\pi_2\).}
We derive the second premise of \(\pi\) by following its three gates in order, using \(H_a|0\>_a=|+\>_a\) and \(H_a|1\>_a=|-\>_a\).
The intermediate roles are written explicitly in the following two \textsc{Seq} steps:
\begin{equation}
    \inferrule*[right=\textsc{(Seq)},RIGHT={\caseproofmark{\pi_2}.}]{
        H_a;\mathrm{CNOT}:\State(Y)\ixrightarrow{I_\cS}\frac{|+\>_a\otimes I_\cS+i|-\>_a\otimes Z_q}{\sqrt{2}}\caseproofmark{\pi_{2,4}}
        \\
        H_a:\frac{|+\>_a\otimes I_\cS+i|-\>_a\otimes Z_q}{\sqrt{2}}\ixrightarrow{I_\cS}\Choice(iZ)\caseproofmark{\pi_{2,3}}
    }{
        H_a;\mathrm{CNOT};H_a:\State(Y)\ixrightarrow{I_\cS}\Choice(iZ)
    }
    \label{eq:case-prepare}
\end{equation}
\[
    \inferrule*[right=\textsc{(Seq)},RIGHT={\caseproofmark{\pi_{2,4}}.}]{
        H_a:\State(Y)\ixrightarrow{I_\cS}\frac{|+\>_a+i|-\>_a}{\sqrt{2}}\otimes I_\cS\caseproofmark{\pi_{2,1}}
        \\
        \mathrm{CNOT}:\frac{|+\>_a+i|-\>_a}{\sqrt{2}}\otimes I_\cS\ixrightarrow{I_\cS}\frac{|+\>_a\otimes I_\cS+i|-\>_a\otimes Z_q}{\sqrt{2}}\caseproofmark{\pi_{2,2}}
    }{
        H_a;\mathrm{CNOT}:\State(Y)\ixrightarrow{I_\cS}\frac{|+\>_a\otimes I_\cS+i|-\>_a\otimes Z_q}{\sqrt{2}}
    }
\]
Both intermediate roles are isometries, since \(|+\>_a\) and \(|-\>_a\) are orthonormal and \(Z_q\) is unitary.
It remains to derive the three individual-gate judgments \(\pi_{2,1}\), \(\pi_{2,2}\), and \(\pi_{2,3}\).

\paragraph{The first Hadamard: \(\pi_{2,1}\).}
Apply \textsc{Glue} with \(A_0=I_\cS/\sqrt{2}\) and \(A_1=iI_\cS/\sqrt{2}\):
\[
    \inferrule*[right=\textsc{Glue},RIGHT={\caseproofmark{\pi_{2,1}}.}]{
        H_a:|0\>_a\otimes I_\cS\ixrightarrow{I_\cS}|+\>_a\otimes I_\cS\caseproofmark{\pi_{2,1,+}}
        \\
        H_a:|1\>_a\otimes I_\cS\ixrightarrow{I_\cS}|-\>_a\otimes I_\cS\caseproofmark{\pi_{2,1,-}}
    }{
        H_a:\State(Y)\ixrightarrow{I_\cS}\frac{|+\>_a+i|-\>_a}{\sqrt{2}}\otimes I_\cS
    }
\]
The coefficients combine the input basis states into \(|Y\>_a\) and their Hadamard images into \((|+\>_a+i|-\>_a)/\sqrt{2}\).
For \(b=0,1\), each premise follows from \textsc{Tensor}, with \(+\) corresponding to \(b=0\) and \(-\) to \(b=1\):
\[
    \inferrule*[right=\textsc{Tensor},RIGHT={\caseproofmark{\pi_{2,1,\pm}}.}]{
        H_a:|b\>_a\ixrightarrow{1}|\pm\>_a
        \\
        I_\cS:I_\cS\ixrightarrow{I_\cS}I_\cS\caseproofmark{\pi_5(I_\cS)}
    }{
        H_a:|b\>_a\otimes I_\cS\ixrightarrow{I_\cS}|\pm\>_a\otimes I_\cS
    }
\]
The ancilla premises are the two instances of \textsc{Ax-H-Anc}, and \(H_a\) in the conclusion includes the identity on \(\cS\).

\paragraph{The middle CNOT: \(\pi_{2,2}\).}
Since \(X_a|+\>_a=|+\>_a\) and \(X_a|-\>_a=-|-\>_a\), a CNOT preserves either target state and implements \(I_\cS\) or \(Z_q\), respectively, on the control.
We derive these two judgments below and combine them using \textsc{Glue}, with coefficients \(I_\cS/\sqrt{2}\) and \(iI_\cS/\sqrt{2}\):
\[
    \inferrule*[right=\textsc{Glue},RIGHT={\caseproofmark{\pi_{2,2}}.}]{
        \mathrm{CNOT}:|+\>_a\otimes I_\cS\ixrightarrow{I_\cS}|+\>_a\otimes I_\cS\caseproofmark{\pi_{2,2,+}}
        \\
        \mathrm{CNOT}:|-\>_a\otimes I_\cS\ixrightarrow{Z_q}|-\>_a\otimes I_\cS\caseproofmark{\pi_{2,2,-}}
    }{
        \mathrm{CNOT}:\frac{|+\>_a+i|-\>_a}{\sqrt{2}}\otimes I_\cS\ixrightarrow{I_\cS}\frac{|+\>_a\otimes I_\cS+i|-\>_a\otimes Z_q}{\sqrt{2}}
    }
\]
The output role includes \(Z_q\) on the \(|-\>_a\) branch, as prescribed by \textsc{Glue}.
To expand the premises, use the same coefficients \(c_j\) and primary gates \(D_j\) as in \(\pi_1\), now with
\[
\begin{aligned}
    (L_1,L_2,L_3,L_4)&=(I_a,I_a,X_a,X_a),\\
    (\mu_{1,+},\mu_{2,+},\mu_{3,+},\mu_{4,+})&=(1,1,1,1),\\
    (\mu_{1,-},\mu_{2,-},\mu_{3,-},\mu_{4,-})&=(1,1,-1,-1).
\end{aligned}
\]
The two \textsc{Sum} steps are
\[
    \inferrule*[right=\textsc{Sum},RIGHT={\caseproofmark{\pi_{2,2,+}},}]{
        L_j\otimes D_j:|+\>_a\otimes I_\cS\ixrightarrow{\mu_{j,+}D_j}|+\>_a\otimes I_\cS\caseproofmark{\pi_{2,2,+,j}}
        \\ j=1,\ldots,4
    }{
        \mathrm{CNOT}:|+\>_a\otimes I_\cS\ixrightarrow{I_\cS}|+\>_a\otimes I_\cS
    }
\]
\[
    \inferrule*[right=\textsc{Sum},RIGHT={\caseproofmark{\pi_{2,2,-}}.}]{
        L_j\otimes D_j:|-\>_a\otimes I_\cS\ixrightarrow{\mu_{j,-}D_j}|-\>_a\otimes I_\cS\caseproofmark{\pi_{2,2,-,j}}
        \\ j=1,\ldots,4
    }{
        \mathrm{CNOT}:|-\>_a\otimes I_\cS\ixrightarrow{Z_q}|-\>_a\otimes I_\cS
    }
\]
For either choice of sign, the premises follow from \textsc{Tensor}:
\[
    \inferrule*[right=\textsc{Tensor},RIGHT={\caseproofmark{\pi_{2,2,\pm,j}}.}]{
        L_j:|\pm\>_a\ixrightarrow{\mu_{j,\pm}}|\pm\>_a
        \\
        D_j:I_\cS\ixrightarrow{D_j}I_\cS
    }{
        L_j\otimes D_j:|\pm\>_a\otimes I_\cS\ixrightarrow{\mu_{j,\pm}D_j}|\pm\>_a\otimes I_\cS
    }
\]
The \textsc{Sum} steps use the identities
\[
\begin{aligned}
    \sum_{j=1}^{4}c_j(L_j\otimes D_j)
        &=\tfrac12\bigl(I_a\otimes I_\cS+I_a\otimes Z_q+X_a\otimes I_\cS-X_a\otimes Z_q\bigr)=\mathrm{CNOT},\\
    \sum_{j=1}^{4}c_j\mu_{j,+}D_j
        &=\tfrac12\bigl(I_\cS+Z_q+I_\cS-Z_q\bigr)=I_\cS,\\
    \sum_{j=1}^{4}c_j\mu_{j,-}D_j
        &=\tfrac12\bigl(I_\cS+Z_q-I_\cS+Z_q\bigr)=Z_q.
\end{aligned}
\]
The ancilla premises are \(\pi_{3,\pm}\) for \(j=1,2\) and \(\pi_{4,\pm}\) for \(j=3,4\), with the same sign used throughout each branch; the primary premises are \(\pi_5(D_j)\).
Thus each physical gate and functionality in the premises and conclusions is unitary.

\paragraph{The final Hadamard: \(\pi_{2,3}\).}
For the last gate, apply \textsc{Glue} with coefficients \(I_\cS/\sqrt{2}\) and \(iZ_q/\sqrt{2}\):
\[
    \inferrule*[right=\textsc{Glue},RIGHT={\caseproofmark{\pi_{2,3}}.}]{
        H_a:|+\>_a\otimes I_\cS\ixrightarrow{I_\cS}|0\>_a\otimes I_\cS\caseproofmark{\pi_{2,3,+}}
        \\
        H_a:|-\>_a\otimes I_\cS\ixrightarrow{I_\cS}|1\>_a\otimes I_\cS\caseproofmark{\pi_{2,3,-}}
    }{
        H_a:\frac{|+\>_a\otimes I_\cS+i|-\>_a\otimes Z_q}{\sqrt{2}}\ixrightarrow{I_\cS}\Choice(iZ)
    }
\]
These coefficients combine the output basis states into \(\Choice(iZ)\).
Derive each premise by \textsc{Tensor}, with \(+\) corresponding to \(b=0\) and \(-\) to \(b=1\):
\[
    \inferrule*[right=\textsc{Tensor},RIGHT={\caseproofmark{\pi_{2,3,\pm}}.}]{
        H_a:|\pm\>_a\ixrightarrow{1}|b\>_a\caseproofmark{\pi_{7,\pm}}
        \\
        I_\cS:I_\cS\ixrightarrow{I_\cS}I_\cS\caseproofmark{\pi_5(I_\cS)}
    }{
        H_a:|\pm\>_a\otimes I_\cS\ixrightarrow{I_\cS}|b\>_a\otimes I_\cS
    }
\]
The two ancilla premises \(\pi_{7,+}\) and \(\pi_{7,-}\) are derived below from the computational-basis Hadamard axioms.
These three gate derivations complete \(\pi_2\) once their shared local premises are expanded as follows.

\paragraph{The three ancilla identity premises: \(\pi_{3,Y}\) and \(\pi_{3,\pm}\).}
The Use branch needs the identity on \(|Y\>_a\), while the CNOT branches need the identity on \(|+\>_a\) and \(|-\>_a\).
For \(|Y\>_a\), apply \textsc{Glue} with \(A_0=1/\sqrt{2}\) and \(A_1=i/\sqrt{2}\):
\[
    \inferrule*[right=\textsc{Glue},RIGHT={\caseproofmark{\pi_{3,Y}}.}]{
        I_a:|0\>_a\ixrightarrow{1}|0\>_a\caseproofmark{\pi_{3,0}}
        \\
        I_a:|1\>_a\ixrightarrow{1}|1\>_a\caseproofmark{\pi_{3,1}}
    }{
        I_a:|Y\>_a\ixrightarrow{1}|Y\>_a
    }
\]
For \(|+\>_a\) and \(|-\>_a\), the coefficients are \(A_0=1/\sqrt{2}\) and \(A_1=\pm1/\sqrt{2}\):
\[
    \inferrule*[right=\textsc{Glue},RIGHT={\caseproofmark{\pi_{3,\pm}}.}]{
        I_a:|0\>_a\ixrightarrow{1}|0\>_a\caseproofmark{\pi_{3,0}}
        \\
        I_a:|1\>_a\ixrightarrow{1}|1\>_a\caseproofmark{\pi_{3,1}}
    }{
        I_a:|\pm\>_a\ixrightarrow{1}|\pm\>_a
    }
\]
The two basis-state premises, for \(b=0,1\), come directly from two \textsc{Ax-X-Anc} instances and \textsc{Seq}:
\[
    \inferrule*[right=\textsc{(Seq)},RIGHT={\caseproofmark{\pi_{3,b}}.}]{
        X_a:|b\>_a\ixrightarrow{1}|b\oplus1\>_a
        \\
        X_a:|b\oplus1\>_a\ixrightarrow{1}|b\>_a
    }{
        I_a:|b\>_a\ixrightarrow{1}|b\>_a
    }
\]
Here the physical circuit in the conclusion is \(X_a;X_a=I_a\).

\paragraph{The \(iY_a\) premise on \(|Y\>_a\): \(\pi_{4,Y}\).}
The Use branch requires \(iY_a:|Y\>_a\ixrightarrow{i}|Y\>_a\), with \(i\) in the functionality.
Its last rule is a single-term \textsc{Sum} with coefficient \(i\):
\[
    \inferrule*[right=\textsc{Sum},RIGHT={\caseproofmark{\pi_{4,Y}}.}]{
        Y_a:|Y\>_a\ixrightarrow{1}|Y\>_a\caseproofmark{\pi_{4,Y,1}}
        \\ c_1=i
    }{
        iY_a:|Y\>_a\ixrightarrow{i}|Y\>_a
    }
\]
The premise follows by \textsc{Glue} with \(A_1=1\):
\[
    \inferrule*[right=\textsc{Glue},RIGHT={\caseproofmark{\pi_{4,Y,1}}.}]{
        Y_a:|Y\>_a\ixrightarrow{-i}i|Y\>_a\caseproofmark{\pi_{4,Y,2}}
        \\ A_1=1
    }{
        Y_a:|Y\>_a\ixrightarrow{1}|Y\>_a
    }
\]
This output role is \((i|Y\>_a)(-i)=|Y\>_a\).
Finally, the preceding \textsc{Sum} step has coefficient \(-i\):
\[
    \inferrule*[right=\textsc{Sum},RIGHT={\caseproofmark{\pi_{4,Y,2}}.}]{
        iY_a:|Y\>_a\ixrightarrow{1}i|Y\>_a\caseproofmark{\pi_6}
        \\ c_1=-i
    }{
        Y_a:|Y\>_a\ixrightarrow{-i}i|Y\>_a
    }
\]
The remaining premise \(\pi_6\) is expanded gate by gate below, using \(iY_a=X_a;H_a;X_a;H_a\).

\paragraph{The two \(X_a\) premises: \(\pi_{4,+}\) and \(\pi_{4,-}\).}
For the \(|+\>_a\) branch, \textsc{Glue} with \(A_0=A_1=1/\sqrt{2}\) gives
\[
    \inferrule*[right=\textsc{Glue},RIGHT={\caseproofmark{\pi_{4,+}}.}]{
        X_a:|0\>_a\ixrightarrow{1}|1\>_a
        \\
        X_a:|1\>_a\ixrightarrow{1}|0\>_a
    }{
        X_a:|+\>_a\ixrightarrow{1}|+\>_a
    }
\]
Both premises are \textsc{Ax-X-Anc} instances.
For the \(|-\>_a\) branch, the required functionality is \(-1\), so the last step is \textsc{Sum} with coefficient \(-1\):
\[
    \inferrule*[right=\textsc{Sum},RIGHT={\caseproofmark{\pi_{4,-}}.}]{
        -X_a:|-\>_a\ixrightarrow{1}|-\>_a\caseproofmark{\pi_{4,-,1}}
        \\ c_1=-1
    }{
        X_a:|-\>_a\ixrightarrow{-1}|-\>_a
    }
\]
Its premise is obtained by the following \textsc{Glue} and \textsc{Sum} steps:
\[
    \inferrule*[right=\textsc{Glue},RIGHT={\caseproofmark{\pi_{4,-,1}},}]{
        -X_a:|-\>_a\ixrightarrow{-1}-|-\>_a\caseproofmark{\pi_{4,-,2}}
        \\ A_1=1
    }{
        -X_a:|-\>_a\ixrightarrow{1}|-\>_a
    }
\]
\[
    \inferrule*[right=\textsc{Sum},RIGHT={\caseproofmark{\pi_{4,-,2}}.}]{
        X_a:|-\>_a\ixrightarrow{1}-|-\>_a\caseproofmark{\pi_{4,-,3}}
        \\ c_1=-1
    }{
        -X_a:|-\>_a\ixrightarrow{-1}-|-\>_a
    }
\]
The first of these steps has output role \((-|-\>_a)(-1)=|-\>_a\).
The remaining premise uses \textsc{Glue} with \(A_0=1/\sqrt{2}\) and \(A_1=-1/\sqrt{2}\):
\[
    \inferrule*[right=\textsc{Glue},RIGHT={\caseproofmark{\pi_{4,-,3}}.}]{
        X_a:|0\>_a\ixrightarrow{1}|1\>_a
        \\
        X_a:|1\>_a\ixrightarrow{1}|0\>_a
    }{
        X_a:|-\>_a\ixrightarrow{1}-|-\>_a
    }
\]
Its output role is \((|1\>_a-|0\>_a)/\sqrt{2}=-|-\>_a\), and both premises are again \textsc{Ax-X-Anc} instances.
All phases in these steps have modulus one, so the displayed gates, functionalities, and roles have the required types.

\paragraph{The primary-qubit premises: \(\pi_5\).}
The primary-qubit axioms and \textsc{Seq} derive the two judgments required above:
\[
    \inferrule*[right=\textsc{(Seq)},RIGHT={\caseproofmark{\pi_5(I_\cS)},}]{
        X_q:I_\cS\ixrightarrow{X_q}I_\cS
        \\
        X_q:I_\cS\ixrightarrow{X_q}I_\cS
    }{
        I_\cS:I_\cS\ixrightarrow{I_\cS}I_\cS
    }
\]
\[
    \inferrule*[right=\textsc{(Seq)},RIGHT={\caseproofmark{\pi_5(Z_q)}.}]{
        \inferrule*[right=\textsc{(Seq)}]{
            H_q:I_\cS\ixrightarrow{H_q}I_\cS
            \\
            X_q:I_\cS\ixrightarrow{X_q}I_\cS
        }{
            H_q;X_q:I_\cS\ixrightarrow{H_q;X_q}I_\cS
        }
        \\
        H_q:I_\cS\ixrightarrow{H_q}I_\cS
    }{
        Z_q:I_\cS\ixrightarrow{Z_q}I_\cS
    }
\]
Every \(X_q\) leaf is an instance of \textsc{Ax-X}, and every \(H_q\) leaf is an instance of \textsc{Ax-H}.
The conclusions use \(X;X=I\) and \(H;X;H=Z\).

\paragraph{The four gates implementing \(iY_a\): \(\pi_6\).}
We now derive the remaining judgment \(iY_a:|Y\>_a\ixrightarrow{1}i|Y\>_a\) by following \(X_a;H_a;X_a;H_a\).
The last \textsc{Seq} step combines the first two gates with the last two:
\[
    \inferrule*[right=\textsc{(Seq)},RIGHT={\caseproofmark{\pi_6}.}]{
        X_a;H_a:|Y\>_a\ixrightarrow{1}\frac{i|+\>_a+|-\>_a}{\sqrt{2}}\caseproofmark{\pi_{6,L}}
        \\
        X_a;H_a:\frac{i|+\>_a+|-\>_a}{\sqrt{2}}\ixrightarrow{1}i|Y\>_a\caseproofmark{\pi_{6,R}}
    }{
        iY_a:|Y\>_a\ixrightarrow{1}i|Y\>_a
    }
\]
Each two-gate premise is another \textsc{Seq} step:
\[
    \inferrule*[right=\textsc{(Seq)},RIGHT={\caseproofmark{\pi_{6,L}},}]{
        X_a:|Y\>_a\ixrightarrow{1}\frac{i|0\>_a+|1\>_a}{\sqrt{2}}\caseproofmark{\pi_{6,1}}
        \\
        H_a:\frac{i|0\>_a+|1\>_a}{\sqrt{2}}\ixrightarrow{1}\frac{i|+\>_a+|-\>_a}{\sqrt{2}}\caseproofmark{\pi_{6,2}}
    }{
        X_a;H_a:|Y\>_a\ixrightarrow{1}\frac{i|+\>_a+|-\>_a}{\sqrt{2}}
    }
\]
\[
    \inferrule*[right=\textsc{(Seq)},RIGHT={\caseproofmark{\pi_{6,R}}.}]{
        X_a:\frac{i|+\>_a+|-\>_a}{\sqrt{2}}\ixrightarrow{1}\frac{i|+\>_a-|-\>_a}{\sqrt{2}}\caseproofmark{\pi_{6,3}}
        \\
        H_a:\frac{i|+\>_a-|-\>_a}{\sqrt{2}}\ixrightarrow{1}i|Y\>_a\caseproofmark{\pi_{6,4}}
    }{
        X_a;H_a:\frac{i|+\>_a+|-\>_a}{\sqrt{2}}\ixrightarrow{1}i|Y\>_a
    }
\]
For the first \(X_a\), apply \textsc{Glue} to the computational-basis axioms with coefficients \(1/\sqrt{2}\) and \(i/\sqrt{2}\):
\[
    \inferrule*[right=\textsc{Glue},RIGHT={\caseproofmark{\pi_{6,1}}.}]{
        X_a:|0\>_a\ixrightarrow{1}|1\>_a
        \\
        X_a:|1\>_a\ixrightarrow{1}|0\>_a
    }{
        X_a:|Y\>_a\ixrightarrow{1}\frac{i|0\>_a+|1\>_a}{\sqrt{2}}
    }
\]
For the following \(H_a\), the coefficients are \(i/\sqrt{2}\) and \(1/\sqrt{2}\):
\[
    \inferrule*[right=\textsc{Glue},RIGHT={\caseproofmark{\pi_{6,2}}.}]{
        H_a:|0\>_a\ixrightarrow{1}|+\>_a
        \\
        H_a:|1\>_a\ixrightarrow{1}|-\>_a
    }{
        H_a:\frac{i|0\>_a+|1\>_a}{\sqrt{2}}\ixrightarrow{1}\frac{i|+\>_a+|-\>_a}{\sqrt{2}}
    }
\]
These two premises are \textsc{Ax-H-Anc} instances.
For the second \(X_a\), use \(\pi_{4,+}\) and \(\pi_{4,-}\), with coefficients \(i/\sqrt{2}\) and \(1/\sqrt{2}\):
\[
    \inferrule*[right=\textsc{Glue},RIGHT={\caseproofmark{\pi_{6,3}}.}]{
        X_a:|+\>_a\ixrightarrow{1}|+\>_a\caseproofmark{\pi_{4,+}}
        \\
        X_a:|-\>_a\ixrightarrow{-1}|-\>_a\caseproofmark{\pi_{4,-}}
    }{
        X_a:\frac{i|+\>_a+|-\>_a}{\sqrt{2}}\ixrightarrow{1}\frac{i|+\>_a-|-\>_a}{\sqrt{2}}
    }
\]
Finally, apply \textsc{Glue} to the two Hadamard judgments needed for \(\pi_{2,3}\), now with coefficients \(i/\sqrt{2}\) and \(-1/\sqrt{2}\):
\[
    \inferrule*[right=\textsc{Glue},RIGHT={\caseproofmark{\pi_{6,4}}.}]{
        H_a:|+\>_a\ixrightarrow{1}|0\>_a\caseproofmark{\pi_{7,+}}
        \\
        H_a:|-\>_a\ixrightarrow{1}|1\>_a\caseproofmark{\pi_{7,-}}
    }{
        H_a:\frac{i|+\>_a-|-\>_a}{\sqrt{2}}\ixrightarrow{1}i|Y\>_a
    }
\]
The output is \((i|0\>_a-|1\>_a)/\sqrt{2}=i|Y\>_a\), which completes \(\pi_6\) once the two Hadamard premises below are supplied.

\paragraph{The Hadamard premises on \(|+\>_a\) and \(|-\>_a\): \(\pi_{7,\pm}\).}
The \(|+\>_a\) premise uses \textsc{Glue} with \(A_0=A_1=1/\sqrt{2}\):
\[
    \inferrule*[right=\textsc{Glue},RIGHT={\caseproofmark{\pi_{7,+}}.}]{
        H_a:|0\>_a\ixrightarrow{1}|+\>_a
        \\
        H_a:|1\>_a\ixrightarrow{1}|-\>_a
    }{
        H_a:|+\>_a\ixrightarrow{1}|0\>_a
    }
\]
The \(|-\>_a\) premise instead uses \(A_0=1/\sqrt{2}\) and \(A_1=-1/\sqrt{2}\):
\[
    \inferrule*[right=\textsc{Glue},RIGHT={\caseproofmark{\pi_{7,-}}.}]{
        H_a:|0\>_a\ixrightarrow{1}|+\>_a
        \\
        H_a:|1\>_a\ixrightarrow{1}|-\>_a
    }{
        H_a:|-\>_a\ixrightarrow{1}|1\>_a
    }
\]
The output identities are \((|+\>_a+|-\>_a)/\sqrt{2}=|0\>_a\) and \((|+\>_a-|-\>_a)/\sqrt{2}=|1\>_a\).
All four leaves are \textsc{Ax-H-Anc} instances, so every branch of the case-study derivation \(\pi\) now reaches the axioms.

\subsection{Uncomputation Synthesis}
\label{app:case-synthesis}
We now apply Table~\ref{tab:uncomputation-construction} recursively to the same numbered derivation trees.
Since the last rule of \(\pi\) is \textsc{Seq}, the first extraction step is
\[
    V(\pi)=V(\pi_2);V(\pi_1).
\]
The target is to restore \(\State(Y)\), the input role of \(\pi\).

\paragraph{Follow the Use subtree \(\pi_1\).}
The last rule of \(\pi_1\) is \textsc{Sum}, so extraction may follow any one of its four premises.
Choose \(\pi_{1,1}\), the identity tensor-product branch, whose last rule is \textsc{Tensor}:
\[
\begin{aligned}
    V(\pi_1)&=V(\pi_{1,1})
        &&\text{by \textsc{Sum}},\\
    V(\pi_{1,1})&=V(\pi_{3,Y})\otimes V(\pi_5(I_\cS))
        &&\text{by \textsc{Tensor}}.
\end{aligned}
\]
The last rule of \(\pi_{3,Y}\) is \textsc{Glue} for the physical identity \(I_a\), so it returns \(I_a^\dagger\) directly.
The tree \(\pi_5(I_\cS)\) consists of \textsc{Seq} applied to two \textsc{Ax-X} leaves, each of which returns \(I_\cS\).
Consequently,
\[
    V(\pi_{3,Y})=I_a,
    \qquad V(\pi_5(I_\cS))=I_\cS;I_\cS=I_\cS,
    \qquad V(\pi_1)=I_a\otimes I_\cS.
\]
No other branch of the \textsc{Sum} in \(\pi_1\) needs to be traversed, and the logical \(S_q\) implemented by this subtree is retained.

\paragraph{Follow the Prepare subtree \(\pi_2\).}
The last rules of \(\pi_2\) and \(\pi_{2,4}\) are both \textsc{Seq}, so extraction follows the same two-level structure in reverse order:
\[
\begin{aligned}
    V(\pi_2)&=V(\pi_{2,3});V(\pi_{2,4}),\\
    V(\pi_{2,4})&=V(\pi_{2,2});V(\pi_{2,1}).
\end{aligned}
\]
Each individual-gate subtree ends with \textsc{Glue}, which returns the inverse of that gate without descending into its premises.
Since \(H_a\) and CNOT are self-inverse,
\[
    V(\pi_{2,1})=H_a,
    \qquad V(\pi_{2,2})=\mathrm{CNOT},
    \qquad V(\pi_{2,3})=H_a.
\]
Hence the extracted Prepare cleanup is
\[
    V(\pi_2)=V(\pi_{2,3});V(\pi_{2,2});V(\pi_{2,1})
    =H_a;\mathrm{CNOT};H_a.
\]
Substituting the results for \(\pi_1\) and \(\pi_2\) into the root step gives
\begin{equation}
    V(\pi)=V(\pi_2);V(\pi_1)
    =H_a;\mathrm{CNOT};H_a.
    \label{eq:case-extracted-uncomputation}
\end{equation}
Thus the derivation-directed construction produces exactly the three-gate uncomputation in Fig.~\ref{fig:gate-teleportation}, as illustrated below.

\begin{figure}[htbp]
    \centering
    \begingroup
    \newcommand{\caseframe}[4]{%
        \gategroup[wires=#2,steps=#3,background,style={cs#1,rounded corners=2pt,inner xsep=2pt,inner ysep=2pt},label style={cs#1label,label position=above,anchor=south,yshift=2pt}]{\footnotesize #4}}
    \begin{quantikz}[row sep={0.65cm,between origins},column sep=0.38cm]
        \lstick{\(q\)} & \ctrl{1}\caseframe{use}{2}{1}{Use} &[8pt] \qw\caseframe{comp}{2}{3}{Prepare} & \ctrl{1} & \qw &[10pt] \qw\caseframe{clean}{2}{3}{Uncomputation} & \ctrl{1} & \qw & \qw \\
        \lstick{\(a:|Y\>\)} & \gate{iY} & \gate{H} & \targ{} & \gate{H} & \gate{H} & \targ{} & \gate{H} & \rstick{\(|Y\>\)}
    \end{quantikz}
    \endgroup
    \caption{Uncomputation extracted from Eq.~\eqref{eq:case-complete-derivation}. The Use and Prepare blocks together equal the original first CNOT. The extracted Uncomputation restores \(|Y\>_a\) while preserving the functionality \(S_q\).}
    \Description{A two-qubit circuit with a controlled-iY Use block, a Hadamard--CNOT--Hadamard Prepare block, and an identical three-gate Uncomputation block. The ancilla starts and ends in the magic state Y.}
    \label{fig:case-extracted-uncomputation}
\end{figure}
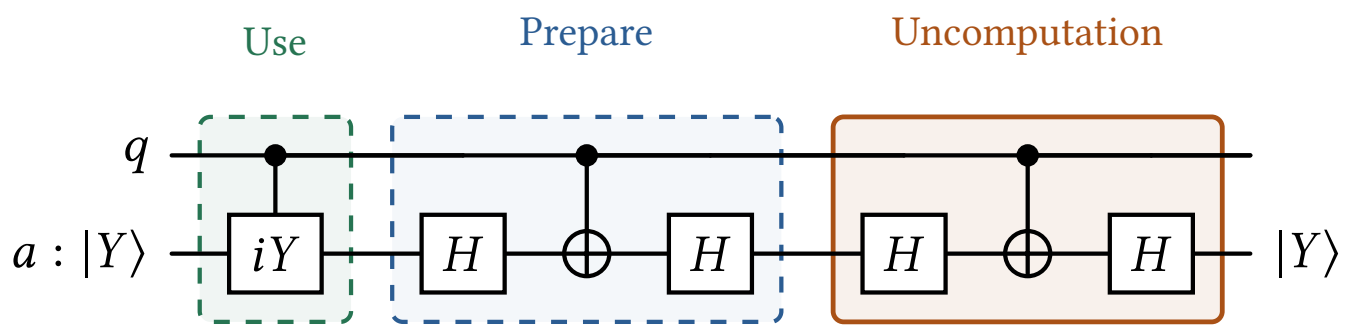

\paragraph{Check the resulting specification.}
The three extracted gates restore the intermediate roles in reverse order.
Using the two CNOT judgments on \(|+\>_a\otimes I_\cS\) and \(|-\>_a\otimes I_\cS\), together with \(Z_q^2=I_\cS\), gives
\[
\begin{aligned}
    V(\pi)\Choice(iZ)
    &=H_a\mathrm{CNOT}H_a\Choice(iZ)\\
    &=H_a\mathrm{CNOT}\frac{|+\>_a\otimes I_\cS+i|-\>_a\otimes Z_q}{\sqrt{2}}\\
    &=H_a\left(\frac{|+\>_a+i|-\>_a}{\sqrt{2}}\otimes I_\cS\right)\\
    &=\State(Y).
\end{aligned}
\]
Thus \(V(\pi):\Choice(iZ)\ixRightarrow{I_\cS}\State(Y)\), and the complete circuit satisfies
\[
    \mathrm{CNOT};H_a;\mathrm{CNOT};H_a:
    \State(Y)\ixRightarrow{S_q}\State(Y).
\]
Equivalently, its output on every input \(|Y\>_a\otimes|\psi\>_q\) is \(|Y\>_a\otimes S_q|\psi\>_q\).

\endgroup

\section{Dependency-Aware Extension of the Mid-Circuit Strategy}
\label{app:dependency-aware-strategy}
This appendix extends Algorithm~\ref{alg:mid-circuit} to account for dependencies between ancillas and evaluates the extension on a larger benchmark suite.
The extension changes the order and placement of uncomputation while retaining the Prepare/Use classification of Algorithm~\ref{alg:use-first}.

\subsection{Choosing the Order and Position of Uncomputation}
\paragraph{Respecting ancilla dependencies.}
An ancilla should remain available until the ancillas that depend on it have been uncomputed.
We record these dependencies in a directed graph built from the input circuit \(U\).
An edge \(a\to b\) is added when an input gate acts on both ancillas, preserves the computational-basis value of \(a\), and may change that of \(b\), as when \(a\) controls a gate targeting \(b\).
The edge therefore schedules \(b\) before \(a\).
At each iteration, we consider only the remaining ancillas with no outgoing edge to another remaining ancilla.
Among them, we select the ancilla whose last participation occurs earliest in the current circuit, counting both control and target positions and including previously inserted gates.
The dependency graph remains fixed; only the set of remaining ancillas and their last participation positions change.
If no ancilla is eligible, the remaining graph contains a cycle and the strategy returns \emph{unsupported}.

\paragraph{Choosing the insertion point.}
For the selected ancilla \(a\), first run Use-first on the prefix ending at its last participating gate \(U'_i\).
Let \(U_{\mathrm{prep}}\) be the product, in execution order, of the gates that this call marks as {\color{csRoleBlue}Prepare} and that act entirely on the original primary system \(\cS\).
We then inspect the consecutive gates immediately after \(U'_i\), stopping as soon as a gate involves any original ancilla.
If these primary-system gates cancel \(U_{\mathrm{prep}}\) up to a global phase, we move the insertion point to the earliest such position and rerun Use-first on the enlarged prefix.
Otherwise, we keep the original insertion point.
Here \(\cS=\dom U\setminus\cA\) always refers to the primary system of the original input, even though each Use-first call designates only \(\{a\}\) as its ancilla system.
Algorithm~\ref{alg:dependency-aware} summarizes this scheduling procedure.

\begin{algorithm}[t]
    \caption{Dependency-aware scheduling with the \textbf{Use-first} strategy.}
    \label{alg:dependency-aware}
    \DontPrintSemicolon
    \KwIn{Computation circuit \(U\) and set \(\cA\) of ancillas initialized to zero.}
    \KwOut{A candidate circuit \(U'\) with uncomputations, or \emph{unsupported}.}
    \(\cS\gets\dom U\setminus\cA;\quad \cA'\gets\cA;\quad U'\gets U\)\;
    Build the ancilla dependency graph from \(U\), with edges \(a\to b\) as described above\;
    \While{\(\cA'\neq\emptyset\)}{
        \(E\gets\{a\in\cA':\text{ no edge }a\to b\text{ with }b\in\cA'\}\)\;
        \If{\(E=\emptyset\)}{\Return{\emph{unsupported}}\;}
        Choose \(a\in E\) whose last participation in \(U'\) is earliest; let \(i\) be that gate's index\;
        \(V\gets\textbf{Use-first}(U'_1;\ldots;U'_i,\{a\})\) using Algorithm~\ref{alg:use-first}\;
        \(U_{\mathrm{prep}}\gets\) product, in execution order, of this call's {\color{csRoleBlue}Prepare} gates acting only on \(\cS\)\;
        Find the smallest \(j\ge i\) such that \(U'_{i+1},\ldots,U'_j\) act only on \(\cS\) and \(U'_j\cdots U'_{i+1}U_{\mathrm{prep}}\propto I_\cS\); use \(j=i\) if no such index exists\;
        \If{\(j>i\)}{
            \(V\gets\textbf{Use-first}(U'_1;\ldots;U'_j,\{a\})\)\;
        }
        Insert \(V\) after \(U'_j\) in \(U'\)\;
        \(\cA'\gets\cA'\setminus\{a\}\)\;
    }
    \Return{\(U'\)}\;
\end{algorithm}

The product of an empty sequence is the identity, so an identity \(U_{\mathrm{prep}}\) leaves the insertion point at \(i\).
For an ancilla that does not participate in the current circuit, use \(i=0\) and an empty prefix; position zero denotes the point before the first gate.
Whenever the insertion point moves, the inserted \(V\) is the full uncomputation returned by the new Use-first call.
This extension is a scheduling heuristic; the evaluation below independently verifies each resulting circuit.

\subsection{Experimental Evaluation}
\paragraph{Benchmark scope.}
Table~\ref{tab:dependency-aware} contains the 19 original instances from Table~\ref{tab:benchmark} and 41 additional instances with expanded parameters.
For the first six original instances and gate teleportation, we apply Algorithm~\ref{alg:use-first} once to the full ancilla register.
For the remaining 53 instances, we use Algorithm~\ref{alg:dependency-aware}.
The reported \(n_A\) is the number of ancillas allocated in the input circuit, rather than the peak number simultaneously in use.

\paragraph{Reqomp and Unqomp.}
Both tools retain the input qubit interface.
Reqomp receives the full input ancilla budget and runs in lazy mode (\texttt{force\_lazy=True}), with ancilla reuse and relative-phase Toffoli optimization disabled (\texttt{reuse\_ancillas=False} and \texttt{optimize\_ccx=False}).
Unqomp runs with ancilla linking disabled (\texttt{linking=False}).
For Adder and the expanded Multiplier instance, it retains the authors' existing carry helpers as atomic \texttt{qfree} gates~\cite[Sec.~6.2]{unqomp}.
These settings evaluate synthesis at the supplied ancilla allocation; they do not measure Reqomp's trade-off between space and gate count under smaller ancilla budgets.

\paragraph{RwUn.}
We use the authors' public implementation associated with artifact v2~\cite{RwUnArtifactV2}, with the same configuration as in Table~\ref{tab:benchmark}.
The modes Sequential, Reverse, Jointly, and Lifetime are tried in that order, stopping at the first mode that returns a circuit.
The released recursive gate expansion and native optimization remain enabled, and we add no further rotation decomposition or encoder rewrite.
We independently verify the selected circuit; a verification failure does not trigger a search through the remaining modes.

\paragraph{Verification and gate counts.}
A run succeeds only if numerical verification confirms that, for every logical input with the ancillas initially zero, the resulting circuit implements the intended logical operation up to a single global phase and returns all ancillas to zero.
The global phase must be common to all logical inputs.
A check mark denotes a verified success, and gate counts are shown only for these runs.
Gate counts include both the computation and synthesized uncomputation, with each atomic instruction counted as one gate.
We apply no additional gate cancellation to the tool outputs.
The controlled-\(S\) decomposition and atomic carry-helper conventions are detailed below the table, since these affect comparisons between methods.

\paragraph{Results.}
Our implementation succeeds on 59 of the 60 instances, including all 53 instances evaluated with the dependency-aware extension.
The sole failure is gate teleportation, whose first CNOT combines nontrivial functionality with an ancilla-role change, as discussed in Section~\ref{sec:case}.
The corresponding success totals for Reqomp, Unqomp, and RwUn are 51, 48, and 36, respectively.

\begingroup
\small
\captionsetup{singlelinecheck=false}
\setlength{\tabcolsep}{0pt}
\setlength{\aboverulesep}{0pt}
\setlength{\belowrulesep}{0pt}
\renewcommand{\arraystretch}{1}
\setbox\strutbox=\hbox{\vrule height10.5pt depth5.5pt width0pt}
\setlength{\arrayrulewidth}{0.25pt}
\setlength{\LTleft}{0pt}
\setlength{\LTright}{\fill}
\newcommand{\depbenchheading}[1]{\raisebox{-8.15pt}[0pt][0pt]{\parbox[c]{0.38\textwidth}{\centering\bfseries #1}}}
\newcommand{\depbenchpair}[1]{\multicolumn{2}{@{}c@{\hspace{6pt}}}{\strut\begin{adjustbox}{max width=\dimexpr0.155\textwidth-6pt\relax}#1\end{adjustbox}}}
\newcommand{\depbenchexample}[2]{\rule[-8.5pt]{0pt}{22pt}\begin{minipage}[c]{\dimexpr\linewidth-5pt\relax}\raggedright\begin{adjustbox}{max width=\linewidth}#1\end{adjustbox}\\[-1pt]\begin{adjustbox}{max width=\linewidth}\scriptsize #2\end{adjustbox}\end{minipage}}
\begin{longtable}{@{}>{\raggedright\arraybackslash}m{0.38\textwidth}*{4}{>{\centering\arraybackslash}m{\dimexpr0.0775\textwidth-4.5pt\relax}@{\hspace{3pt}}>{\centering\arraybackslash}m{\dimexpr0.0775\textwidth-4.5pt\relax}@{\hspace{6pt}}}@{}}
\caption{Synthesis success and gate counts for the 19 original and 41 expanded instances.}\label{tab:dependency-aware}\\
\toprule
\depbenchheading{Examples} & \depbenchpair{\textbf{Our Strategy}} & \depbenchpair{\textbf{Reqomp~\cite{reqompArtifact}}} & \depbenchpair{\textbf{Unqomp~\cite{unqomp}}} & \depbenchpair{\textbf{RwUn~\cite{RwUn}}} \\
\cmidrule(lr){2-3}\cmidrule(lr){4-5}\cmidrule(lr){6-7}\cmidrule(lr){8-9}
 & \depbenchpair{\footnotesize\textbf{Success}\hspace{3pt}\textbf{\# gates}} & \depbenchpair{\footnotesize\textbf{Success}\hspace{3pt}\textbf{\# gates}} & \depbenchpair{\footnotesize\textbf{Success}\hspace{3pt}\textbf{\# gates}} & \depbenchpair{\footnotesize\textbf{Success}\hspace{3pt}\textbf{\# gates}} \\
\midrule
\endfirsthead
\multicolumn{9}{l}{\textit{Table~\ref{tab:dependency-aware} continued.}}\\
\toprule
\depbenchheading{Examples} & \depbenchpair{\textbf{Our Strategy}} & \depbenchpair{\textbf{Reqomp~\cite{reqompArtifact}}} & \depbenchpair{\textbf{Unqomp~\cite{unqomp}}} & \depbenchpair{\textbf{RwUn~\cite{RwUn}}} \\
\cmidrule(lr){2-3}\cmidrule(lr){4-5}\cmidrule(lr){6-7}\cmidrule(lr){8-9}
 & \depbenchpair{\footnotesize\textbf{Success}\hspace{3pt}\textbf{\# gates}} & \depbenchpair{\footnotesize\textbf{Success}\hspace{3pt}\textbf{\# gates}} & \depbenchpair{\footnotesize\textbf{Success}\hspace{3pt}\textbf{\# gates}} & \depbenchpair{\footnotesize\textbf{Success}\hspace{3pt}\textbf{\# gates}} \\
\midrule
\endhead
\midrule\multicolumn{9}{r}{\textit{Continued on the next page.}}\\
\endfoot
\bottomrule
\endlastfoot
\multicolumn{9}{l}{\textit{Original instances}}\\*
\depbenchexample{Fractional query~\cite{fractional-query}}{paper instance, $n_A=1$} & $\checkmark$ & 5 & $\times$ & \textemdash & $\times$ & \textemdash & $\times$ & \textemdash \\
\depbenchexample{Fourier-basis phase kickback~\cite{Fourier-basis}}{$3$ data, $3$ ancilla, $n_A=3$} & $\checkmark$ & 13 & $\times$ & \textemdash & $\times$ & \textemdash & $\times$ & \textemdash \\
\depbenchexample{Controlled-\(T\)~\cite{controlled-T}}{paper instance, $n_A=1$} & $\checkmark$ & 3 & $\times$ & \textemdash & $\times$ & \textemdash & $\times$ & \textemdash \\
\depbenchexample{Relative-phase-Toffoli $C^3X$~\cite{relative-phase-toffoli}}{paper instance, $n_A=1$} & $\checkmark$ & 19 & $\times$ & \textemdash & $\times$ & \textemdash & $\times$ & \textemdash \\
\depbenchexample{QSVT projector phase~\cite{QSVT}}{paper instance, $n_A=1$} & $\checkmark$ & 7 & $\times$ & \textemdash & $\times$ & \textemdash & $\times$ & \textemdash \\
\depbenchexample{Single-clock HHL~\cite{single-clock-HHL}}{paper instance, $n_A=1$} & $\checkmark$ & 7 & $\times$ & \textemdash & $\times$ & \textemdash & $\times$ & \textemdash \\
\depbenchexample{Adder~\cite{reqompArtifact}}{$n=2$, $n_A=2$} & $\checkmark$ & 12 & $\checkmark$ & 10 & $\checkmark$ & 8 & $\checkmark$ & 11 \\
\depbenchexample{WeightedAdder~\cite{reqompArtifact}}{$n=1,\;w=[3]$, $n_A=1$} & $\checkmark$ & 5 & $\checkmark$ & 5 & $\checkmark$ & 5 & $\checkmark$ & 6 \\
\depbenchexample{Multiplier~\cite{reqompArtifact}}{$n=1$, $n_A=2$} & $\checkmark$ & 4 & $\checkmark$ & 4 & $\checkmark$ & 4 & $\checkmark$ & 4 \\
\depbenchexample{IntegerComparator~\cite{reqompArtifact}}{$n=2,\;x\geq1$, $n_A=1$} & $\checkmark$ & 3 & $\checkmark$ & 3 & $\checkmark$ & 3 & $\checkmark$ & 8 \\
\depbenchexample{Clean Incrementer~\cite{RwUn}}{$n=3$, $n_A=1$} & $\checkmark$ & 5 & $\checkmark$ & 5 & $\checkmark$ & 5 & $\checkmark$ & 8 \\
\depbenchexample{MCX~\cite{reqompArtifact}}{$c=3$, $n_A=1$} & $\checkmark$ & 3 & $\checkmark$ & 3 & $\checkmark$ & 3 & $\checkmark$ & 3 \\
\depbenchexample{MCRY~\cite{reqompArtifact}}{$c=1,\;\theta=\pi/4$, $n_A=1$} & $\checkmark$ & 3 & $\checkmark$ & 3 & $\checkmark$ & 3 & $\times$ & \textemdash \\
\depbenchexample{PiecewiseLinearR~\cite{reqompArtifact}}{$n=1,\;b=1,\;s=o=\pi/4$, $n_A=1$} & $\checkmark$ & 7 & $\checkmark$ & 7 & $\checkmark$ & 7 & $\times$ & \textemdash \\
\depbenchexample{PolynomialPauliR~\cite{reqompArtifact}}{$n=2,\;d=2,\;a_j=\pi/4$, $n_A=1$} & $\checkmark$ & 6 & $\checkmark$ & 6 & $\checkmark$ & 6 & $\times$ & \textemdash \\
\depbenchexample{ConditionalCleanMCS~\cite{RwUn,ConditionalCleanMCS}}{$c=6$, $n_A=2$} & $\checkmark$ & 25 & $\checkmark$ & 17 & $\times$ & \textemdash & $\checkmark$ & 26 \\
\depbenchexample{Deutsch--Jozsa~\cite{reqompArtifact}}{$n=3$, $n_A=1$} & $\checkmark$ & 11 & $\checkmark$ & 11 & $\checkmark$ & 11 & $\checkmark$ & 9 \\
\depbenchexample{Grover~\cite{reqompArtifact}}{$n=3,\;s=1$, $n_A=2$} & $\checkmark$ & 41 & $\checkmark$ & 41 & $\checkmark$ & 41 & $\checkmark$ & 37 \\
\depbenchexample{Gate teleportation~\cite{gate-teleportation}}{paper instance, $n_A=1$} & $\times$ & \textemdash & $\times$ & \textemdash & $\times$ & \textemdash & $\times$ & \textemdash \\
\midrule
\multicolumn{9}{l}{\textit{Expanded parameter instances}}\\*
\depbenchexample{Adder~\cite{reqompArtifact}$^{\dagger}$}{$n=3$, $n_A=3$} & $\checkmark$ & 22 & $\checkmark$ & 18 & $\checkmark$ & 14 & $\checkmark$ & 35 \\
\depbenchexample{Adder~\cite{reqompArtifact}$^{\dagger}$}{$n=4$, $n_A=4$} & $\checkmark$ & 32 & $\checkmark$ & 26 & $\checkmark$ & 20 & $\checkmark$ & 104 \\
\depbenchexample{WeightedAdder~\cite{reqompArtifact}}{$n=2,\;w=[1,1]$, $n_A=2$} & $\checkmark$ & 8 & $\checkmark$ & 8 & $\checkmark$ & 8 & $\checkmark$ & 11 \\
\depbenchexample{WeightedAdder~\cite{reqompArtifact}}{$n=1,\;w=[7]$, $n_A=3$} & $\checkmark$ & 19 & $\checkmark$ & 17 & $\times$ & \textemdash & $\checkmark$ & 24 \\
\depbenchexample{WeightedAdder~\cite{reqompArtifact}}{$n=3,\;w=[1,1,1]$, $n_A=3$} & $\checkmark$ & 12 & $\checkmark$ & 12 & $\checkmark$ & 12 & $\checkmark$ & 18 \\
\depbenchexample{WeightedAdder~\cite{reqompArtifact}}{$n=4,\;w=[1,1,1,1]$, $n_A=12$} & $\checkmark$ & 36 & $\checkmark$ & 36 & $\checkmark$ & 36 & $\checkmark$ & 84 \\
\depbenchexample{WeightedAdder~\cite{reqompArtifact}}{$n=2,\;w=[7,7]$, $n_A=10$} & $\checkmark$ & 64 & $\checkmark$ & 56 & $\times$ & \textemdash & $\checkmark$ & 102 \\
\depbenchexample{Multiplier~\cite{reqompArtifact}$^{\dagger}$}{$n=2$, $n_A=8$} & $\checkmark$ & 30 & $\checkmark$ & 26 & $\checkmark$ & 22 & $\checkmark$ & 37 \\
\depbenchexample{IntegerComparator~\cite{reqompArtifact}}{$n=3,\;x\geq1$, $n_A=2$} & $\checkmark$ & 5 & $\checkmark$ & 5 & $\checkmark$ & 5 & $\checkmark$ & 17 \\
\depbenchexample{IntegerComparator~\cite{reqompArtifact}}{$n=4,\;x\geq1$, $n_A=3$} & $\checkmark$ & 7 & $\checkmark$ & 7 & $\checkmark$ & 7 & $\checkmark$ & 34 \\
\depbenchexample{IntegerComparator~\cite{reqompArtifact}}{$n=5,\;x\geq1$, $n_A=4$} & $\checkmark$ & 9 & $\checkmark$ & 9 & $\checkmark$ & 9 & $\checkmark$ & 67 \\
\depbenchexample{IntegerComparator~\cite{reqompArtifact}}{$n=6,\;x\geq1$, $n_A=5$} & $\checkmark$ & 11 & $\checkmark$ & 11 & $\checkmark$ & 11 & $\checkmark$ & 132 \\
\depbenchexample{Clean Incrementer~\cite{RwUn}}{$n=4$, $n_A=2$} & $\checkmark$ & 8 & $\checkmark$ & 8 & $\checkmark$ & 8 & $\checkmark$ & 20 \\
\depbenchexample{Clean Incrementer~\cite{RwUn}}{$n=5$, $n_A=3$} & $\checkmark$ & 11 & $\checkmark$ & 11 & $\checkmark$ & 11 & $\checkmark$ & 44 \\
\depbenchexample{Clean Incrementer~\cite{RwUn}}{$n=6$, $n_A=4$} & $\checkmark$ & 14 & $\checkmark$ & 14 & $\checkmark$ & 14 & $\checkmark$ & 92 \\
\depbenchexample{MCX~\cite{reqompArtifact}}{$c=4$, $n_A=2$} & $\checkmark$ & 5 & $\checkmark$ & 5 & $\checkmark$ & 5 & $\checkmark$ & 6 \\
\depbenchexample{MCX~\cite{reqompArtifact}}{$c=5$, $n_A=3$} & $\checkmark$ & 7 & $\checkmark$ & 7 & $\checkmark$ & 7 & $\checkmark$ & 10 \\
\depbenchexample{MCX~\cite{reqompArtifact}}{$c=6$, $n_A=4$} & $\checkmark$ & 9 & $\checkmark$ & 9 & $\checkmark$ & 9 & $\checkmark$ & 15 \\
\depbenchexample{MCX~\cite{reqompArtifact}}{$c=8$, $n_A=6$} & $\checkmark$ & 13 & $\checkmark$ & 13 & $\checkmark$ & 13 & $\checkmark$ & 28 \\
\depbenchexample{MCRY~\cite{reqompArtifact}}{$c=3,\;\theta=\pi/4$, $n_A=2$} & $\checkmark$ & 5 & $\checkmark$ & 5 & $\checkmark$ & 5 & $\times$ & \textemdash \\
\depbenchexample{MCRY~\cite{reqompArtifact}}{$c=4,\;\theta=\pi/4$, $n_A=3$} & $\checkmark$ & 7 & $\checkmark$ & 7 & $\checkmark$ & 7 & $\times$ & \textemdash \\
\depbenchexample{MCRY~\cite{reqompArtifact}}{$c=5,\;\theta=\pi/4$, $n_A=4$} & $\checkmark$ & 9 & $\checkmark$ & 9 & $\checkmark$ & 9 & $\times$ & \textemdash \\
\depbenchexample{MCRY~\cite{reqompArtifact}}{$c=6,\;\theta=\pi/4$, $n_A=5$} & $\checkmark$ & 11 & $\checkmark$ & 11 & $\checkmark$ & 11 & $\times$ & \textemdash \\
\depbenchexample{PiecewiseLinearR~\cite{reqompArtifact}}{$n=2,\;b=1,\;s=o=\pi/4$, $n_A=2$} & $\checkmark$ & 13 & $\checkmark$ & 13 & $\checkmark$ & 13 & $\times$ & \textemdash \\
\depbenchexample{PiecewiseLinearR~\cite{reqompArtifact}}{$n=3,\;b=1,\;s=o=\pi/4$, $n_A=3$} & $\checkmark$ & 19 & $\checkmark$ & 19 & $\checkmark$ & 19 & $\times$ & \textemdash \\
\depbenchexample{PiecewiseLinearR~\cite{reqompArtifact}}{$n=4,\;b=1,\;s=o=\pi/64$, $n_A=4$} & $\checkmark$ & 25 & $\checkmark$ & 25 & $\checkmark$ & 25 & $\times$ & \textemdash \\
\depbenchexample{PiecewiseLinearR~\cite{reqompArtifact}}{$n=5,\;b=1,\;s=o=\pi/64$, $n_A=5$} & $\checkmark$ & 31 & $\checkmark$ & 31 & $\checkmark$ & 31 & $\times$ & \textemdash \\
\depbenchexample{PolynomialPauliR~\cite{reqompArtifact}$^{\dagger}$}{$n=3,\;d=2,\;a_j=\pi/4$, $n_A=3$} & $\checkmark$ & 13 & $\checkmark$ & 13 & $\checkmark$ & 13 & $\times$ & \textemdash \\
\depbenchexample{PolynomialPauliR~\cite{reqompArtifact}$^{\dagger}$}{$n=3,\;d=3,\;a_j=\pi/4$, $n_A=5$} & $\checkmark$ & 18 & $\checkmark$ & 18 & $\checkmark$ & 18 & $\times$ & \textemdash \\
\depbenchexample{PolynomialPauliR~\cite{reqompArtifact}}{$n=3,\;d=2,\;a_j=\pi/64$, $n_A=3$} & $\checkmark$ & 13 & $\checkmark$ & 13 & $\checkmark$ & 13 & $\times$ & \textemdash \\
\depbenchexample{PolynomialPauliR~\cite{reqompArtifact}}{$n=3,\;d=3,\;a_j=\pi/64$, $n_A=5$} & $\checkmark$ & 18 & $\checkmark$ & 18 & $\checkmark$ & 18 & $\times$ & \textemdash \\
\depbenchexample{PolynomialPauliR~\cite{reqompArtifact}}{$n=4,\;d=2,\;a_j=\pi/64$, $n_A=6$} & $\checkmark$ & 23 & $\checkmark$ & 23 & $\checkmark$ & 23 & $\times$ & \textemdash \\
\depbenchexample{PolynomialPauliR~\cite{reqompArtifact}}{$n=4,\;d=3,\;a_j=\pi/256$, $n_A=14$} & $\checkmark$ & 43 & $\checkmark$ & 43 & $\checkmark$ & 43 & $\times$ & \textemdash \\
\depbenchexample{ConditionalCleanMCS~\cite{RwUn,ConditionalCleanMCS}$^{\ddagger}$}{$c=7$, $n_A=2$} & $\checkmark$ & 37 & $\times$ & \textemdash & $\times$ & \textemdash & $\checkmark$ & 26 \\
\depbenchexample{ConditionalCleanMCS~\cite{RwUn,ConditionalCleanMCS}$^{\ddagger}$}{$c=8$, $n_A=2$} & $\checkmark$ & 37 & $\times$ & \textemdash & $\times$ & \textemdash & $\checkmark$ & 26 \\
\depbenchexample{Deutsch--Jozsa~\cite{reqompArtifact}}{$n=4$, $n_A=2$} & $\checkmark$ & 15 & $\checkmark$ & 15 & $\checkmark$ & 15 & $\checkmark$ & 11 \\
\depbenchexample{Deutsch--Jozsa~\cite{reqompArtifact}}{$n=5$, $n_A=3$} & $\checkmark$ & 19 & $\checkmark$ & 19 & $\checkmark$ & 19 & $\checkmark$ & 13 \\
\depbenchexample{Deutsch--Jozsa~\cite{reqompArtifact}}{$n=6$, $n_A=4$} & $\checkmark$ & 23 & $\checkmark$ & 23 & $\checkmark$ & 23 & $\checkmark$ & 15 \\
\depbenchexample{Grover~\cite{reqompArtifact}}{$n=4,\;s=1$, $n_A=9$} & $\checkmark$ & 84 & $\checkmark$ & 84 & $\checkmark$ & 84 & $\checkmark$ & 66 \\
\depbenchexample{Grover~\cite{reqompArtifact}}{$n=5,\;s=1$, $n_A=20$} & $\checkmark$ & 143 & $\checkmark$ & 143 & $\checkmark$ & 143 & $\checkmark$ & 103 \\
\depbenchexample{Grover~\cite{reqompArtifact}}{$n=6,\;s=1$, $n_A=42$} & $\checkmark$ & 260 & $\checkmark$ & 260 & $\checkmark$ & 260 & $\checkmark$ & 176 \\
\midrule
\textbf{Verified successes} & 59/60 & \textemdash & 51/60 & \textemdash & 48/60 & \textemdash & 36/60 & \textemdash \\
\end{longtable}
\endgroup

\begingroup
\footnotesize
\noindent\textit{Parameters.}
\(n_A\) denotes the number of allocated ancillas.
For piecewise rotations, \(b\) is the breakpoint, and \(s\) and \(o\) are the slope and offset.
For polynomial rotations, \(d\) is the degree, and every coefficient has the displayed value \(a_j\).
\par\smallskip
\noindent\(\dagger\)~At least one uncomputation round inserts no gate, has no reachable state in which the selected ancilla differs from \(|0\>\), or implements no nonidentity Use operation.
These instances are retained as regression tests.
\par\smallskip
\noindent\(\ddagger\)~For ConditionalCleanMCS with seven or eight controls, the input retains the gate order produced by the constructor because Reqomp's input-preparation step rejects these parameter values.
The Reqomp failures on these two instances arise from an implementation bug: its control-value comparison stops after the first candidate control, which does not affect the six-control instance (\(c=6\)) because the controlled gates that temporarily modify and restore its primary qubits are single-control \(\mathrm{CNOT}\) gates.
\par\smallskip
\noindent\textit{Gate-set conventions.}
For every ConditionalCleanMCS instance, RwUn receives the equivalent decomposition of controlled-\(S\) into \(\mathrm{CNOT}\), \(T\), and \(T^\dagger\) gates; the other methods retain atomic controlled-\(S\) gates.
Each atomic instruction counts as one gate, so these entries use different gate sets.
For Adder and the expanded Multiplier instance, each carry helper retained by Unqomp also counts as one instruction~\cite[Sec.~6.2]{unqomp}.
\par
\endgroup

\end{document}